\documentclass[fleqn,usenatbib]{mnras}

\usepackage{newtxtext,newtxmath}

\usepackage[T1]{fontenc}

\DeclareRobustCommand{\VAN}[3]{#2}
\let\VANthebibliography\thebibliography
\def\thebibliography{\DeclareRobustCommand{\VAN}[3]{##3}\VANthebibliography}

\usepackage{graphicx}	
\usepackage{amsmath}	
\usepackage{dirtytalk} 
\usepackage{booktabs}
\usepackage{adjustbox} 
\usepackage{array}
\usepackage{ragged2e}

\RequirePackage{listings} 
\definecolor{template_blue}{HTML}{002db3}   
\definecolor{subtitle}{cmyk}{0,0,0,0}       
\definecolor{template_text}{HTML}{434655}   
\definecolor{template_lightgrey}{HTML}{A8AABC}  
\definecolor{template_mediumgrey}{HTML}{747687}
\definecolor{template_pink}{HTML}{BFA5A6}
\definecolor{template_lightred}{HTML}{A54653}
\definecolor{template_darkred}{HTML}{6C0F26}
\definecolor{template_beige}{HTML}{857555}
\definecolor{template_yellow}{HTML}{D2A517}
\hypersetup{colorlinks,breaklinks,linkcolor=black,citecolor=template_blue,urlcolor=template_darkred}

\usepackage{algorithm}

\usepackage{algpseudocode}

\title[Choosing the right MCMC sampler]{Choosing the right MCMC sampler: a systematic benchmark of gradient-free methods}

\author[C. M. Poppelaars \& M. P. van Daalen]{
Colin M. Poppelaars$^{1}$\thanks{E-mail: colin.poppelaars@gmail.com}
\& Marcel P. van Daalen$^{1}$
\\
$^{1}$Leiden Observatory, Leiden University, PO Box 9513, NL-2300 RA Leiden, The Netherlands}

\date{Accepted XXX. Received YYY; in original form ZZZ}

\pubyear{\the\year{}}

\begin{document}
\label{firstpage}
\pagerange{\pageref{firstpage}--\pageref{lastpage}}
\maketitle

\begin{abstract}
    We present a set of metrics and methods for testing and comparing a range of modern gradient-free Markov Chain Monte Carlo (MCMC) samplers against the commonly used Metropolis-Hastings (MH) algorithm. The goal is to quantify key performance metrics, including sampler ergodicity, robustness and overall likelihood performance. To provide a controlled and interpretable testbed, we use the Rosenbrock function and Neal's funnel as representative unimodal cases, while Gaussian random likelihood landscapes in three, five, and eight dimensions serve as multimodal test scenarios. The samplers considered include affine-invariant moves from the literature, such as the stretch and walk moves, the differential evolution move, and the snooker move. We additionally introduce two novel variations: a modified stretch move that incorporates a Principal Component Analysis (PCA) transformation, and a hybrid blend move that combines features of both differential evolution and stretch dynamics. Beyond sampler evaluation, we demonstrate reconstructing likelihood landscapes from sampled points using a quadtree algorithm. Additionally, we explore the use of optimisation algorithms to refine the best parameter set in terms of its likelihood, and find consistent improvements in log-likelihood values, with the post-sampling gain becoming more significant in higher-dimensional problems. Our comparative results of sampler testing show that the differential evolution algorithm, when tuned to a target acceptance fraction of 25\%, consistently outperforms all other samplers in terms of ergodicity, robustness, and likelihood performance.
\end{abstract}

\begin{keywords}
    methods: data analysis -- methods: numerical -- methods: statistical 
\end{keywords}



\section{Introduction}\label{sec: introduction}

\begin{figure}
    \centering
    \includegraphics[width=0.5\linewidth, trim=40mm 6mm 40mm 6mm]{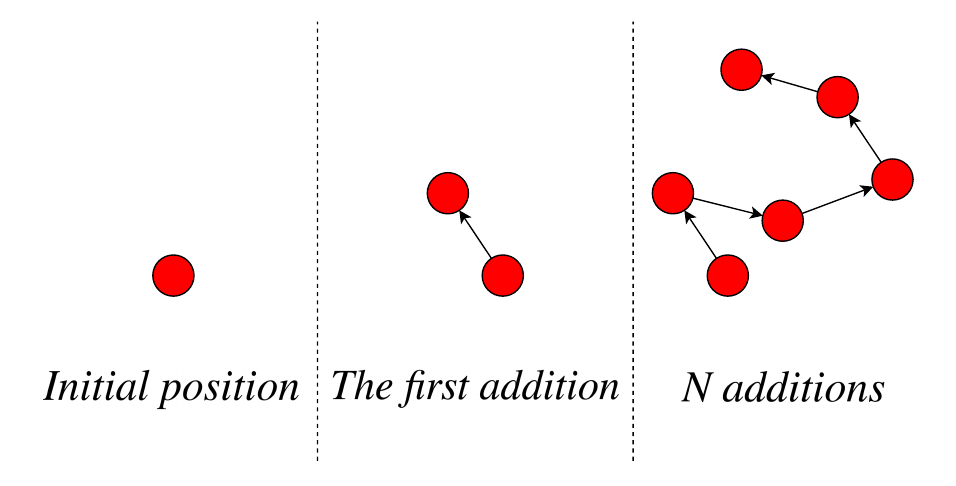}
    \caption{An illustration of a Markov chain. The chain starts from an initial position, from which proposals are created. Once a proposal is accepted based on the acceptance criterion, the sample is added to the chain. This process is repeated for $N$ times until the chain length is reached or if another termination cause has been reached. Here $N=5$.}
    \label{fig:markov_chain_illustrated}
\end{figure}

Markov Chain Monte Carlo (MCMC) methods are a class of algorithms used for sampling complex probability distributions. They are particularly useful in Bayesian inference, where one seeks to explore the posterior distribution over model parameters given observed data. This is often required when the posterior is not analytically tractable, particularly in models with many parameters. Rather than returning a single point estimate, MCMC methods generate a sequence of correlated samples from the target distribution. These samples collectively approximate the posterior and allow for robust estimation of not only the most probable parameter values but also their associated uncertainties. This sets MCMC apart from standard optimisation techniques, which typically aim to optimise a loss function to obtain a best-fit solution without quantifying uncertainty. MCMC is particularly powerful in exploring high-dimensional parameter spaces, where grid-search-based optimisation is computationally prohibitive, as the number of possible parameter sets grows infinitely large. The name “Markov Chain Monte Carlo” reflects its two fundamental components: a Markov Chain and a Monte Carlo part. The Markov chain refers to the sequence of samples generated by the algorithm. Each sample depends only on the previous sample. This property is known as the Markov property. It implies that the states in the chain are memoryless. Therefore, future states depend only on the current state and are conditionally independent of all earlier states. We illustrate the idea of a Markov chain in figure~\ref{fig:markov_chain_illustrated}. It starts from an initial position, after which it adds new elements, resulting in a chain. The Monte Carlo part highlights the use of random sampling techniques to perform numerical approximations. Monte Carlo simulations involve repeated random draws to probabilistically estimate quantities that are often analytically intractable. In MCMC, this is used for stochastic exploration of the parameter space through random proposals and acceptance criteria. 

The core idea of MCMC is to construct a Markov chain whose equilibrium distribution matches the target posterior distribution. This is achieved by evaluating the likelihood function at successive points in parameter space and accepting or rejecting proposed samples according to a criterion that ensures that the chain satisfies detailed balance. Detailed balance means that the transition probability from a state $\vec{X}$ to $\vec{X}^{\text{new}}$ is equal to the reverse transition probability. When detailed balance holds, the posterior remains unchanged under the chain’s dynamics. By satisfying detailed balance, the Markov chain is ergodic. This implies that the long-run frequency with which each state is visited becomes proportional to its posterior probability. In practice, however, the chain must first move from its initial state toward regions of high posterior probability. This is known as the burn-in period. During this initial phase the samples are strongly influenced by the starting point and may therefore not yet represent the target distribution. To avoid biasing statistical estimates, the samples generated during the burn-in are discarded before analysing the chain. After this initial phase, the chain is expected to have converged sufficiently close to the stationary distribution, such that the remaining samples can be considered representative draws from the posterior.

MCMC methods have become a central tool in modern astrophysics, particularly in exploring high-dimensional parameter spaces. Semi-analytic galaxy formation models such as L-Galaxies \citep{LGAL2020Henriques} often contain dozens of coupled parameters describing processes like star formation, feedback processes, gas cooling and hydrodynamics. This makes efficient sampling techniques essential for parameter calibration and uncertainty estimation. Beyond galaxy formation, MCMC techniques are also widely used in exoplanet science \citep[e.g.][]{DE_Nelson_2014}, where they enable the inference of orbital parameters, planetary masses, and atmospheric properties from noisy observational data. Many other examples of uses of MCMC methods in other astrophysical fields, such as cosmological analysis, can be found in e.g.\ \citet{Trotta2017,Sharma2017}. Improvements in sampling strategies, therefore, have direct relevance for a broad range of astrophysical applications where robust statistical inference is required.

Broadly speaking, MCMC samplers fall into two categories: those that use gradient information and those that do not. Gradient-based samplers are generally more efficient in high-dimensional spaces due to their ability to follow the structure of the posterior. However, their applicability is limited to models where gradient information is available. Consequently, gradient-free methods are more widely applicable, but these typically suffer from reduced sampling efficiency. A foundational algorithm of a gradient-based sampler is Hamiltonian Monte Carlo (HMC), which uses Hamiltonian dynamics to propose new samples. By simulating trajectories along the gradient of the posterior, HMC enables efficient traversal of the parameter space, reducing random walk behaviour and improving convergence to the target posterior. This makes it especially effective in high-dimensional problems \citep[e.g.][]{Neal2011HMC}. Beyond HMC, recent innovations in this class include non-reversible samplers based on piecewise deterministic Markov processes (PDMPs). These samplers introduce deterministic motions interrupted by stochastic events. The idea is that between events, the trajectories of sampling are deterministically understood. This leads to continuous-time dynamics that improve mixing and eliminate trajectory tuning. Notable methods include the Zig-Zag sampler \citep{BierkensRobertsZigZag2019}, the Boomerang sampler \citep{Bierkens2020}, and the Bouncy Particle Sampler \citep{BouchardCoteVollmerDoucet2018}.

In parallel, development in gradient-free MCMC methods has also advanced considerably since the foundational work of \citet{Metropolis1953}. Many efforts have aimed to improve robustness, sampling efficiency, and scalability in the absence of gradient information. A promising direction in this context involves the use of machine learning techniques to overcome classical limitations of gradient-free MCMC. One example is Posterior Contouring Monte Carlo \citep[\texttt{pocoMC},][]{Karamanis2022}, which combines normalising flows with traditional MCMC sampling. By learning an invertible transformation of the posterior into a simpler latent space, pocoMC enables efficient proposals that adapt to complex distributions, including those that are multimodal or highly anisotropic.

A related approach is flow-based MCMC, where normalising flows are trained to flatten the geometry of the posterior. These flows are often incorporated into a Metropolis–Hastings framework as flexible proposal distributions, preserving the correctness of the chain while benefiting from the expressivity of the learned transformation. Hybrid strategies such as surrogate-accelerated MCMC further extend this idea by introducing emulators, such as neural networks or Gaussian processes, that approximate the likelihood function. This is especially useful when model evaluations are computationally expensive.

There has also been development in adaptive and ensemble-based samplers. Adaptive methods incorporate feedback mechanisms that adjust proposal scales or directions in response to the evolving behaviour of the chain \citep[e.g.][]{Haario1999, Haario2001}. Ensemble samplers, by contrast, operate with a population of (parallel) chains, using their collective state to inform new proposals. A widely used implementation is the \texttt{EMCEE} package, based on the affine-invariant ensemble sampler from \citet{goodman2010}, and commonly employed in astrophysical parameter inference \citep{Foreman_Mackey_2013}. In addition to \texttt{EMCEE}'s stretch move, there exists a whole collection of ensemble methods. Examples include the differential evolution algorithm of \citet{Braak2006, DE_Nelson_2014}, the snooker sampler of \citet{gilkssnooker1994, terBraak2008}, the walk move \citep{goodman2010}, and the t-walk move \citep{ChristenFox2007}.

The distinction between MCMC algorithms has profound implications for algorithmic efficiency and scalability. Gradient-based samplers leverage the gradient of the log posterior distribution to inform their proposal mechanisms. By exploiting local geometric information, they can traverse the parameter space more efficiently than their gradient-free counterparts. Gradient information allows the sampler to follow the contours of regions with high posterior density, rather than taking uninformed random walks typical of gradient-free methods \citep{Betancourt2017HMC}. In continuous sample spaces, the gradient can be interpreted as defining a vector field that is aligned with the posterior landscape. If this field is well-aligned with the underlying geometry of the target distribution, it can guide the sampler towards high-probability regions, replacing inefficient random walk behaviour with directionally-biased movement. Repeated application of such informed steps generates trajectories that are more efficient for exploring the posterior. This often results in such samplers requiring fewer samples to capture the target distribution accurately \citep[e.g.][]{Neal2011HMC, Betancourt2017HMC}. However, to effectively exploit gradient information, these methods typically require additional geometric structure to remove dependencies on the specific parametrisation of the model. This is necessary because the raw gradient alone is often not sufficient; its sensitivity must be correctly aligned with the local curvature of the posterior \citep{Betancourt2017HMC}. Additionally, they are also more complex to implement; improper settings can lead to inefficient sampling or instability. They may suffer from drifting, which is the result of numerical error accumulation. This drift refers to the growing discrepancy between the true posterior trajectory and the numerical approximation produced by the sampler. Such drift becomes more severe in very high-dimensional spaces and can degrade the sampler’s ability to accurately represent the target distribution.

Contrary to gradient-based samplers, gradient-free methods do not require derivative information. Instead, they generate proposals through random-walk behaviour. Gradient-free methods are much simpler to implement since they require only a black-box evaluation of the posterior. However, these methods suffer from several challenges, which are particularly apparent in high-dimensional spaces. The random-walk proposals of these methods tend to scale poorly with dimensionality, which leads to high autocorrelation and slow exploration from the chains \citep[e.g.][]{Roberts1997}. Additionally, these methods often require careful tuning of the proposal distribution width, which can be non-trivial in multimodal posteriors or parameter spaces with narrow, high-curvature regions \citep[e.g.][]{Christen2005}. Such posteriors don't have a single value that fits well for all their space. Adaptive schemes can alleviate some of these issues, but they often lag behind the efficiency of gradient-based methods in smooth, high-dimensional settings.

When gradient information is not directly available, one possible strategy is to approximate the posterior with a surrogate model that is differentiable. \citet{Christen2005} propose a two-stage Metropolis-Hastings algorithm in which an inexpensive surrogate is used to approximate the posterior and pre-screen proposed samples before evaluating the full, expensive posterior. While their original method does not require gradients, this framework naturally allows the use of a differentiable surrogate. By choosing a differentiable surrogate function, approximate gradient information can be extracted and used to inform proposal mechanisms such as Hamiltonian Monte Carlo. In this way, gradient-free samplers can be enhanced with surrogate-derived gradient information without modifying the underlying expensive model. However, such modifications can be technically demanding and computationally expensive. Consequently, the gains of switching to a gradient-based sampler might not outweigh the costs. 

The aim of the current work is to to define key performance metrics for gradient-free MCMC samplers, in order to determine which samplers converge most quickly or reliably for a given likelihood landscape, and/or can most efficiently explore it. This allows for a more informed sampler choice in situations where no gradient is available, and can lead to order-of-magnitude improvements in runtime in high-dimensional settings, which are not uncommon in astronomy. \\\\
This paper is structured as follows. In section \ref{sec:CH2}, we introduce the foundational algorithm of MCMC sampling, the Metropolis-Hastings algorithm \citet{Hastings1970}. Following this, we shall focus on variations of this algorithm as well as the collection of \textit{affine-invariant ensemble samplers}. After introducing all samplers included in this paper, we turn to our toy models in section \ref{sec:CH3}. These are used to compare and rank the samplers. We explore two kinds of toy models: unimodal and multimodal models. Our unimodal models consist of the Rosenbrock \citet{rosenbrock1960automatic} and Neal's funnel \citet{Neal2003Slice} functions. For our multimodal test cases, we introduce the concept of $N$-dimensional Gaussian landscapes. Also in section \ref{sec:CH3}, we demonstrate the reconstruction of our multimodal likelihood landscapes through an application of a quadtree algorithm. Next, in section \ref{sec:CH4}, we put our samplers to the test using various metrics regarding the ergodicity, convergence, performance and robustness. We introduce an ergodicity metric that aims to compare cell occupation from sampling our multimodal landscapes to the expectation values of these cells. Additionally, we also introduce performance and robustness metrics concerning the likelihood estimations from the samplers. The performance metric maps the median performance over many random seed runs, whereas the robustness metric maps the discrepancy between the best and worst performing chain populations of samplers over these runs. Finally, we summarise and discuss our findings in section \ref{sec:CH5}, where we also outline future directions of the research field. 

\section{MCMC algorithms}\label{sec:CH2}
\subsection{Classic MCMC}
\subsubsection{The Metropolis-Hastings (MH) algorithm}
\label{sec:MH_algorithm}
The basic algorithm of MCMC sampling is the Metropolis-Hastings (MH) algorithm, which followed from the development of the foundational \citet{Metropolis1953} algorithm. The MH algorithm proposes samples from a target distribution using random proposals based on the current state of the Markov chain. Steps are either accepted or rejected based on their consistency with the detailed balance condition. The Metropolis algorithm assumes symmetric proposal distributions, allowing for a simple acceptance ratio that only depends on the posterior:
\begin{equation}\label{eq:Metropolis_acceptance_criterion}
    \alpha = \min \left(1, \frac{P(Y \mid \vec{X_{i}}^{\text{new}}) P(\vec{X_{i}}^{\text{new}})}{P(Y \mid \vec{X}_{i}) P(\vec{X}_{i}) } \right) = \text{min}\left(1,\;\frac{\pi(\vec{X}_{i}^{\text{new}})}{\pi(\vec{X}_{i})}\right),
\end{equation}
where $\vec{X}_{i}$ represents the model parameters and $Y$ the data against which the model is tested. $P(\vec{X}_{i})$ represents our prior information about the model parameters, whereas $\pi$ represents the posterior distribution. $P(Y\;|\;\vec{X_{i}})$ is the likelihood of observing data $x$ given the current model $\vec{X_{i}}$.

A uniform random number between 0 and 1 can then be drawn and compared to $\alpha$, in order to either accept or reject the proposed sample. \citet{Hastings1970} later generalised this to allow for \emph{asymmetric} proposal distributions, leading to the MH algorithm. This includes a correction for the asymmetry via the ratio of proposal probabilities, as shown in equation~\eqref{eq:MH_acceptance_criterion}.

\begin{equation}\label{eq:MH_acceptance_criterion}
    \alpha = \text{min}\left(1,\;\frac{\pi(\vec{X}_{i}^{\text{new}})}{\pi(\vec{X_{i}})}\frac{q(\vec{X_{i}}\;|\;\vec{X_{i}}^{\text{new}})}{q(\vec{X_{i}}^{\text{new}}\;|\;\vec{X_{i}})}\right).
\end{equation}

The loop inside the MH algorithm defines the core of all MCMC methods. The algorithm starts with initialisation, followed by proposal generation, acceptance testing, and repetition of the generation and acceptance testing. The chain is typically terminated after a predefined number of steps, a fixed computation time, or when a convergence diagnostic is satisfied.

\begin{figure}
    \centering
    \includegraphics[width=0.5\linewidth, trim=6mm 6mm 6mm 6mm]{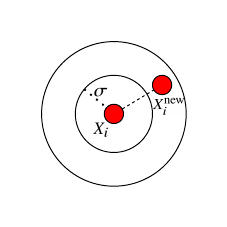}
    \caption{A Gaussian Metropolis-Hastings move for current chain $\vec{X_{i}}$. The move is based on centring a Gaussian around the current position $\vec{X_{i}}$ with a standard deviation of $\sigma$. From this, a new point is sampled, $\vec{X_{i}}^{\text{new}}$.}
    \label{fig:Ch2_MH_move_illustrated}
\end{figure}

Typically, a Gaussian proposal distribution is used for proposals in the MH algorithm. This is centred at the current position $\vec{X_{i}}$, with standard deviation $\sigma$. This symmetric choice simplifies the acceptance ratio by removing the proposal terms, reducing the acceptance probability to the original Metropolis form. We illustrate this move in figure~\ref{fig:Ch2_MH_move_illustrated}. The choice of $\sigma$ significantly affects the efficiency of the sampling process. If $\sigma$ is too small, proposed moves deviate slightly from the current position, leading to slow exploration and thus inefficient sampling. If $\sigma$ is too large, most proposed moves land in low-probability regions and are rejected, again reducing efficiency. Both extremes are undesired. Ultimately, we aim to find a proposal distribution that maximises the algorithm's efficiency. Unfortunately, this fine-tuning is problem-specific, and it isn't easy to find the best proposal width or distribution. A rich body of literature addresses the tuning of this parameter and distribution, see for instance \citet{Gelman1996, Roberts1997, Haario1999, Roberts2001, Haario2001, Atchade2005, brooks2011handbook}.

Despite the simplicity of the MH algorithm, it struggles in high-dimensional spaces, where it is difficult to identify a proposal distribution that effectively balances acceptance and exploration. This challenge is often referred to as the \emph{curse of dimensionality}. Additionally, MH may perform poorly in scenarios involving strong correlations among parameters, as its proposal width is uniform across all parameters. These limitations have motivated the development of more specialised sampling strategies, including adaptive and ensemble-based methods.

\subsubsection{Metropolis-Hastings with standardised scaling (MH SS)}
In high-dimensional problems, isotropic Gaussian proposals (or other symmetric forms) assume similar scaling across dimensions. As such, parameters with small ranges might dominate the proposed moves, while large-range parameters receive poor exploration. To reduce the sensitivity to the different scales on which parameters are sampled, it is beneficial to scale every parameter prior to the same range before applying an MCMC algorithm. Often, the unit interval $[0, 1]$ is chosen. This approach is known as standardised scaling and improves the efficiency of the sampling procedure. By rescaling, we ensure uniform proposal behaviour across all dimensions, making it easier to select a single scalar proposal width. It also mitigates issues with tuning, as proposals in the [0,1] space tend to behave more consistently. Although the proposal width still needs to be tuned, typical values in the range of $0.05$ to $0.2$ tend to work well for most applications. 

To apply standardised scaling, we transform each parameter in $\vec{X_{i}}$ using its prior bounds:
\begin{equation}
X_{i,\; \text{standardised}} = \frac{X_{i} - P(X_{i})_{\text{lower}}}{P(X_{i})_{\text{upper}} - P(X_{i})_{\text{lower}}},
\end{equation}
where $X_i$ is the current position in the parameter space, and $P(X_{i})_{\text{lower}}$ and $P(X_{i})_{\text{upper}}$ are the prior lower and upper limits for parameter $X_{i}$.

\subsubsection{The Adaptive-Metropolis (AM) algorithm}
An adaptive version of the Metropolis algorithm is the AM algorithm from \citet{Haario2001}. It extends the classic Metropolis algorithm by adapting the proposal distribution during sampling. By using a normal distribution with a covariance for the proposal distribution, the covariance can be continuously updated based on the history of the Markov chain. Through adaptation, it improves sampling. The key idea of the AM algorithm is to alleviate the difficulty of choosing a well-tuned proposal distribution by adapting to it. The adaptivity is both spatial and orientational, meaning that the proposal distribution gradually aligns itself with the geometry of the posterior. 

Importantly, this adaptation makes the algorithm non-Markovian, as each proposal now depends not only on the current state but also on all previous states. This makes it difficult to use parallelisation strategies. While multiple adaptive chains can be run in parallel, their trajectories cannot be directly combined into a single long Markov chain. Consequently, one must run a single long chain to match the performance of methods that can use parallelisation. We will describe how parallelisation works for our ensemble methods in section \ref{sec:ensembles}.

Despite this, it is shown that under certain regularity conditions, the algorithm maintains the ergodicity required for correct sampling from the target posterior distribution.

\begin{figure}
    \centering
    \includegraphics[width=0.5\linewidth, trim=16mm 8mm 16mm 4mm]{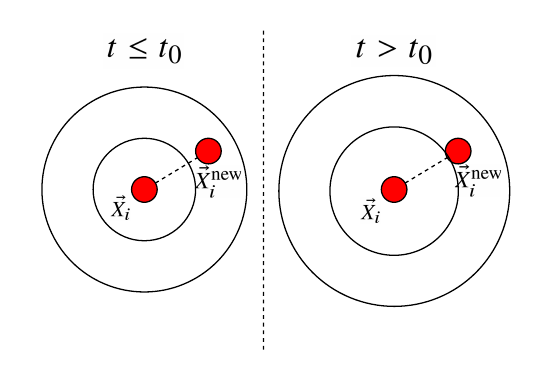}
    \caption{An adaptive Metropolis move for chain $\vec{X_{i}}$. Initially, before $t=t_{0}$, we sample the $\vec{X_{i}}^{\text{new}}$ from the initial Gaussian distribution with covariance $C_{0}$. Starting from $t_{0}$, we use the empirical covariance $C_{t}$. This adaptive covariance can increase or decrease the width of the Gaussian distribution. For this figure, we present an increased width.}
    \label{fig:Ch2_AM_move_illustrated}
\end{figure}

The AM move is written as:
\begin{equation}\label{eq:AM_walk_eq}
    \vec{X}^{\text{new}}_{i} = \vec{X_{i}} + \mathcal{N}(0, C_{t}),
\end{equation}
where $\vec{X_{i}}$ denotes the current position of chain $i$ and $\mathcal{N}(0, C_{t})$ the Gaussian proposal distribution from which a random sample is drawn with a covariance matrix $C_{t}$. We illustrate the AM move in figure~\ref{fig:Ch2_AM_move_illustrated}.

To ensure stable behaviour, the AM chain collects sufficient samples initially for a reliable covariance estimate. During this initial phase, the proposal covariance remains fixed at a user-specified matrix $C_0$, typically the identity matrix. The updates on the covariance matrix are applied during each step, but switching to this covariance matrix only occurs after a fixed period of length $t_0$. After $t_0$, the covariance matrix continues to be updated.

The recursive update formula used to update the empirical covariance of the chain is shown in equation~\eqref{eq:AM_cov}:
\begin{equation}\label{eq:AM_cov}
   C_{t+1} = \frac{t-1}{t}C_{t} + \frac{S_{d}}{t}(t\bar{X}_{t-1}\bar{X}_{t-1}^{T}-(t+1)\bar{X}_{t}\bar{X}_{t}^{T}+X_{t}X_{t}^{T}+\varepsilon I_{d}),
\end{equation}
where $S_{d}=2.4^{2}/d$ is a dimension-dependent scaling factor, as recommended by \citet{Gelman1996} to optimize the efficiency of Gaussian proposals. Additionally, $\bar{X}_{t}$ represents the empirical mean of the chain up to length $t$, and $\varepsilon > 0$ is a small regularisation constant that ensures the covariance matrix remains positive definite. This is especially important in the presence of multimodal or ill-conditioned posteriors. Lastly, $I_{d}$ represents the $d$-dimensional identity matrix. This ensures that the regularisation constant is uniformly applied to all dimensions. After $t_0$, the covariance matrix is continued to be updated using the recursive formula from equation~\eqref{eq:Covariance_AM_move}. 
\begin{equation}\label{eq:Covariance_AM_move}
    C_{t}=
    \begin{cases}
        C_{0}, \;\;\;\;\;\;\;\;\;\;\;\;\;\;\;\;\;\;\;\;\;\;\;\;\;\;\;\;\;\;\;\;\;\;\;\;\;\;\;\;\;\;\;t\leq t_{0}
        \\ s_{d}\;\mathrm{cov}(X_{0}, \dots, X_{t-1})+s_{d}\varepsilon I_{d}, \;\;t > t_{0}
    \end{cases}
\end{equation}

\subsection{Parallel affine-invariant ensemble samplers}\label{sec:ensembles}

\subsubsection{Ensembles}
Ensemble MCMC methods are an alternative to traditional single-chain algorithms. Instead of evolving a single chain, these methods employ a collection of chains that together form an \textit{ensemble}. Formally, the ensemble is the set $S=\{\vec{X_{i}}:1\leq i \leq M\}$, where chain $i$ is denoted as $\vec{X_{i}}$. Unlike standard MCMC, proposals in ensemble methods are not generated independently; the proposal for any chain $i$ depends not only on its state but also on the states of the other walkers. 

This coupling allows ensemble methods to explore the posterior distribution more effectively. Since proposals are defined relative to other walkers, the algorithm inherently adapts to the local geometry of the posterior, reducing the need for manual tuning of global proposal scales. This makes ensemble MCMC particularly effective in poorly preconditioned or anisotropic parameter spaces. The simultaneous evolution of multiple walkers enables broad exploration of the parameter space, thereby improving sampling efficiency. Additionally, the interaction between walkers helps reduce the probability of them getting trapped in local modes, making ensemble methods suitable for sampling from complex, multimodal posteriors.

The ensemble is constructed such that the walkers are marginally independent samples from the same target posterior distribution $\pi$. The joint target distribution for the ensemble is the product of the individual marginal densities:
\begin{equation}
    \Pi(S)=\Pi(\vec{X_{1}}, \vec{X_{2}}, \dots, \vec{X_{M}}) = \pi(\vec{X_{1}})\pi(\vec{X_{2}})\dots\pi(\vec{X_{M}}).
\end{equation}
This structure allows the ensemble sampler to update each walker by conditioning on the others while preserving the overall stationary distribution $\Pi$. That is, as long as the update rule for walker $\vec{X_{i}}$ leaves its marginal distribution $\pi(\vec{X_{i}})$ invariant, the full ensemble distribution remains invariant. This condition mirrors the transition kernel condition,
\begin{equation}
    \pi(\vec{X_{i}}) = \int K(\vec{X_{i}}^{\text{new}}|\vec{X_{i}})\pi(\vec{X_{i}})d\vec{X_{i}}
\end{equation}
for each chain $\vec{X_{i}}$. This condition is achieved using detailed balance \citep{goodman2010}. 

\subsubsection{Parallelisation of ensembles}
Parallelisation can greatly reduce the required computation time for algorithms. For classic MCMC, it can effectively cut down the process of serial running into simultaneous execution since a chain has no references to other chains. This, unfortunately, does not apply to ensemble methods since the updates of chains refer to one another. This breaks the possibility of running all chains in parallel. However, \citet{Foreman_Mackey_2013} give a solution to this problem. Instead of evolving all chains within the entire ensemble ($S$) simultaneously, it is better to divide the collection into two ensembles of equal size. These ensembles are defined as $S^{0} = \{\vec{X_{i}}, \forall \; i = 1, \dots, S/2\}$ and $S^{1}= \{\vec{X_{i}}, \forall \; i = S/2 + 1, \dots, S\}$. From these, we can simultaneously update all walkers of one ensemble, using proposals based on the other complementary ensemble. This allows us to parallelise half of the complete ensemble, giving a large boost in performance compared to a sequential calculation. To update all chains, we switch between each ensemble, i.e. first ensemble $S^{0}$ is updated based on $S^{1}$ and then ensemble $S^{1}$ is updated based on $S^{0}$. This continues until termination.

To further improve this method, reshuffling is used for the two ensembles after each global iteration. This reallocates the chains to the ensembles after both have been updated once. Reshuffling allows walkers to interact with a broader and more varied subset of the ensemble. As a result, the algorithm benefits from improved mixing and reduced risk of local optima. This strategy is implemented in the \texttt{EMCEE}\footnote{\href{https://emcee.readthedocs.io/en/stable/}{https://emcee.readthedocs.io/en/stable/}} package \citep{Foreman_Mackey_2013}, which served as the inspiration for incorporating this additional performance enhancement to our ensemble-based MCMC algorithms.

\subsubsection{Affine invariance}
An affine transformation is defined as an invertible transformation of the form $y=Ax+b$, where $A\in\mathbb{R}^{n\times n}$ is an invertible matrix and $b\in\mathbb{R}^{n}$ is a vector. They preserve the linear structure of the parameter space. An affine-invariant MCMC sampler is one whose performance remains unchanged under an affine transformation of the parameter space. It implies that the sampler performs equally well on a target $\pi$ as well as the transformed target $\hat{\pi}=\pi(A^{-1}(y-b))$ without the need of manually tuning the proposal distribution to the transformed geometry of $\pi$ \citep{HuijserGoodmanBrewer2017}. 

The samplers we consider in this work all contain the affine-invariance property. As highlighted by \citet{goodman2010, Foreman_Mackey_2013}, having this property is beneficial since the performance does not degrade when the target distribution is stretched, rotated, or highly correlated. This implies that such samplers adapt automatically to the geometry of the target distribution.

\subsubsection{The PAIES algorithm}
To generalise affine-invariant ensemble methods, we present our description of such samplers in algorithm~\ref{Algorithm:AIES}. This is based on the parallelised formulation introduced by \citet{Foreman_Mackey_2013}, and called the parallelised affine-invariant ensemble sampling algorithm, or PAIES in short. The algorithm begins with an initial ensemble $S$ consisting of $N$ walker positions. At each iteration, the ensemble is randomly shuffled and divided into two equally sized subsets, $S^{1}$ and $S^{2}$, respectively. During the updating of a set, one of these subsets is designated as the active set $S_i$, while the other, denoted as $S'$, is fixed and is used to inform the proposals for the active half. For each walker $\vec{X_i}$ in the active ensemble $S_i$, a new position $\vec{X_i}^{\text{new}}$ is proposed using an affine-invariant update rule $R$ that depends on $\vec{X_i}$ and the complementary ensemble $S'$. The proposal is accepted with probability $\alpha$, calculated using the posterior distribution ($\pi$) and the proposal distribution ($q(\cdot|\cdot)$). Accepted proposals replace the current walker positions in the active half. Rejected proposals are discarded, and the corresponding walkers retain their current positions in the active set. After updating the current active half, roles are switched, and the originally fixed half is now updated. Once both halves have been updated, the full ensemble is stored in the chain collection $C$. The chain collection contains all individual chains; the chains are not combined into a single chain; they are each stored individually. Only the current ensemble is modified during each iteration, while $C$ contains the entire Markov chain history for each walker. The process repeats until the target chain length $T$ is reached.

\begin{algorithm}
\caption{The Affine-Invariant Ensemble Algorithm}
\label{Algorithm:AIES}
\begin{algorithmic}[1]
\State \textbf{Input:} Initial ensemble $S$ of $N$ walkers, target chain length $T$, affine-invariant update rule $R$
\State Initialise chain collection $C \leftarrow S$
\For{$t = 1, \dots, T$}
    \State Randomly shuffle and split $S$ into two halves: $S^1$ and $S^2$
    \For{$S_{i}\in  \{S^{1}, S^{2}\}$}
        \State Let $S' \gets S \setminus S_i$ \Comment{Complementary ensemble}
        \For{each walker $\vec{X_{i}}\in S_{i}$}
            \State Propose $\vec{X}^{\text{new}}_{i}$ using $R$: 
                \State$\vec{X}^{\text{new}}_{i} \leftarrow R( \vec{X_{i}}, S^{'})$
            \\
            \State Compute the acceptance probability:
            \\ \State $
            \alpha = \min\left(1, \frac{\pi(\vec{X}^{\text{new}}_{i}) \, q(\vec{X}_{i}\mid \vec{X}^{\text{new}}_{i})}{\pi(\vec{X_{i}}) \, q(\vec{X}^{\text{new}}_{i} \mid \vec{X_{i}})}\right)
            $
            \\ \State Sample $u \sim \mathcal{U}(0, 1)$
            \If{$u < \alpha$}
                \State Accept the proposal: set $\vec{X_{i}} = \vec{X}^{\text{new}}_{i}$ 
            \Else
                \State Reject the proposal: retain $\vec{X_{i}}$
            \EndIf
        \EndFor
    \EndFor
    \State Append updated ensemble $S$ to $C$.
\EndFor
\State \Return $C$
\end{algorithmic}
\end{algorithm}

\subsubsection{The stretch move}\label{sec:stretch_move_theory}
Given that we have now discussed how ensemble samplers work and what affine invariance implies, we will start by introducing the many affine-invariant moves that can be used as an update rule $R$ for proposing a new set of parameters. We begin with the introduction of the stretch move from the well-known \texttt{EMCEE} package from \citet{Foreman_Mackey_2013}. 

\begin{figure}
    \centering
    \includegraphics[width=0.5\linewidth, trim=16mm 6mm 16mm 8mm]{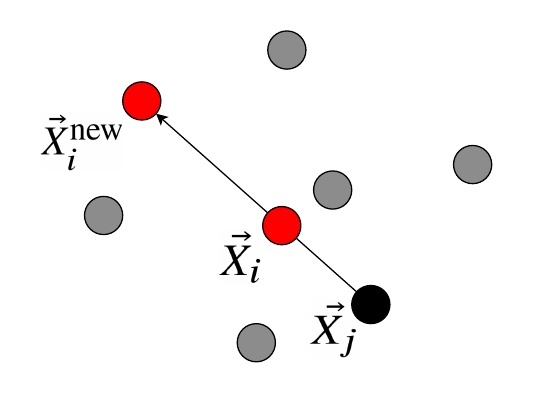}
    \caption{A stretch move for current chain $\vec{X_{i}}$. The black dot represents the reference walker $\vec{X_{j}}$ that is taken from the complementary ensemble of walkers. The stretch move stretches the difference between walkers $\vec{X_{i}}$ and $\vec{X_{j}}$ by a factor of $Z$, as seen in equation~\eqref{eq:stretch_move}, where $Z$ is obtained by equation~\eqref{eq:sampling_z_stretch_move}. From this move, we arrive at the new position, $\vec{X_{i}}^{\text{new}}$.}
    \label{fig:Ch2_Stretch_move_illustrated}
\end{figure}

The stretch move was introduced by \citet{goodman2010}, following the walk move from \citet{ChristenFox2007}, with a modification to make it affine-invariant. The stretch move proposal is given by:
\begin{equation}\label{eq:stretch_move}
    \vec{X_{i}}^{\text{new}} = \vec{X_{j}} + Z(\vec{X_{i}}-\vec{X_{j}}),
\end{equation}
in which $\vec{X_{j}}$ is a walker that is randomly sampled from the complementary ensemble, and $Z$ is the stretch factor. We illustrate the stretch move in figure~\ref{fig:Ch2_Stretch_move_illustrated}. The stretch move \textit{stretches} the difference vector from chains $i$ and $j$ with a factor of $Z$ to put chain $\vec{X_{i}}$ in a different location within the parameter space.

To use the stretch move, one must sample the stretch factor $Z$. This sampling is done through:
\begin{equation}\label{eq:sampling_z_stretch_move}
    Z = u^{2}\left(a+\frac{1}{a}-2\right)+2u\left(1-\frac{1}{a}\right)+\frac{1}{a},
\end{equation}
where $u\sim \mathcal{U}(0,1)$ and $a=2$. We use a value of $a=2$ as recommended by \citet{goodman2010}. This results, for $u=0$, in $Z=1/2$, and for $u=1$ in $Z=2$, therefore, we see that $1/2\leq Z\leq 2$, as described by \citet{goodman2010}. We present the derivation of this formula in appendix~\ref{appendix:sampling_z}. Additionally, we present the proof of affine invariance for the stretch move in appendix~\ref{appendix:stretch_AI}. To satisfy detailed balance for the stretch move, we require a modification of the Metropolis-Hastings acceptance criterion. From \citet{goodman2010}, a $Z^{N-1}$ factor is introduced to ensure this condition is met. This results in: 
\begin{equation}\label{eq:stretch_criteria}
\alpha = \min\left(1, Z^{N-1}\frac{\pi(\vec{X_{i}}^{\text{new}})}{\pi(\vec{X_{i}})}\right),
\end{equation}
where $N$ is the dimensionality of the parameter space. We present the proof of this equation in appendix~\ref{appendix:detailed_balance_stretch_move}.

\subsubsection{The walk move}
\begin{figure}
    \centering
    \includegraphics[width=0.5\linewidth, trim=16mm 4mm 16mm 8mm]{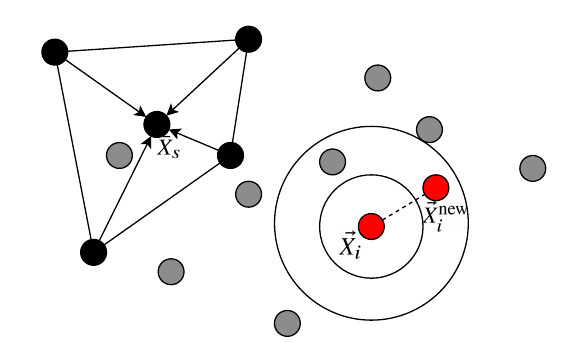}
    \caption{A walk move for current chain $\vec{X_{i}}$. The black and grey dots represent the complementary ensemble. The black dots mark the chains used from the complement for this walk move, i.e. $|S|=4$. The sample mean of these points is given by $\bar{X}_{s}$. From this, we construct the covariance matrix $C$, which is used to propose the new point $\vec{X_{i}}^{\text{new}}$ by centring a normal distribution about $\vec{X_{i}}$ with covariance $C$. }
    \label{fig:Ch2_Walk_move_illustrated}
\end{figure}

In addition to the stretch move, \citet{goodman2010} also introduced the walk move. This move is illustrated in figure~\ref{fig:Ch2_Walk_move_illustrated}. The walk move proposes the next set of parameters for a walker $i$, using its current position $\vec{X_{i}}$ and the covariance between a subset of walkers from the complementary ensemble. The proposal is given by: 
\begin{equation}\label{eq:walk_move_eq}
    \vec{X}^{\text{new}}_{i} = \vec{X_{i}} + \mathcal{N}(0, C),
\end{equation}
where $C$ represents the covariance matrix of the complementary ensemble, and is given by:
\begin{equation}
    C = \frac{1}{|S|}\sum_{\vec{X_{j}}\in S}(\vec{X_{j}}-\bar{X}_{s})(\vec{X_{j}}-\bar{X}_{s})^T,
\end{equation}
in which $|S|$ represents the size of the complementary ensemble, required to be at least size 2. The ensemble mean position is given by:
\begin{equation}
    \bar{X}_{s} = \frac{1}{|S|}\sum_{\vec{X_{j}}\in S}\vec{X_{j}}.
\end{equation}

\subsubsection{The differential evolution (DE) move}\label{sec:DE_move}
\begin{figure}
    \centering
    \includegraphics[width=0.5\linewidth,
    trim=16mm 8mm 16mm 4mm]{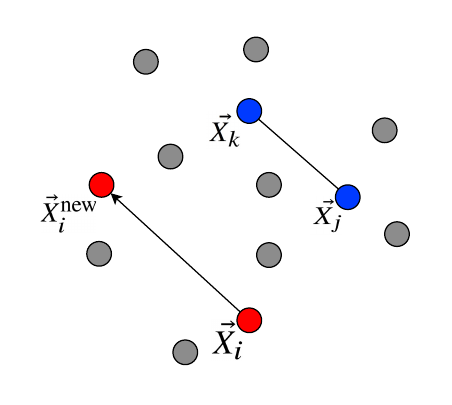}
    \caption{A Differential Evolution (DE) move for current walker $\vec{X_{i}}$. The blue and grey dots represent the walkers from the complementary ensemble. The blue dots are the reference walkers that are sampled at random from the ensemble, from which we determine the displacement that is scaled and added to $\vec{X_{i}}$ to arrive at the proposal $\vec{X_{i}}^{\text{new}}$.}
    \label{fig:Ch2_DE_move_illustrated}
\end{figure}

The differential evolution (DE) move \citep{DE_Nelson_2014} is a sampling strategy that resembles the stretch move in structure, but differs in how the proposal is constructed and adapted. The move originates from the differential evolution approach introduced by \citet{Braak2006}, with the modification of making it affine-invariant. The move is written as:
\begin{equation}\label{eq:DE_move}
    \vec{X_{i}}^{\text{new}} = \vec{X_{i}} + \gamma (\vec{X_{k}} - \vec{X_{j}}),
\end{equation}
where $\vec{X_i}$ is the current state of walker $i$, and $\vec{X_j}$ and $\vec{X_k}$ are two walkers drawn at random from the complementary ensemble. We illustrate the differential evolution move in figure~\ref{fig:Ch2_DE_move_illustrated}. Compared to the stretch move, $\gamma$ acts as the stretching factor $Z$, however, unlike the stretch factor, $\gamma$ is partially sampled, i.e. although it uses a sampled $z$, $\gamma$ is constructed by multiplying this $z$ with $\gamma_0$:
\begin{equation}
    \gamma = \gamma_0 (1 + z),
\end{equation}
with
\begin{equation}
    z \sim \mathcal{N}(0, \sigma_\gamma), \quad \text{and} \quad \gamma_0 = \frac{2.38}{\sqrt{2 n_{\text{dim}}}},
\end{equation}
where $n_{\text{dim}}$ denotes the dimensionality of the parameter space, and $\sigma_\gamma$ is a control parameter for the amount of noise added to the scaling factor.  \citet{DE_Nelson_2014} recommend choosing $\sigma_\gamma$ between $10^{-4}$ and $10^{-1}$. In our implementation, however, we use $\sigma_\gamma = 0.5$ to encourage more diverse proposals. Based on the optimal scaling analysis of Metropolis algorithms by \citet{Roberts2001}, the factor $2.38/\sqrt{2n_{\text{dim}}}$ is used to set the initial scaling factor of the differential evolution method.

A strength of the DE move is its built-in mechanism to adapt the proposal scale based on the chain's performance. DE adjusts $\gamma_0$ in response to the current acceptance fraction of all chains. After each global step, the algorithm computes the average acceptance fraction across all walkers, as seen in equation~\eqref{eq:mean_acc}, and uses this to rescale $\gamma_{0}$, as per equation~\eqref{eq:DE_gamma_up}.
\begin{equation}\label{eq:mean_acc}
    \overline{\text{Acceptance fraction}} = \frac{1}{|S|} \sum_{i=1}^{S} \left(1 - \frac{\text{\# Rejections chain}_i}{\text{Chain length}_{i}}\right)
\end{equation} 
\begin{equation}\label{eq:DE_gamma_up}
    \gamma_0 \leftarrow \gamma_0 \sqrt{\frac{\overline{\text{Acceptance fraction}}}{\text{Target acceptance fraction}}}.
\end{equation}
The size of the typical stretch step is thus based on how much the current mean acceptance fraction deviates from the target. If the acceptance rate is too high, the proposals are too conservative, and the algorithm increases $\gamma_0$ to encourage exploration. In contrast, if the acceptance rate is too low, the proposals are likely too aggressive, and the algorithm reduces $\gamma_0$ to make smaller moves.

By constantly steering the sampler towards a target acceptance fraction, the algorithm automatically balances exploration and exploitation. This dynamic adaptation aids the sampler to remain efficient. This is beneficial when the optimal scale is unknown beforehand or when it varies across the parameter space[, and allows the algorithm to be adaptive to local geometries. \citet{DE_Nelson_2014} suggest a target acceptance fraction of 0.25. This is in agreement with the recommended acceptance fraction range of 0.2–0.5 from \citet{goodman2010} and \citet{Foreman_Mackey_2013}. Choosing a target near the lower end favours broader proposals, which can be advantageous in complex or multimodal posterior landscapes. The higher end favours smaller proposals, resulting in more exploitative behaviour. The DE move accepts proposals based on the Metropolis acceptance criterion, equation~\eqref{eq:Metropolis_acceptance_criterion}, ensuring that it satisfies detailed balance. This differs from the acceptance criterion used for the stretch move. The reason for this is that there is no dependence on the current point $\vec{X_{i}}$ in the stretching term. Additionally, as $\gamma$ is drawn from a symmetric distribution, the two result in the proposal densities ($q(\cdot|\cdot)$) being symmetric; therefore, no adjustment is required in the acceptance criterion. We present the proof of affine invariance of the DE move in appendix~\ref{sec:DE_affine_proof}.

\subsubsection{The snooker move}

\begin{figure}
    \centering
    \includegraphics[width=0.5\linewidth,
    trim=22mm 8mm 22mm 4mm]{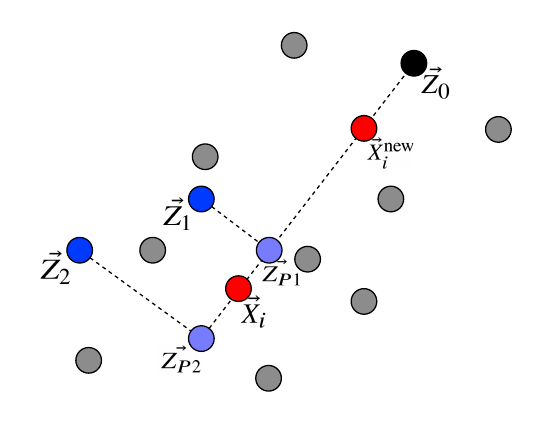}
    \caption{A snooker move for current chain $\vec{X_{i}}$. The red dots mark the initial and proposed position of walker $i$. The black dot marks the reference walker $\vec{Z_{0}}$, which is used to define a line between itself and $\vec{X_{i}}$. The dark blue dots mark the positions of the walkers, $\vec{Z_{1}}$ and $\vec{Z_{2}}$, that are orthogonally projected onto this line. The light blue dots mark the positions of the orthogonal positions, $\vec{Z_{P1}}$ and $\vec{Z_{P2}}$. The new point, $\vec{X_{i}^{\text{new}}}$ is the sum of the initial position $X_{i}$ and the scaled vector between $\vec{Z_{P1}}$ and $\vec{Z_{P2}}$.}
    \label{fig:Ch2_snooker_move_illustrated}
\end{figure}

The snooker move is a type of differential evolution move as introduced by \citet{terBraak2008}. It is based on the DE algorithm of \citet{Braak2006}, but finds its origin in \citet{gilkssnooker1994}, \citet{ROBERTS1994287} and \citet{Liang2001}. We illustrate the snooker move in figure~\ref{fig:Ch2_snooker_move_illustrated}.

The snooker move updates a current walker $\vec{X_{i}}$ based on itself and the addition of two other walkers ($\vec{Z_{1}}$ and $\vec{Z_{2}}$) that are orthogonally projected on the line through $\vec{X_{i}}$ and another walker $\vec{Z_{0}}$. These other walkers are sampled from the complementary ensemble at random. The snooker move's proposal is given by:
\begin{equation}
    \vec{X_{i}}^{\text{new}} = \vec{X_{i}} + \gamma_{s}(\vec{Z_{P1}} - \vec{Z_{P2}}),
\end{equation}
where $\gamma_{s}$ represents the scaling factor, similar to the stretch factor $Z$ in the stretch move. The $\gamma_{s}$ factor is taken to be equal to 1.7, based on $\gamma=2.38/\sqrt{2d}$ with $d=1$ taken from \citet{Roberts2001}. \citet{terBraak2008} argue that the projection step of the snooker move reduces the variance of the difference enough, and that it takes care of the dimensionality, allowing us to pick $d=1$ for $\gamma$. In addition, they also argue that it can be beneficial to take $\gamma_{s}\sim\mathcal{U}(1.2, 2.2)$ such that it is centred on 1.7, but with the possibility of having different sizes, which can improve exploration.

Contrary to \citet{terBraak2008}, we run the algorithm without mixing it with another sampler, i.e. not \textit{blending} the move. The authors use this move for $10\%$ of their updates, in addition to the $90\%$ of their differential evolution move. This is done to diversify the differential moves, but they do not argue why the snooker move alone would presumably underperform. 

Given the similarity in the proposal move to the stretch move from section \ref{sec:stretch_move_theory}, we will not present the proof of affine invariance for the snooker move. To ensure detailed balance, the snooker move requires an adjustment to the Metropolis acceptance criteria, in a similar sense as for the stretch move, since its proposals are not symmetric. For the snooker move, we use: 
\begin{equation}
    \frac{q(\vec{X_{i}}|\vec{X_{i}}^{\text{new}})}{q(\vec{X_{i}}^{\text{new}}|\vec{X_{i}})} = \frac{||\vec{X_{i}}^{\text{new}}-\vec{Z_{0}}||^{n-1}}{||\vec{X_{i}}-\vec{Z_{0}}||^{n-1}},
\end{equation}
implying that instead of $Z^{N-1}$ for the stretch move, we multiply the ratio of posterior values by the ratio of the norms between the old and new point and the reference walker $\vec{Z_{0}}$. This results in the following acceptance criterion:
\begin{equation}
    \alpha = \text{min}\left(1, \frac{||\vec{X_{i}}^{\text{new}}-\vec{Z_{0}}||^{n-1}}{||\vec{X_{i}}-\vec{Z_{0}}||^{n-1}}\frac{\pi(\vec{X_{i}}^{\text{new}})}{\pi(\vec{X_{i}})}\right).
\end{equation}

\subsubsection{The blend move}
\begin{figure}
    \centering
    \includegraphics[width=0.5\linewidth,
    trim=22mm 5mm 22mm 4mm]{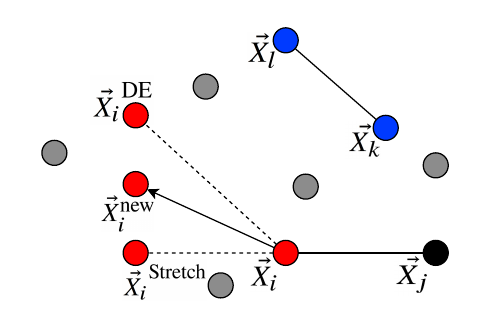}
    \caption{A blend move for current chain $\vec{X_{i}}$. The blend move combines the proposal moves of the stretch, equation~\eqref{eq:stretch_move}, and the differential evolution (DE) move, equation~\eqref{eq:DE_move}. The grey, blue, and black dots correspond to the complementary ensemble from which we sample the random walkers $j,\;k$, and $l$. From walker $j$, we obtain the stretch proposal; from $k$ and $l$ we obtain the DE proposal. These proposals are then blended, using a $\beta=0.5$, into a new position $\vec{X_{i}}^{\text{new}}$.}
    \label{fig:Ch2_blend_move_illustrated}
\end{figure}

Taking inspiration from \citet{terBraak2008}, we introduce the \textit{blend} move, which combines the differential evolution algorithm of \citet{DE_Nelson_2014} and the stretch move of \citet{goodman2010}. We take these two moves based on their empirical success in the area of astrophysics, e.g. \citet{Sanders_2013, reis2013, Akeret_2013, 10.1093/mnras/stx2109, hubble_constant_w_EMCEE, EHT_colab_DE_application2019}. The idea is that through utilising two kinds of moves, we are returned with a potentially better proposal that can explore parameter spaces efficiently. Contrary to \citet{terBraak2008}, we do not pick separate moves with a fixed probability, but we combine the two moves with some fraction of $\beta\in[0,1]$, which controls how much we favour the stretch move over the differential evolution move. The blend move is formulated as:
\begin{equation}
    \vec{X_{i}}^{\text{new}} = \beta \vec{X_{i}}^{\text{Stretch}} + (1-\beta)\vec{X_{i}}^{\text{DE}} 
\end{equation}
\begin{equation*}
    = \beta(\vec{X_{j}}+Z(\vec{X_{i}}-\vec{X_{j}})) + (1-\beta)(\vec{X_{i}}+\gamma(\vec{X_{l}}-\vec{X_{k}})),
\end{equation*}
where we use three randomly sampled walkers of the complementary ensemble, $j,\;k$ and $l$. We illustrate the blend move in figure~\ref{fig:Ch2_blend_move_illustrated}.

Just like DE, we utilise a target acceptance fraction; therefore, the algorithm adjusts its stepping size of $\gamma$. To ensure detailed balance, we must adjust the $Z^{N-1}$ factor by the amount of stretching that we use, i.e. $\beta$. This gives us the following acceptance criteria:
\begin{equation}
    \alpha =\text{min}\left(1, \;\beta Z^{N-1}\frac{\pi(\vec{X_{i}}^{\text{new}})}{\pi(\vec{X_{i}})}\right).
\end{equation} 

\subsubsection{The PCA stretch move}\label{sec:PCA_stretch_move}

In addition to the blend move, we also introduce a Principal Component Analysis (PCA) stretch move. This is a variation of the stretch move from \citet{goodman2010}. This move is designed to enhance exploration in directions of the highest variance within the complementary ensemble. The idea is to perform the stretch move in a rotated coordinate system defined by the principal components of the complementary ensemble. This coordinate system thereby aligns the sampling directions with the dominant axes of variation. In this context, PCA acts as a rotation because it finds a new coordinate system, i.e. a new basis in which the axes align with the directions of maximum variance in the data. 

\begin{figure}
    \centering
    \includegraphics[width=0.75\linewidth,
    trim=33mm 8mm 33mm 2mm]{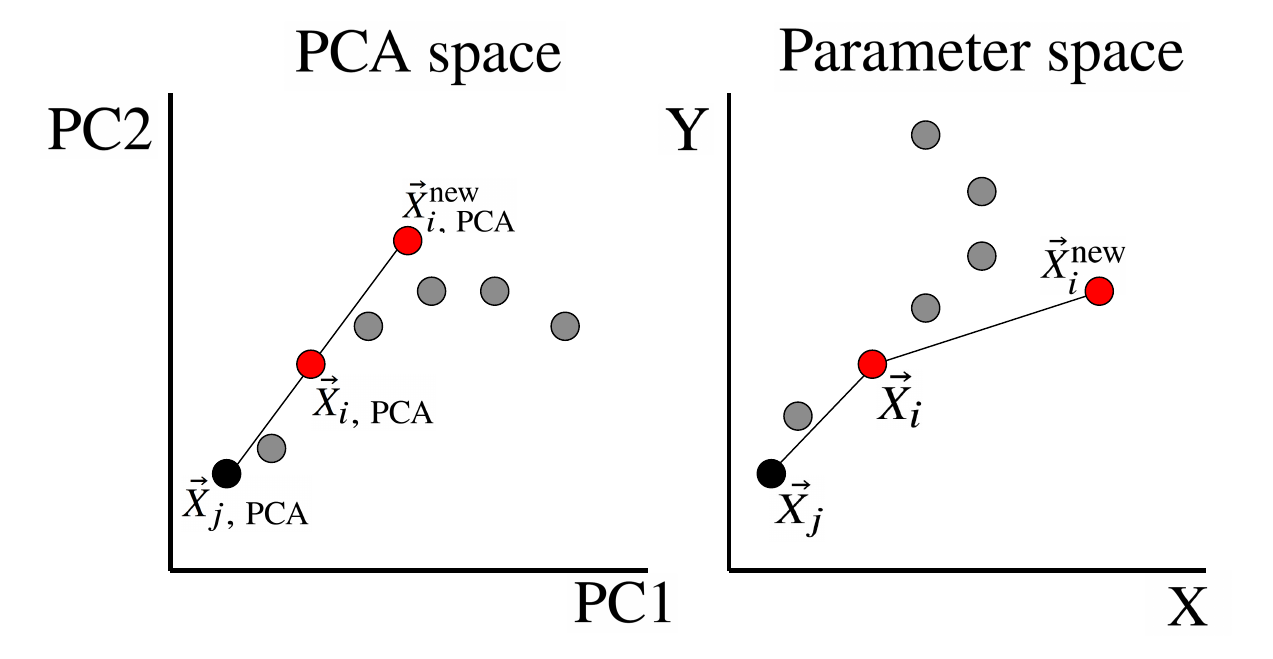}
    \caption{A Principal Component Analysis (PCA) stretch move for current chain $\vec{X_{i}}$. The PCA stretch move transforms the complementary ensemble to the PCA space and picks a random walker $j$ to propose the new position $\vec{X}_{i,\;\mathrm{PCA}}^{\text{new}}$, using the transformed $\vec{X_{i}}$. This new position is then transformed back into the parameter space by the inverse PCA transform, such that we obtain $\vec{X}_{i}^{\text{new}}.$}
    \label{fig:Ch2_PCA_Stretch_illustrated}
\end{figure}

While PCA is typically used for dimensionality reduction, here it is applied with the full number of dimensions. This ensures that the PCA transformation is a full-rank affine rotation and scaling, thereby preserving affine invariance. The move transforms the complementary ensemble of walkers into PCA space, after which it transforms the current walker of interest $i$ to this space, and picks a random complementary ensemble walker $j$. Then, following the stretch move, we obtain the proposed position in PCA space, $\vec{X}_{i,\; \mathrm{PCA}}^{\text{new}}$. To obtain the proposed position in the parameter space, we transform this position back through the inverse transformation. We use the \texttt{scikit-learn}\footnote{\href{https://scikit-learn.org/stable/}{https://scikit-learn.org/stable/}} implementation of PCA for this transformation. The PCA stretch move is given by:
\begin{equation}\label{eq:PCA_stretch_move}
    \vec{X_{i}}^{\text{new}} = T^{-1}(\vec{X}_{i, \; \mathrm{PCA}}^{\text{new}}) = T^{-1}(\vec{X}_{j,\;\mathrm{PCA}}+ Z(\vec{X}_{i,\;\mathrm{PCA}}-\vec{X}_{j,\;\mathrm{PCA}}))
\end{equation}
where $T^{-1}$ represents the inverse PCA transformation. We illustrate the PCA stretch move in figure~\ref{fig:Ch2_PCA_Stretch_illustrated}. Since PCA is a linear and invertible transformation when using all dimensions, the Jacobian determinant of the change of variables is constant and cancels out in the Metropolis-Hastings ratio. Therefore, the standard acceptance probability from the original stretch move still applies.

\section{Likelihood landscapes}\label{sec:CH3}
\subsection{The need for toy models}
To compare the performance of the samplers introduced in section \ref{sec:CH2}, we require a controlled test environment. For this purpose, we utilise \textit{toy models}: mathematical constructions that define artificial \textit{likelihood landscapes}. A major advantage is their independence from datasets, allowing for a controlled and systematic evaluation of sampler performance. The landscapes remain fixed across experiments, resulting in stability, which makes it possible to discover differences in convergence speed, sampling efficiency, or accuracy solely to the algorithms themselves, without interference from data variability, model assumptions, or noise.

Typical MCMC applications rely on empirical data and a forward model to define the likelihood function (e.g. \citet{brooks2011handbook}). While this is essential for real-world inference, such setups complicate fair comparisons between samplers because performance becomes entangled with factors like data quality, noise characteristics, model adequacy, and domain-specific choices. Toy models avoid these issues and thereby shift the focus of testing from \say{Can we fit this data?} to the more fundamental question: \say{How effectively does the sampler explore the underlying landscape of the posterior?}

To build intuition, the likelihood function can be seen as the description of a multi-dimensional \say{landscape} in which each set of points in the parameter space is associated with a likelihood value. This can be viewed as a topographic map with elevation values for geographic coordinates. The task of the sampler is to explore the landscape efficiently, while spending more time in the high-likelihood regions than in low-likelihood regions.

We begin with \textit{unimodal} landscapes, which feature a single global optimum and no local optima. These provide a straightforward setting for testing how rapidly and accurately samplers converge to the target distribution. We then consider more complex \textit{multimodal} likelihood landscapes, which better resemble realistic inference problems where posteriors contain multiple distinct high-probability regions. Such local optima pose significant challenges for MCMC samplers since they cannot jump easily from one to another. Effective sampling requires not only identifying the modes but also transitioning between them in proportion to their posterior weight. 

Our goal in using toy models is to identify MCMC algorithms that are both efficient and robust. To test this, we systematically apply the algorithms on landscapes of increasing complexity, starting from simple, unimodal cases and progressing to complex, multimodal scenarios.

\subsection{Unimodal landscapes}
\subsubsection{Rosenbrock}\label{subsubsection:Rosenbrock}
As the first instance of our unimodal likelihood toy models, we include the well-known \textit{Rosenbrock function}. Originally introduced by \citet{rosenbrock1960automatic} as a performance test for optimisation algorithms, this function has become a classic benchmark in both optimisation and sampling literature.

The Rosenbrock function is non-convex and presents an interesting challenge for both optimisation and MCMC algorithms since the location of the global minimum is numerically difficult to find. The function's landscape consists of a long, narrow, and curved valley; finding this valley is relatively straightforward, but navigating it to converge on the global minimum is where the difficulty arises. Algorithms must contend with the strong curvature and the significant difference in scale between the valley’s width and length, making the Rosenbrock function an ideal test case for evaluating how well a sampler can handle anisotropic and curved posterior geometries.

The Rosenbrock function is defined as:
\begin{equation}\label{eq:rosenbrock}
f(x, y) = (a - x)^2 + b(y - x^2)^2,
\end{equation}
where the global minimum is located at $(x, y) = (a, a^2)$, for which $f(x, y) = 0$. Standard parameter choices are $a = 1$ and $b = 100$, which we also adopt. These parameters result in a characteristic banana-shaped valley. 

\begin{figure}
    \centering
    \includegraphics[width=0.675\linewidth, trim=30mm 8mm 30mm 2mm]{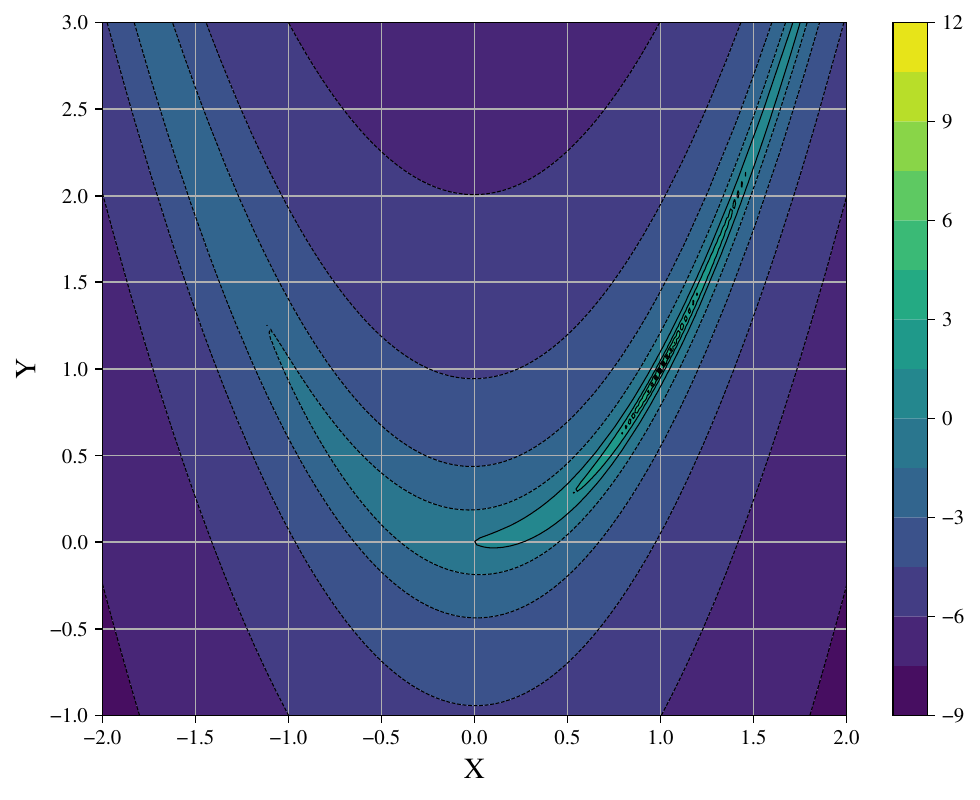}
    \caption{The Rosenbrock likelihood landscape. The landscape exhibits a banana-like shaped region of interest where the log-likelihood is highest. Outside this region, the log-likelihood is lower. The global optimum is located at (1,1).}
    \label{fig:CH3_Rosenbrock}
\end{figure}

In the context of our MCMC models, we use the logarithmic version of this function, resulting in log-likelihood values. Additionally, we take the negative sign value of this logarithmic function to define the probability density. By using the negative version of the Rosenbrock function, we change the valley into a mountain. We do this for consistency across our test cases. All toy models that we will present have regions of interest associated with regions of high likelihood. Furthermore, to prevent the logarithmic calculation of 0, we define this in equation~\eqref{eq:rosenbrock_in_ln}. For numerical stability, we add an addition of -10 to this equation. We present the Rosenbrock likelihood landscape in figure~\ref{fig:CH3_Rosenbrock}.

\begin{equation}\label{eq:rosenbrock_in_ln}
    \mathcal{L}(x,y) = - \text{ln}\left((a - x)^2 + b(y - x^2)^2\right)
\end{equation}

\subsubsection{Neal's funnel}\label{subsubsection:Nealsfunnel}
As our second toy model, we use the \textit{Neal's funnel} distribution, introduced by \citet{Neal2003Slice}. This synthetic distribution is known for its challenging geometry, which can cause significant difficulties for MCMC algorithms. In particular, its funnel-shaped geometry of the probability contours leads to extreme anisotropy and sharp scale changes. 

The distribution of Neal's funnel consists of two variables: an upper-level variable $v$ and a lower-level variable $y$. The variable $v$ is sampled from a normal distribution with zero mean and variance 3. Conditional on $v$, the variable $y$ is sampled from a normal distribution with zero mean and a variance that grows or shrinks exponentially with $v$. The sampling process is given by:
\begin{equation}
    v \sim \mathcal{N}(0\mid3), \quad y \sim \mathcal{N}\left(0\mid \exp\left(\frac{v}{2}\right)\right).
\end{equation}
As a result, the standard deviation of $y$ changes exponentially with $v$: for large positive values of $v$, the variance of $y$ becomes large, creating a wide region of support; for negative values of $v$, the variance collapses, forming a narrow neck in the probability landscape.

\begin{figure}
\centering
\includegraphics[width=0.675\linewidth, trim=30mm 8mm 30mm 2mm]{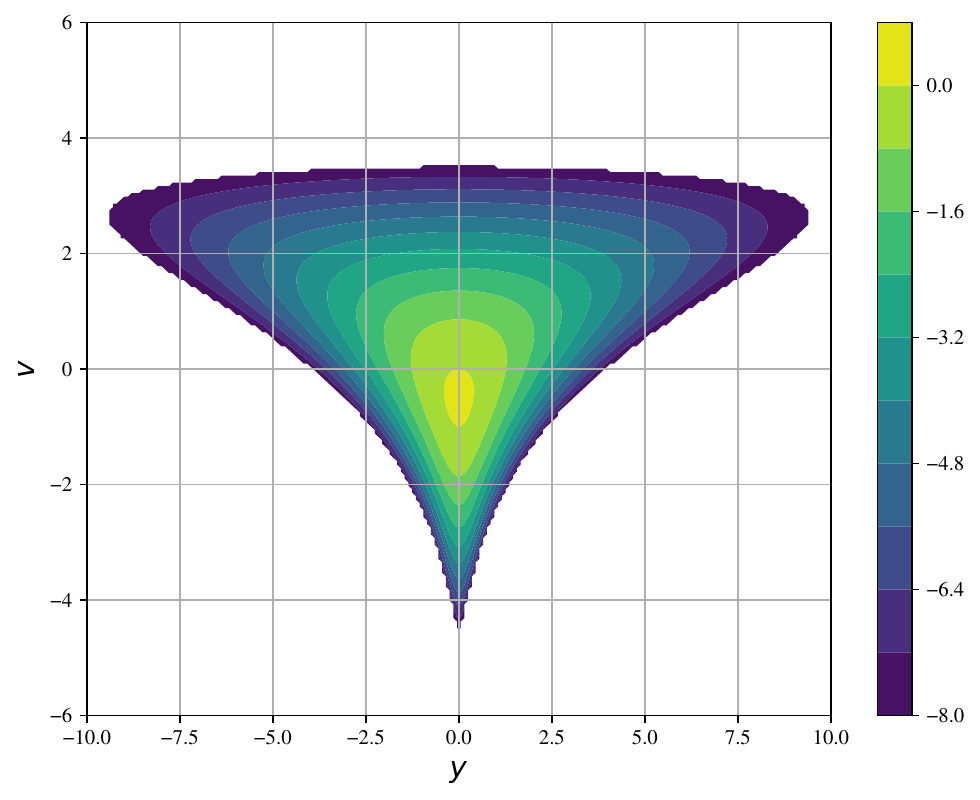}
\caption{The log-likelihood landscape of Neal's funnel for the upper-level variable $v$ and the lower-level variable $y$. The geometry exhibits the characteristic funnel shape, with a narrow neck at low $v$ and an increasingly wide spread for larger values of $v$. We limit the space occupied in the figure to highlight the shape of the landscape. In reality, the landscape is defined for any pair of $(v,\;y)\in\mathbb{R}^{2}$.}
\label{fig:CH3_NealsFunnel}
\end{figure}

The corresponding likelihood function of Neal's funnel is given by: 
\begin{equation}\label{eq:nealsfunnel}
    \mathcal{L}(v, y) = \exp\left(-\frac{1}{2}\left(v^2 + \frac{y^2}{\exp(v)} + v\right)\right).
\end{equation}
We present this likelihood landscape in figure~\ref{fig:CH3_NealsFunnel}.

\subsubsection{Experimental samplers setup}
For all samplers applied to the unimodal landscapes, we define a set of general parameters, summarised in table~\ref{tab:general_settings}. Each sampler operates with an ensemble of 10 walkers and a chain length of 2\,500 steps. To ensure a fair comparison between samplers, the initial positions of the walkers are fixed for each run. This is important because these can significantly influence sampler performance. The initial positions are generated by drawing at random from the prior ranges. This choice ensures that differences in performance arise from the sampling algorithms themselves rather than disparities in basic configuration. 

For our unimodal landscapes, we perform 1\,000 independent runs with different random seeds. By averaging performance over many different seeds, we aim to reduce the variability introduced and obtain a more reliable assessment of each sampler. \\

\begin{table}
\centering
\caption{General sampler settings used across all unimodal landscapes.}
\label{tab:general_settings}
\begin{adjustbox}{width=0.3\textwidth}
\begin{tabular}{@{}>{\raggedright\arraybackslash}p{2cm} p{2cm}@{}}
\toprule
\textbf{Setting} & \textbf{Value} \\
\midrule
Walkers & 10 \\
Chain length & 2\,500 steps \\
Random seeds & 1\,000 \\
\bottomrule
\end{tabular}
\end{adjustbox}
\end{table}
\noindent In addition to these general choices, each sampler requires specific algorithmic settings. We summarise these in table~\ref{tab:specific_settings}. Where possible, recommendations from the literature are followed to ensure comparability and to align with best practices. For some algorithms, such as the snooker move, differential evolution, and blend move, multiple versions are tested to assess the sensitivity of performance to key hyperparameters. Additionally, for the adaptive Metropolis algorithm, we set the initial covariance to the identity matrix, reflecting an absence of prior information on parameter correlations. We utilise the initial covariance matrix until we have updated the running covariance matrix for $10\%$ of the entire chain length, after which we use the updated covariance. The Metropolis-Hastings algorithms use a Gaussian proposal with a modest standard deviation to balance exploration and acceptance rates.

In addition to the algorithms introduced in section~\ref{sec:CH2}, we also include the \texttt{EMCEE} implementation of the stretch move in our experiments. Incorporating this established version serves two purposes: first of all, it allows us to validate the correctness of our implementation by direct comparison; second of all, it provides a well-regarded baseline against which the performance of the other samplers can be evaluated. 

\begin{table*}
\centering
\caption{Sampler-specific settings used in the single and multimodal likelihood landscapes.}
\label{tab:specific_settings}
\begin{adjustbox}{width=\textwidth}
\begin{tabular}{@{}>{\raggedright\arraybackslash}p{4cm} >{\raggedright\arraybackslash}p{5cm} >{\raggedright\arraybackslash}p{6cm}@{}}
\toprule
\textbf{Sampler} & \textbf{Setting} & \textbf{Value/details} \\
\midrule
Metropolis-Hastings (Regular, SS) & Proposal distribution & Normal distribution with standard deviation 0.2 \\
\addlinespace
Adaptive Metropolis & Initial covariance ($C_{0}$) & Identity matrix \\
 & Covariance update start & After 10\% of chain length \\
  & Covariance update frequency & Every step \\
\addlinespace 
Stretch (Regular, PCA) & Stretch factor ($a$) & 2  \\
\addlinespace
Snooker & $\gamma$ value & (1) Fixed at 1.7 \\
        &               & (2) Sampled from $\mathcal{U}(1.2,\; 2.2)$ \\
\addlinespace
Differential Evolution & Target acceptance fraction & (1) Value of 0.25 \\
                        &                           & (2) Value of 0.40 \\
\addlinespace
Walk & Complementary ensemble & Full ensemble used for mean calculation \\
\addlinespace
Blend & Move composition & (1) 80\% Stretch / 20\% DE, acceptance fraction 0.40 \\
      &                   & (2) 80\% Stretch / 20\% DE, acceptance fraction 0.25 \\
      &                   & (3) 20\% Stretch / 80\% DE, acceptance fraction 0.25 \\
\bottomrule
\end{tabular}
\end{adjustbox}
\end{table*}

\subsection{Multimodal landscapes}\label{sec:multi_modal_landscapes}
\subsubsection{Creation of Gaussian landscapes}
In addition to the unimodal landscapes, we present here a methodology for generating multimodal $N$-dimensional likelihood landscapes. To generate such landscapes, we require a fundamental building block that can be naturally extended to any dimensionality of interest. An obvious choice for this building block is the Gaussian function, since it scales seamlessly to any dimensionality. We define an $N$-dimensional Gaussian by:
\begin{equation}\label{eq:NdimGaussian}
A\cdot \text{exp}\left({-\frac{(\vec{r}-\vec{\mu})^{2}}{2\vec{\sigma}^{2}}}\right) = A\cdot \prod_{i=1}^{N}\text{exp}\left({-\frac{({r_{i}}-{\mu_{i}})^{2}}{2\sigma_{i}^{2}}}\right),
\end{equation}
where we use vector notation for the components: $\vec{r}$ represents the positional coordinates, $\vec{\mu}$ the coordinates of the mean and $\vec{\sigma}$ the standard deviation. The dimensionality is given through $N$. The amplitude of the Gaussian is given by $A$.

To construct landscapes, we define the likelihood function as the sum of the values from multiple Gaussians, i.e. for any given $N$-dimensional input vector, this function returns the likelihood value at that point in space. For a landscape composed of $M$ Gaussians, the likelihood function is defined as:

\begin{equation}\label{eq:NdimGaussianLhood}
\mathcal{L}(\vec{r}) = \sum_{m=1}^{M}A_{m}\left(\prod_{i=1}^{N}\text{exp}\left({-\frac{({r_{i}}-{\mu_{m,i}})^{2}}{2\sigma_{m,i}^{2}}}\right) \right).
\end{equation}

Each Gaussian in the landscape is characterised by its centre $\vec{\mu}$, width $\vec{\sigma}$, and amplitude $A$. We define these parameters by randomly drawing them from uniform distributions constrained by the directional sizes of a landscape. To avoid placing Gaussians directly on the edges of these prior ranges, we apply a small buffer equal to the standard deviations ($\sigma$) of our Gaussians. The choices made for the parameters of each landscape will be detailed later.  

To ensure sufficient coverage of the multi-dimensional space, we define the number of Gaussians as a function of the dimensionality $N$ using equation~\eqref{eq:MGaussians}. This allows the complexity of the landscape to grow, but capped at $15\,625$ to keep the computational cost manageable.
\begin{equation}\label{eq:MGaussians}
M=\text{min}\left(5^{N}, 5^6\right).
\end{equation}

\subsubsection{Creating projections of Gaussian landscapes}\label{sec:projection_plots}

To provide interpretability and insight into the structure of the likelihood surfaces described, we turn to visualisation. We visualise an $N$-dimensional Gaussian landscape, by creating projections along all axes. By reducing the dimensionality through these projections, we gain intuitive insights into the structure of high-dimensional landscapes before applying MCMC. 

To generate these projections, we sample points from the likelihood landscape. We do this by creating a meshgrid of equally spaced points across the predefined parameter ranges. This meshgrid provides a structured sampling of the landscape, enabling us to approximate the likelihood across the full domain. Given an $N$-dimensional space with coordinates $(x_{1}, x_{2}, \dots, x_{N})$, we generate a projection for any pair of axes by summing over all other dimensions. The projection values are determined using the likelihood values.

The meshgrid approach, however, has a trade-off: small-scale features of the landscape may be smeared out due to the fixed grid resolution. This limitation arises because the number of meshgrid points increases exponentially with the number of dimensions, a manifestation of the curse of dimensionality. To balance computational cost with visual detail, we limit the number of grid points per axis. Specifically, for $N \leq 5$, we set the number of points equal to: $\text{max}(150 - 3^{N}, \;8)$. For higher dimensions, we fix the number of points per axis to 5. 

\subsubsection{Multimodal toy models}\label{sec:multi_modal_testcases}
In what follows, we shall introduce our multimodal landscapes using the plotting projection method from section~\ref{sec:projection_plots}. For our work, we define three multimodal toy models of increasing dimensionality: a 3D, 5D and 8D case. 

\begin{figure*}
\centering

\includegraphics[width=0.9\linewidth,trim=33mm 8mm 33mm 2mm]{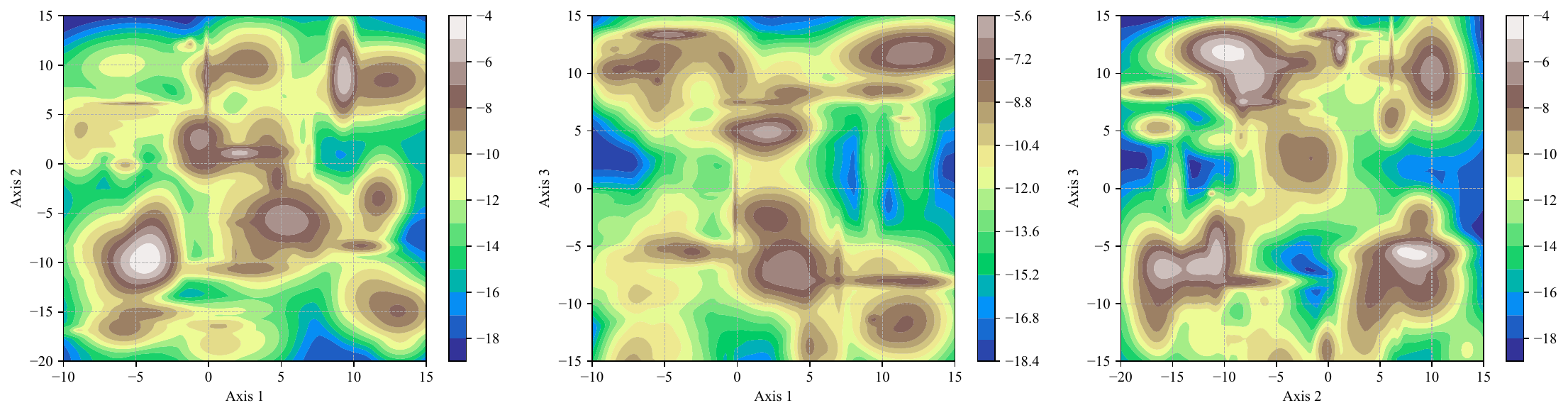}

\caption{The projections of the log-likelihood landscape of the 3D test case. The contours highlight elevation changes in the log-likelihood values. High values are associated with a white colour, whereas low values are represented in dark blue, as indicated by the globally scaled colour bar. In this figure, a total of 123 points per axis were used for the sampling of the landscape.}
\label{fig:Ch3_projections_3D}
\end{figure*}

The first test case we define serves as an intuitive low-dimensional example of the multimodal Gaussian landscapes studied in this work. We present the setup of this landscape in table~\ref{tab:landscape_3D}. The projections of the landscape are presented in figure~\ref{fig:Ch3_projections_3D}. It can be seen that the sum of many Gaussians results in a multimodal landscape of many peaks and valleys, making the identification of an optimal set of parameters a non-trivial problem.

\begin{table}
\centering
\caption{Setup parameters for our 3D Gaussian landscape.}
\label{tab:landscape_3D}
\begin{adjustbox}{width=0.45\textwidth}
\begin{tabular}{@{}p{3.8cm} p{4.5cm}@{}}
\toprule
\textbf{Parameter} & \textbf{Value} \\
\midrule
Dimensionality ($N$) & 3 \\
Number of Gaussians ($M$) & 125 \\
Ranges &
\begin{tabular}[t]{@{}l@{}}
$(-10,15)$, $(-20,15)$, $(-15,15)$
\end{tabular} \\
Gaussian widths ($\vec{\sigma}$) & Uniformly sampled between 0 and 1.5 in each direction \\
Amplitude & Uniformly sampled between 0 and 2 \\
\bottomrule
\end{tabular}
\end{adjustbox}
\end{table}

\begin{figure*}
\centering

\includegraphics[width=0.9\linewidth, trim=33mm 8mm 33mm 2mm]{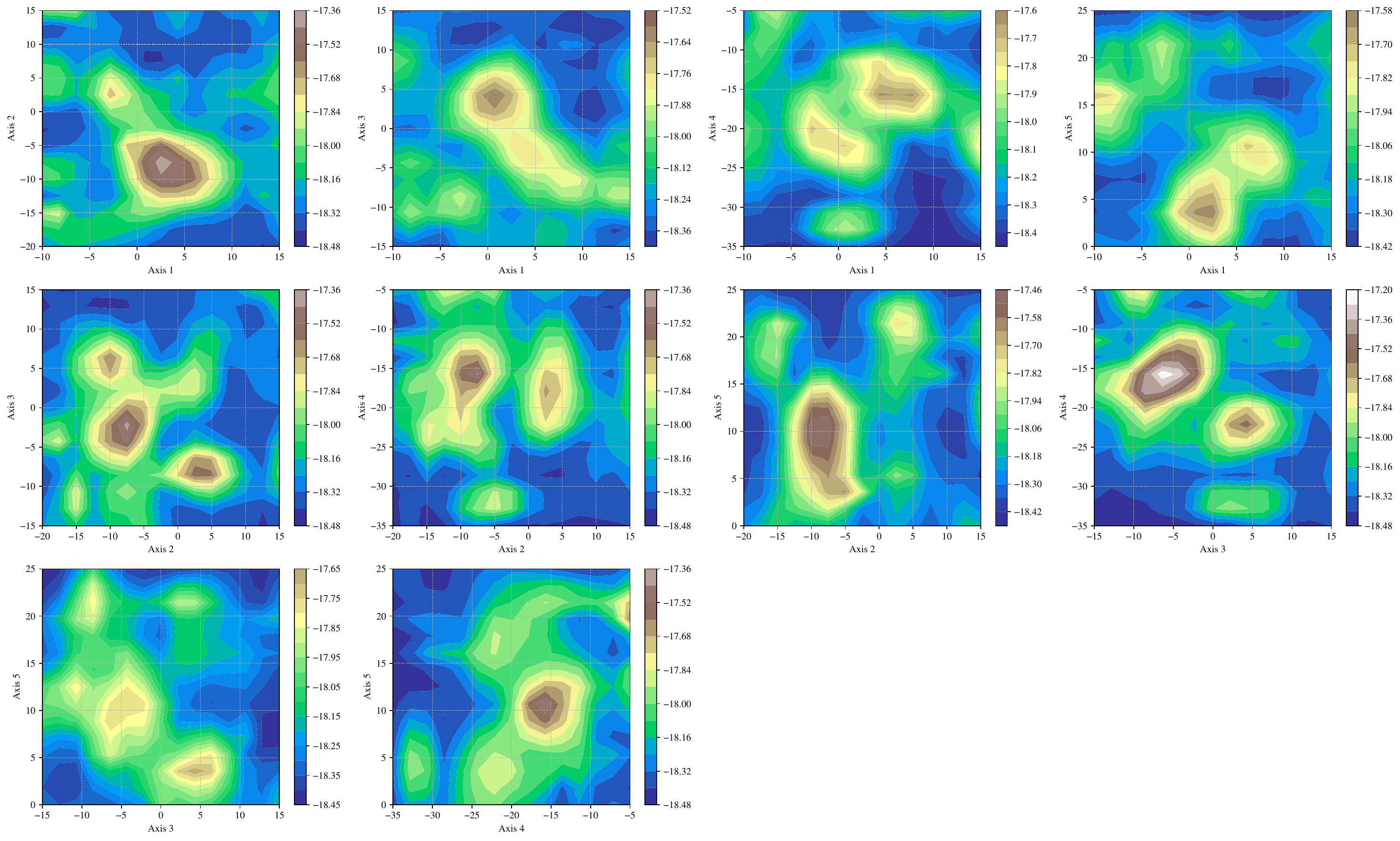}

\caption{As figure~\ref{fig:Ch3_projections_3D}, but for the 5D test case, using 8 points per axis and showing pairwise combinations of dimensions.}
\label{fig:Ch3_projections_5D}
\end{figure*}

\noindent In addition, by introducing two more dimensions, resulting in a testcase of 5 dimensions, the parameter space is enlarged, which provides samplers a more challenging setting. The projections of this landscape are presented in figure~\ref{fig:Ch3_projections_5D} and the setup is shown in table~\ref{tab:landscape_5D}. Contrary to figure~\ref{fig:Ch3_projections_3D}, the 5D projections show less detail, and more islands appear in the sea of the parameter spaces.

\begin{table}
\centering
\caption{Setup parameters for our 5D Gaussian landscape.}
\label{tab:landscape_5D}
\begin{adjustbox}{width=0.45\textwidth}
\begin{tabular}{@{}p{3.8cm} p{4.5cm}@{}}
\toprule
\textbf{Parameter} & \textbf{Value} \\
\midrule
Dimensionality ($N$) & 5 \\
Number of Gaussians ($M$) & 3\,125 \\
Ranges &
\begin{tabular}[t]{@{}l@{}}
$(-10,15)$, $(-20,15)$, $(-15,15)$,\\
$(-35,-5)$, $(0,25)$
\end{tabular} \\
Gaussian widths ($\vec{\sigma}$) & Uniformly sampled between 0 and 1.5 in each direction \\
Amplitude & Uniformly sampled between 0 and 2 \\
\bottomrule
\end{tabular}
\end{adjustbox}
\end{table}

\begin{figure*}
\centering

\includegraphics[width=0.9\linewidth,trim=33mm 8mm 33mm 2mm]{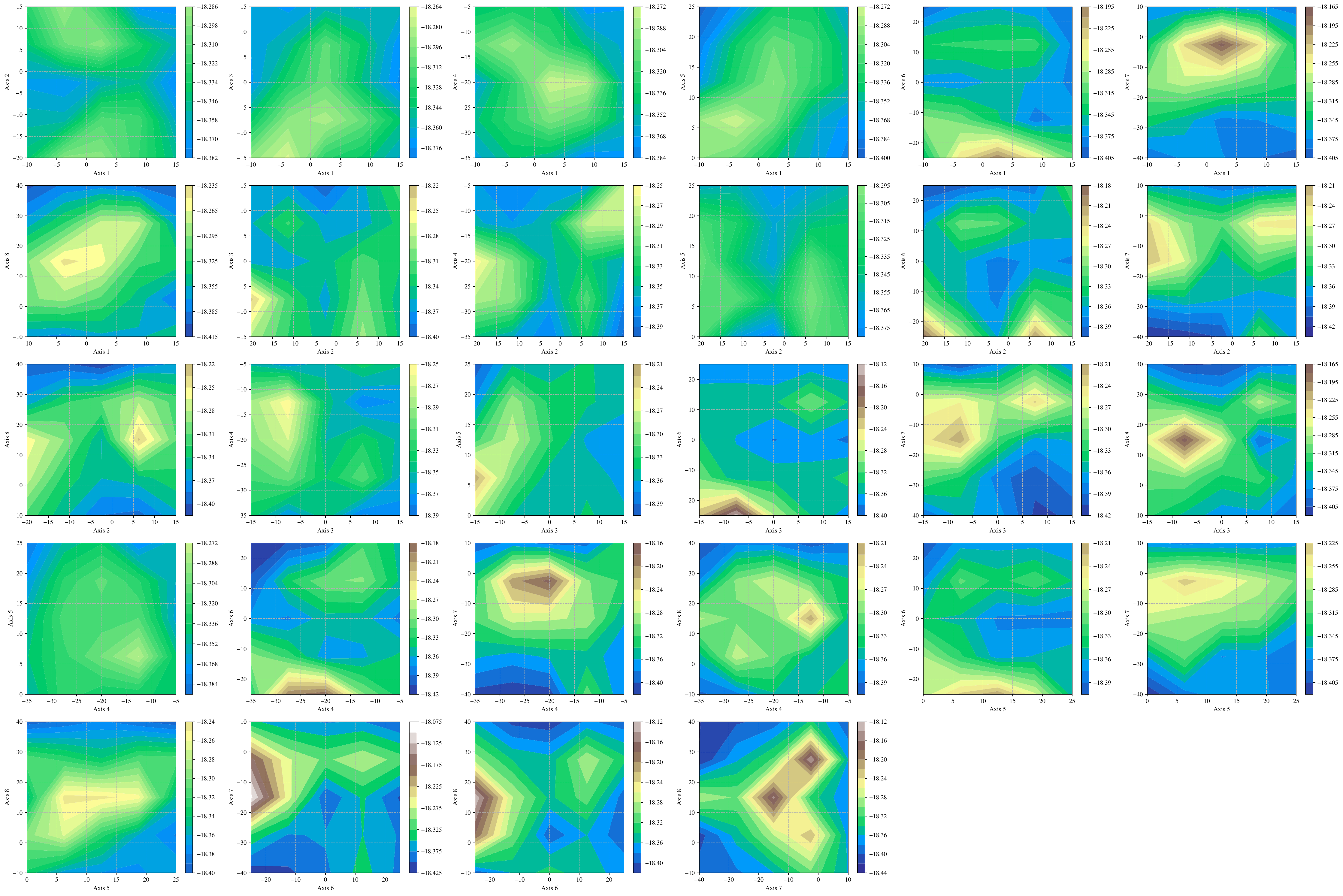}

\caption{As figure~\ref{fig:Ch3_projections_3D}, but for the 8D test case, using 5 points per axis and showing all pairwise axis combinations.}

\label{fig:Ch3_projections_8D}
\end{figure*}

\noindent Lastly, we have the 8D test case, which presents a substantially more complex landscape compared to the previous two cases. For the 8D toy model, we are faced with further reduction in visual interpretation as seen in figure~\ref{fig:Ch3_projections_8D}. From it, we see that the peaks and valleys are smeared out, as there are fewer peaks and valleys in general to be found in the figure compared to figures~\ref{fig:Ch3_projections_3D} and \ref{fig:Ch3_projections_5D}. To support the generation of this higher-dimensional space, the amplitude of the Gaussians is increased compared to the previous cases. This adjustment makes it slightly easier to distinguish local maxima in projection views, allowing us to validate if a potential 8D test case is sufficient for sampler testing. We present the setup of our 8D test case in table~\ref{tab:landscape_8D}. 

\begin{table}
\centering
\caption{Setup parameters for the 8D Gaussian landscape.}
\label{tab:landscape_8D}
\begin{adjustbox}{width=0.45\textwidth}
\begin{tabular}{@{}p{3.8cm} p{4.5cm}@{}}
\toprule
\textbf{Parameter} & \textbf{Value} \\
\midrule
Dimensionality ($N$) & 8 \\
Number of Gaussians ($M$) & 15\,625 \\
Ranges &
\begin{tabular}[t]{@{}l@{}}
$(-10,15)$, $(-20,15)$, $(-15,15)$,\\
$(-35,-5)$, $(0,25)$, $(-25,25)$,\\
$(-40,10)$ and $(-10,40)$
\end{tabular} \\
Gaussian widths ($\vec{\sigma}$) & Uniformly sampled between 0 and 3.5 in each direction \\
Amplitude & Uniformly sampled between 10 and 500 \\
\bottomrule
\end{tabular}
\end{adjustbox}
\end{table}

\subsubsection{Experimental samplers setup}
For all samplers, we define a consistent set of general parameters for each multimodal landscape. We summarise these in table~\ref{tab:general_settings_multi_modal}. We can see that there are some differences between the landscapes, notably the number of chains used is increased compared to the unimodal landscapes as described in table~\ref{tab:general_settings}. For our multimodal landscapes, we run with at least double the number of chains. This choice was made as starting in multiple positions allows us to capture more of the landscape and to better test the dependence of each algorithm's convergence on the initial positions. From table~\ref{tab:general_settings_multi_modal}, we can also see that between our 3D and 5D test cases is the number of chains; instead of 20, the 5D model has 40 chains. The chain length remains the same. However, our 8D test case has an increased chain length due to the complexity of the landscape. By increasing the chain length, we give the samplers more time to explore, as exploration is expected to be more difficult.

In table~\ref{tab:general_settings_multi_modal}, we also provide the number of runs executed for each test case. To ensure a fair comparison between samplers, the initial positions of the walkers are fixed for each run. The initial positions are generated by randomly drawing them from the prior ranges once again. The sampler-specific settings from table~\ref{tab:specific_settings} also apply to our multimodal landscapes. 

Importantly, our 5D and 8D test cases do not include the adaptive-Metropolis algorithm, due to its non-Markovian nature, which limits us in combining chains when we aim to do inference; thereby, its applicability for higher dimensionality is limited, as it is beneficial to have multiple chains with varying starting positions since a single chain is more likely to get trapped in a local optima. Additionally, for the 5D and 8D runs, we only pick a single blend move, the $80\%$ stretch $20\%$ DE move with an acceptance fraction target of 0.40.

\begin{table}
\centering
\caption{General sampler settings used across the multimodal landscapes. The landscapes and initial positions remain fixed under the seeds.}
\label{tab:general_settings_multi_modal}
\begin{adjustbox}{width=0.4\textwidth}
\begin{tabular}{@{}p{2cm} p{1cm} p{1cm} p{1cm}@{}}
\toprule
\textbf{Setting} & \textbf{3D} & \textbf{5D} & \textbf{8D} \\
\midrule
Walkers & 20 & 40 & 40 \\
Chain length & 10\,000 & 10\,000 & 15\,000 \\
Random seeds & 150 & 100 & 100 \\
\bottomrule
\end{tabular}
\end{adjustbox}
\end{table}

\subsection{Landscape reconstruction from MCMC samples}\label{sec:lhood_land_reconstruction}
An interesting application of MCMC samples is the possibility to (re)construct a likelihood landscape. Since each sample in a chain represents a set of parameters with an associated likelihood value that lies somewhere within the landscape, we can use these to map out the projections of the underlying landscape. From sampling in proportion to the posterior, we have high resolution in the regions of high likelihood and lower resolution in the low likelihood regions, which allows us to adaptively zoom in on interesting regions. To create such adaptive resolution images, we apply a \textit{quadtree} algorithm. This method also allows us to map out likelihood landscapes that are not known analytically. From these projections, it is therefore possible to validate the correctness of the individual model components, as well as finding possible degeneracies in underlying likelihood models.

To apply a quadtree in our cases, we select a pair of parameters, as in the projection plots, to define a two-dimensional projection plane. From the full set of MCMC samples, we consider only this pair of dimensions. For each resulting quad, we compute the mean likelihood value of the points inside it. The quadrants are then coloured based on this mean. This visually distinguishes regions of high and low likelihood. In this way, the quadtree provides plots similar in style to those in section~\ref{sec:projection_plots}, but with the added benefit of adaptive resolution that reflects the density of the posterior.

\subsubsection{Application to Gaussian landscapes}

To demonstrate how the quadtree method can reconstruct likelihood landscapes, we apply it to our 3D and 5D multimodal toy models for a single seed run (thereby including 200\,000 and 400\,000 samples respectively). To obtain the reconstructed landscapes, we run the Metropolis-Hastings algorithm with standardised scaling using the settings from table~\ref{tab:general_settings_multi_modal} for each test case. For the sampler, we employ a Gaussian proposal distribution with a width of 0.2. 

\begin{figure*}
\centering
\includegraphics[width=0.9\linewidth,trim=33mm 8mm 33mm 2mm]{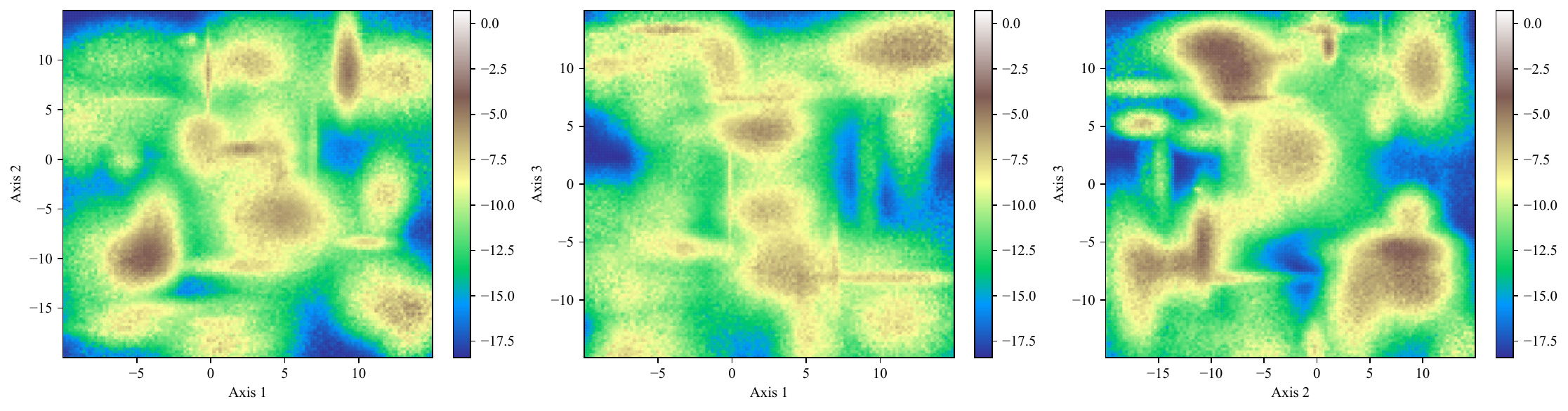}
\caption{Reconstructed log-likelihood landscape of the 3D multiple Gaussian test case as shown in figure~\ref{fig:Ch3_projections_3D}, using a quadtree. The colour scaling reflects the minimum and maximum log-likelihood values from the samples. The maximal depth of the quadtree used here is set at 7. }
\label{fig:CH3_3D_reconstructed}
\end{figure*}

The reconstruction of the 3D landscape is shown in figure~\ref{fig:CH3_3D_reconstructed}. From this figure, we see that the overall shape of the reconstructed landscape closely matches the original projections from figure~\ref{fig:Ch3_projections_3D}. We see that the key features of the original landscape are preserved, with well-defined regions emerging from the sampling. Additionally, the colour bar scaling has shifted from [$\simeq -19$, $-4$] in the original to [$\simeq -18$, $\sim 0.25$] in the reconstruction. This expanded scaling arises from having more estimates of the landscape, which enhances the resolution. Although the scale extends to values above zero, these highest regions are not prominently visible in the figure because they correspond to small, isolated areas within the landscape.

\begin{figure*}
\centering
\includegraphics[width=0.9\linewidth,trim=33mm 8mm 33mm 2mm]{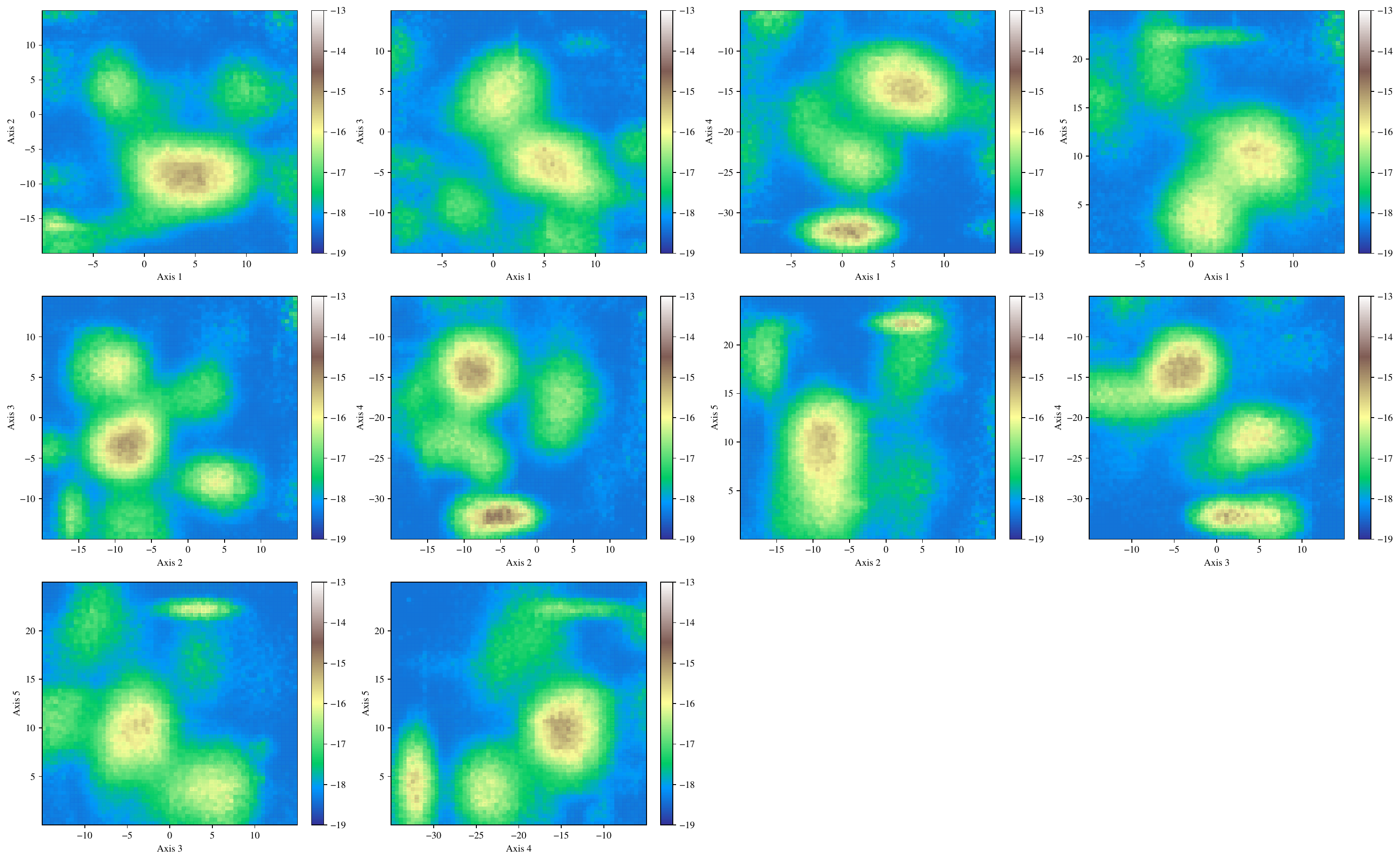}
\caption{As figure~\ref{fig:CH3_3D_reconstructed}, but for the 5D test case with projections shown in figure~\ref{fig:Ch3_projections_5D}. A maximal quadtree depth of 6 is used. The colour scale has been manually adjusted to span values from $-19$ to $-13$ for improved visibility. Isolated islands reflect the multimodal nature of the landscape.}
\label{fig:CH3_5D_reconstructed}
\end{figure*}

The reconstruction of the 5D toy model is presented in figure~\ref{fig:CH3_5D_reconstructed}. Here, the colour bar has been adjusted to improve the visibility of landscape features. While the original colour range was determined by the likelihood values from the samples, the highest-likelihood regions in this case are particularly small and would otherwise be difficult to distinguish. Additionally, it would make the overall structure less visible through the colour scaling. The general structure of the original projections from figure~\ref{fig:Ch3_projections_5D} is preserved, with increased resolution revealing additional details. In particular, features that initially appeared as well-connected regions are now seen as more isolated islands, highlighting the multimodal nature of the landscape. The original figure’s colour scale spanned approximately 1.22 orders of magnitude, while the reconstructed landscape spans a broader range of 6. This discrepancy arises from the MCMC chains exploring the landscape more efficiently compared to the meshgrid method, as well as the number of available points being significantly larger in the quad tree application.

\section{Sampler performance}\label{sec:CH4}
\subsection{Measuring the performance of a sampler}
Having introduced the various samplers in section~\ref{sec:CH2} and the toy models in section~\ref{sec:CH3}, we now shift our focus to assessing and comparing the performance of the samplers. The goal is to understand how effectively and reliably a sampler can explore the target posterior distribution. This involves not only evaluating how well it identifies regions of high probability but also assessing whether the distribution of samples is representative of the underlying landscape. By evaluating the performance of samplers for toy models, we can understand the strengths and limitations of the moves in a controlled setting.

To this end, we propose a set of performance metrics that are broadly applicable across MCMC samplers and likelihood landscapes. First, we introduce an ergodicity metric in section~\ref{sec:ergodicity_metric}, which quantifies how proportionally a sampler explores the posterior distribution. This metric is constructed by comparing the empirical distribution of likelihood values from the samples to the expectation values of the true landscape. It allows us to assess whether a sampler samples the posterior in an ergodic manner. A sampler that fails to exhibit ergodicity is of limited practical use, as it may systematically miss important regions of the posterior. Following this, we move to a performance metric that tracks the evolution of log-likelihood values over chain length, presented in section~\ref{sec:log_likelihood_performance}. By evaluating the chains of multiple runs and visualising the progression of their best median log-likelihood values, we gain insight into how quickly and reliably each sampler finds high-probability regions. This metric highlights differences in convergence speed, exploration strategies, and overall effectiveness. Furthermore, we also consider sampler robustness, as defined in section~\ref{sec:robustness_metric}. This metric captures the variability between chains of a sampler by comparing the best and worst outcomes across multiple runs with different random seeds. A robust sampler should produce consistently good results, while large discrepancies indicate sensitivity to initial conditions and instability. Finally, in section~\ref{sec:improving_best_estimates}, we explore the potential benefit of applying a non-stochastic optimisation algorithm to the best parameter estimates produced by each MCMC sampler. This hybrid approach of combining MCMC exploration with local optimisation can lead to improved likelihood values and more accurate estimates. For models where the best-fit parameters are critical, this step can offer a significant practical advantage.

\subsection{The ergodicity metric}\label{sec:ergodicity_metric}

Being able to determine how ergodic a sampler is helps to understand how effectively the sampler can reproduce the target posterior distribution. Whenever a sampler is ergodic, it samples points in proportion to the posterior, which implies that the number density of points is proportional to the posterior values:
\begin{equation}
    n(\theta) \propto P(\theta|x).
\end{equation}
Using Bayes' theorem, this can be rewritten as a product of the likelihood and the prior functions: 
\begin{equation}
    n(\theta) \propto P(x|\theta)P(\theta),
\end{equation}
or as a function of only the likelihood whenever a prior is used that is flat within its bounds: 
\begin{equation}\label{eq:ergodicity_metric}
    n(\theta) \propto \mathcal{L}(x|\theta).
\end{equation}
This equation implies that when the sampler is ergodic, the number density of points should be proportional to the respective likelihood values of these samples.

To create the ergodicity metric, we divide any likelihood landscape into hypercube segments. We present this method in appendix~\ref{appendix:hypercube_subdivison}. This hypercube division allows us to compare the expectation value of individual cells to the number of samples in these cells from sampling. This detailed approach increases the accuracy of the metric as there is less smearing-out compared to the calculation of the entire landscape. Given the multimodality of our Gaussian toy models, segmenting these landscapes is beneficial as cells can differ dramatically. The obtained likelihood values are calculated following the method described in appendix~\ref{appendix:NDIM_integration}. These are used to determine the expectation value per cell, which represents the expected number of samples in a cell based on the likelihood value. The expectation value of a hypercube cell is given by:
\begin{equation}\label{eq:cell_expect_val}
    \text{Cell expectation value} = \frac{\mathcal{L}_{\text{cell}}}{\mathcal{L}_{\text{total}}}N_{\text{samples}},
\end{equation}
where the likelihood of a cell is given by $\mathcal{L}_{\text{cell}}$, the total likelihood by $\mathcal{L}_{\text{total}}$ and the total number of samples drawn from the landscape by $N_{\text{samples}}$. Both are determined through numerical integration, where the bounds are defined by the cell's bounds and the landscape bounds, respectively. If the sampler is ergodic, the number of samples in each cell should approximately match this expectation value. 

To visualise the expectation values and the number of samples of the cells, we plot the values using overlaying histograms. These allow us to identify whether a sampler over- or underestimates the expectation values of the cells within a landscape. If the heights of the bars in the figure match, then the sampler is said to be ergodic for that regime. The results of our ergodicity metric, applied to our Gaussian test cases, are presented in section~\ref{sec:ergodicity_applied_to)landscapes}. To obtain a sufficient number of bins, we apply the Rice rule to the number of cells that, on average, have a cell occupation above 1, i.e. $N_{\text{bins}}=2n^{1/3}$.

\subsubsection{Application to our multimodal toy models}\label{sec:ergodicity_applied_to)landscapes}

\begin{figure*}
    \centering
    \includegraphics[width=0.85\linewidth]{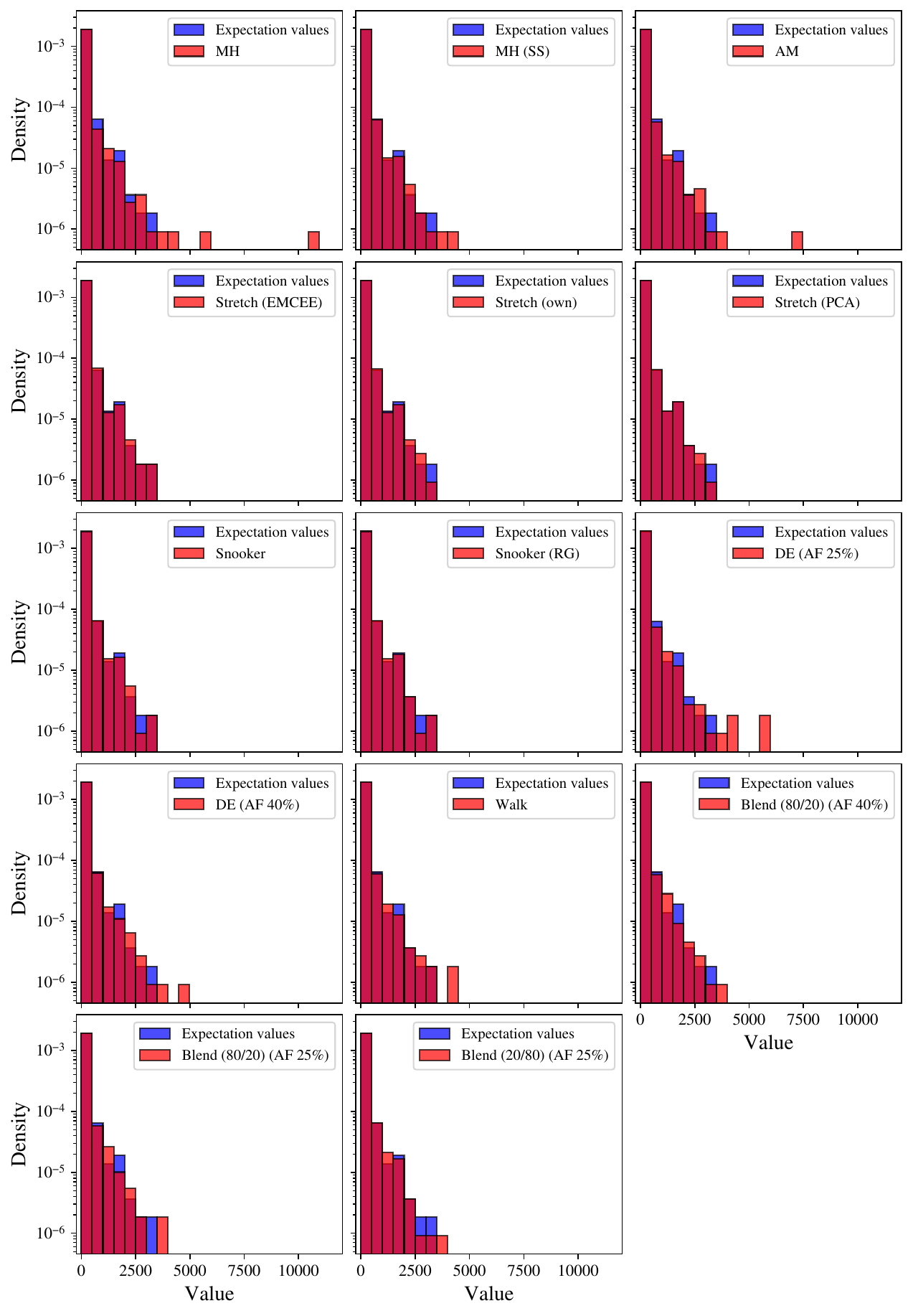}
    \caption{Ergodicity comparison between the expectation values of cells and the number of samples per cell for our 3D toy model. The data have been binned according to the Rice rule. The expectation values are plotted in blue, and the number of samples of each sampler is plotted in red. Each sampler has its unique subfigure, as highlighted by the label of the red colour bins. By overlaying the two, we identify the regimes where the sampler is considered ergodic, resulting in a darker red colour. In this figure, we apply a shared x-axis to compare the samplers effectively. The effective number of non-zero cells was 1\,194. The total number of cells of the landscape was 2\,197.}
    \label{fig:CH4:ergodicity_comparison_3D}
\end{figure*}

Having run according to the general settings in table~\ref{tab:specific_settings} and the landscape-specific settings in table~\ref{tab:general_settings_multi_modal}, we have collected the results of the random-seed runs for the collection of samplers. From these runs, we determine the average cell occupation over all runs, without removing any burn-in, and use it to obtain a more robust estimate of the ergodicity, as it smooths out any potential outlier runs. We present the result of our ergodicity metric applied to the samplers for the 3D Gaussian landscape in figure~\ref{fig:CH4:ergodicity_comparison_3D}. From this, we can see that all samplers follow the expectation values more closely than the Metropolis-Hastings (MH) algorithm. The MH algorithm has some outlying bins, indicating that it oversamples compared to the expectation values for some cells on average. 
Additionally, we see that our stretch move implementation has a comparable ergodicity outcome to the original stretch move from \texttt{EMCEE}. These moves visually follow the expectation values the closest among all samplers. 
Moreover, we also note that the newly introduced methods of the blend move and the PCA stretch move follow similar behaviour to the well-established moves of snooker, differential evolution, stretch and walk moves. 
Notably, however, the differential evolution (DE) algorithm with an acceptance fraction of 0.25 oversamples some cells, whereas the 0.40 target acceptance fraction variant does less oversampling. 
The snooker move gives similar results to the version where $\gamma$ is sampled. However, it can be seen that the oversampled region of the snooker with a fixed $\gamma$ disappears when using the sampled version. Amongst the blend moves, the (20/80) version with an acceptance fraction target of $25\%$ visually follows the expectation value bins the best.


\begin{figure*}
    \centering
    \includegraphics[width=0.85\linewidth]{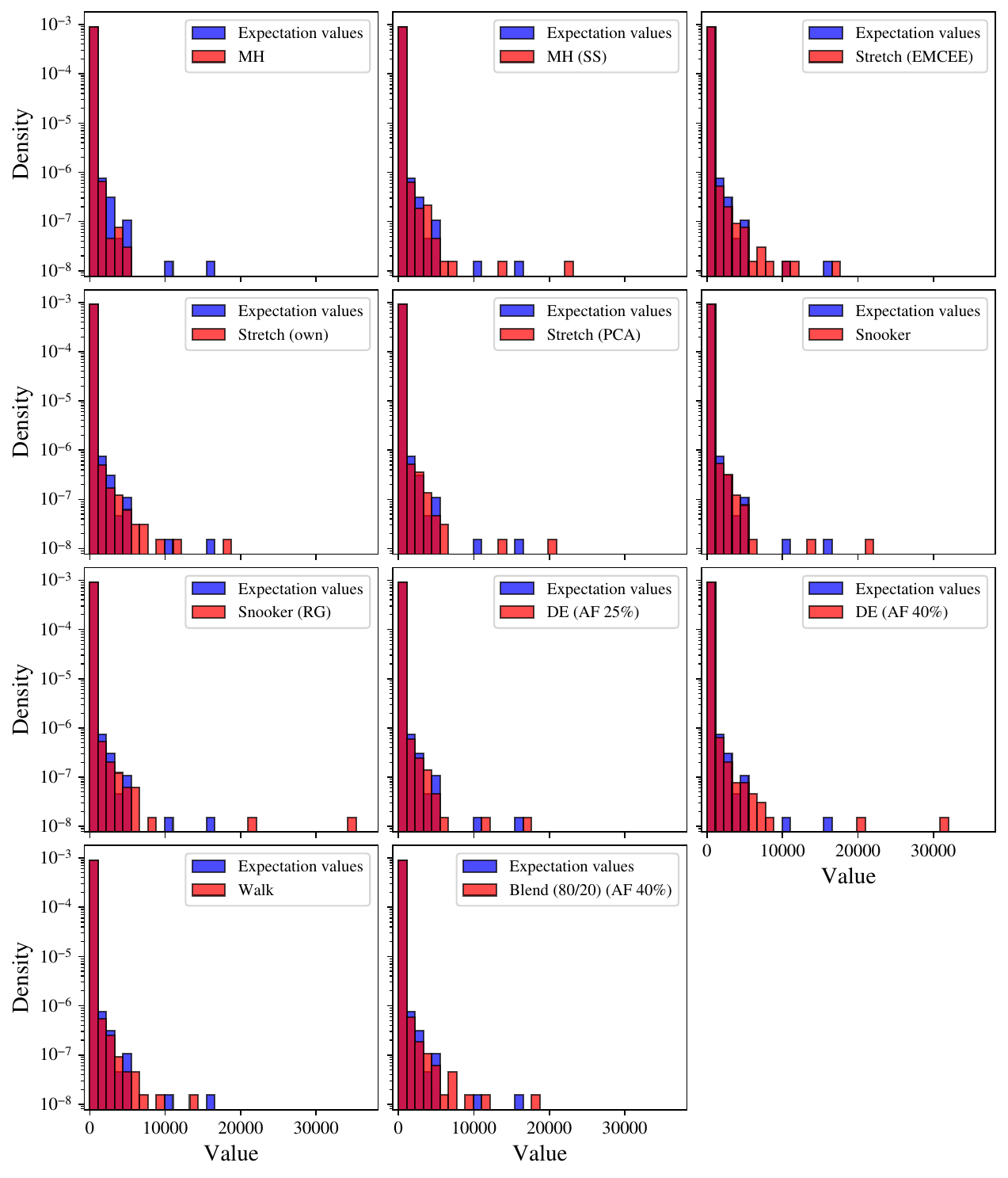}
    \caption{As figure~\ref{fig:CH4:ergodicity_comparison_3D}, but for our 5D Gaussian toy model. The effective number of non-zero cells was 3\,896, and the total number of cells was 59\,049.}
    \label{fig:CH4_ergodicity_5D}
\end{figure*}

We present the 5D result of the ergodicity metric for our samplers in figure~\ref{fig:CH4_ergodicity_5D}. From this figure, we see that the MH sampler undersamples certain regions of the parameter space compared to the ergodic expectations, as seen from the isolated blue bins. The other samplers do not present undersampling. Most samplers oversample the high-expectation valued cells. The DE (AF $25\%$) and walk sampler oversample the expectation value bin around a value of 15\,000 the most, going up to 30\,000-35\,000. Additionally, we see that more bins are not completely overlayed, as opposed to the results of the 3D testcase in figure~\ref{fig:CH4:ergodicity_comparison_3D}.
 
\begin{figure*}
    \centering
    \includegraphics[width=0.85\linewidth]{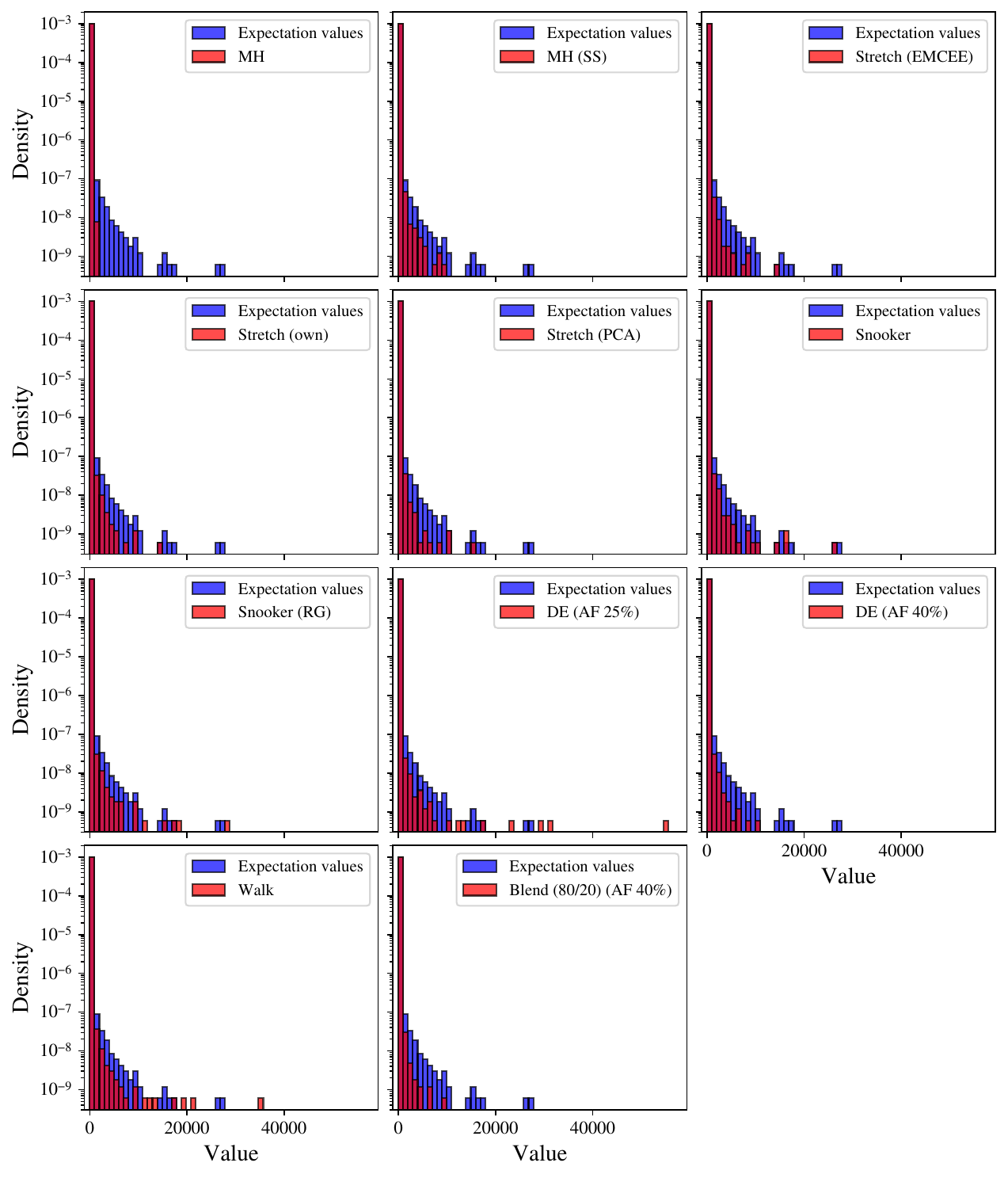}
    \caption{As figure~\ref{fig:CH4:ergodicity_comparison_3D}, but for our 8D Gaussian toy model. The effective number of non-zero cells was 21\,437, and the total number of cells was 1\,679\,616.}
    \label{fig:CH4_ergodicity_8D}
\end{figure*}

Lastly, we present the results of the ergodicity metric for our highest-dimensional landscape, the 8D model, in figure~\ref{fig:CH4_ergodicity_8D}. From this, we see that most samplers undersample high-likelihood regions: a lot of blue is observed. The snooker moves, the DE 25\%, and the walk move cover the most among all the samplers; these samplers seem most ergodic. Interestingly, the DE move with an acceptance of 25\% is the only sampler that overestimates the expectation value of some cells by a large margin, with the largest overestimation of above 60\,000 compared to the expectation value of just below 30\,000.

The MH algorithm has the lowest ergodic coverage, with the fewest overlaps and histogram bars. The stretch moves also have less coverage of the expectation values compared to the snooker and DE moves.

\subsection{The performance of samplers}\label{sec:log_likelihood_performance}
To have a consistent and robust comparison of sampler performance across multiple runs, we introduce a \textit{log-likelihood ratio evolution metric}. This metric quantifies how far each sampler deviates from the best observed sampler over chain length, averaged across chains and runs. After each run, for any sampler, we compute the best likelihood attained by every chain. From this collection of best likelihoods, we select the median chain per run. This provides a representative estimate of the sampler’s typical performance within that run. This median chain is less sensitive to outliers than using the absolute best or worst-performing chain and thus provides a representative reflection of average convergence behaviour. From the selected median chain, we construct a monotonic likelihood trajectory by replacing each point with the best likelihood reached up to that stage in the chain. This ensures that the convergence profiles only improve when a better parameter set is found, making comparisons between samplers more interpretable. Having obtained the median chains across many independent runs, we then average the likelihood trajectories for each sampler over all runs. This yields a smoothed profile of how the sampler tends to evolve over many runs, accounting for variations due to randomness. To compare samplers, we express their performance relative to the overall best sampler by computing likelihood ratios. We define the best sampler as the one that achieves the best median likelihood over chain length. Since our likelihoods are in log-space, these ratios reduce to simple differences:
\[
\log\left(\frac{\mathcal{L}_{\mathrm{best}}}{\mathcal{L}}\right) = \log(\mathcal{L}_{\mathrm{best}}) - \log(\mathcal{L}),
\]
where $\mathcal{L}$ denotes the average likelihood trace of a given sampler and $\mathcal{L}_{\mathrm{best}}$ that of the best one.

Plotting these differences over time gives a set of \textit{log-likelihood ratio curves}, where the vertical axis indicates deviation from the best-performing sampler. A steep decline suggests rapid convergence, whereas a flatter trajectory indicates slower progress. The final offset on the $y$-axis reflects the residual performance gap compared to the best sampler over the full chain length. This visualisation highlights both convergence speed and final performance. It offers a compact summary of how each sampler performs on average for arbitrary seeds. We apply this method across all unimodal and multimodal toy models, with their results presented in sections \ref{sec:log_evo_unimodal} and \ref{sec:log_evo_multimodal}, respectively.

\subsubsection{Application to unimodal landscapes}\label{sec:log_evo_unimodal}

\begin{figure*}
    \centering
    \includegraphics[width=0.8\linewidth]{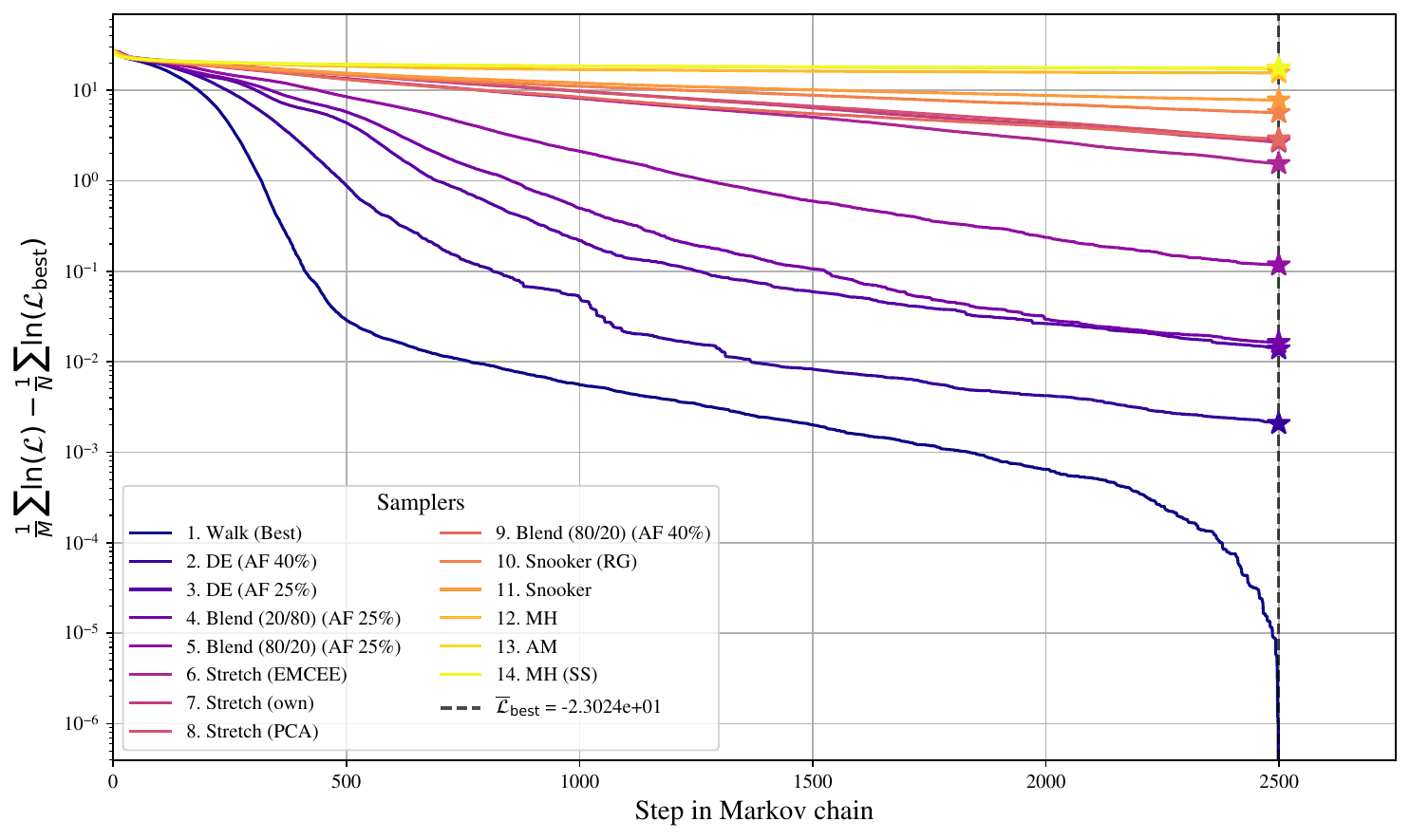}
    \caption{Sampler log-likelihood evolution over chain length compared to the best sampler of the Rosenbrock test case. Each curve is the difference of the average of median log-likelihoods over many runs between a sampler and the best sampler. The samplers are ranked based on their final positions relative to the best-performing sampler. In addition, the final likelihoods are indicated with a star marker. The best sampler's best likelihood is indicated with a dashed line to highlight the position in the chain where this was reached on average, as well as the average median log-likelihood value.}
    \label{fig:CH4_evo_Rosenbrock}
\end{figure*}

\begin{figure*}
    \centering
    \includegraphics[width=0.8\linewidth]{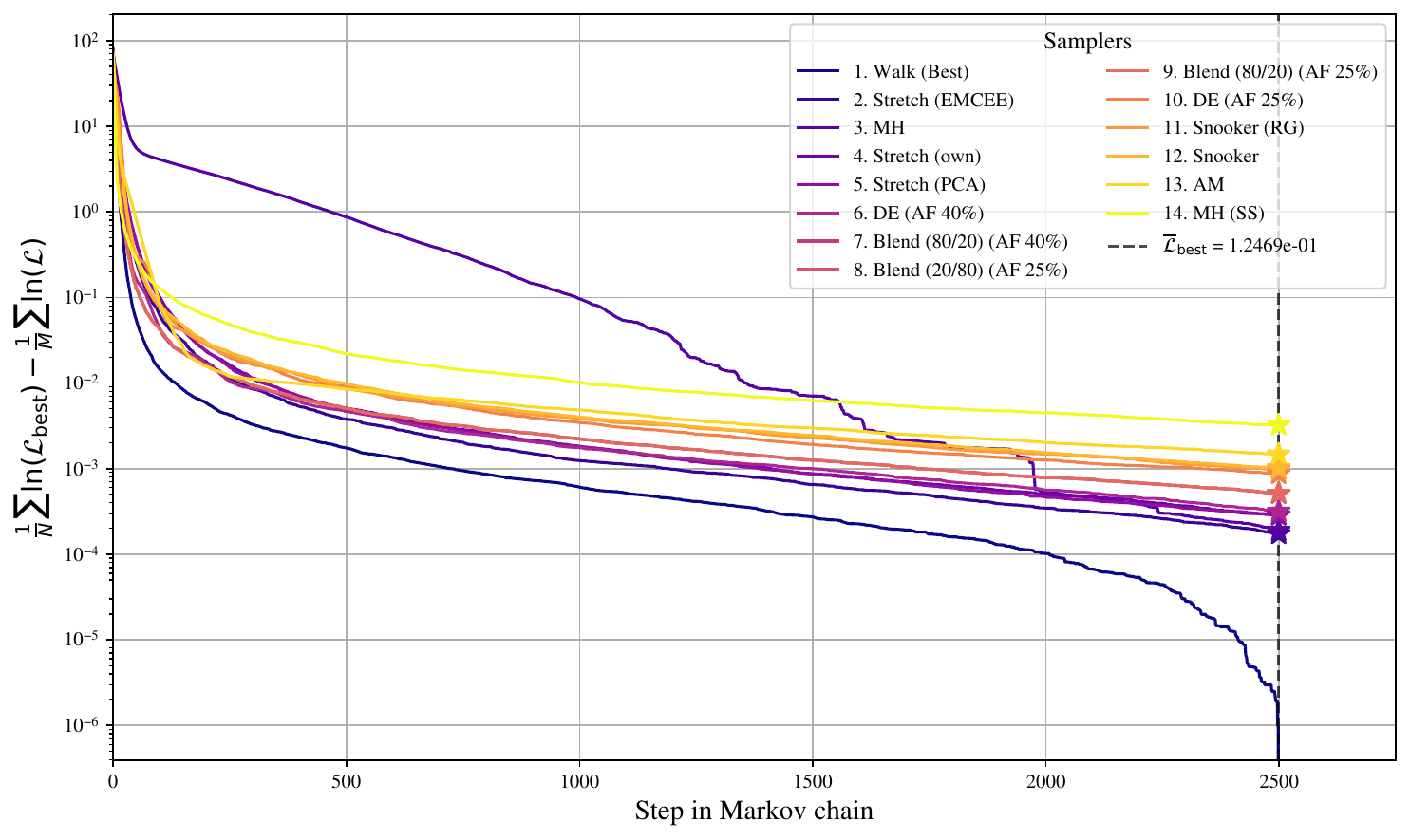}
    \caption{As figure~\ref{fig:CH4_evo_Rosenbrock}, but for Neal's funnel test case.}
    \label{fig:CH4_evo_Neal}
\end{figure*}

We present the results of the log-likelihood evolution for the unimodal landscapes in figures~\ref{fig:CH4_evo_Rosenbrock} and \ref{fig:CH4_evo_Neal}. From these figures, we see that the walk move outperforms the other samplers. 
The MH-based moves perform the worst as seen from their lowest ranking of 12, 13 and 14 in the Rosenbrock case. However, in Neal's funnel case, the MH algorithm itself ranks in the top three samplers. This is also seen from its curve, which is significantly different from all sampler curves. The adaptive Metropolis (AM) and the MH with standardised scaling version remain the worst among the pool of samplers for Neal's funnel. 

Between the two unimodal test cases, we see notable differences in the shape of the curves. In the Rosenbrock case, it takes the samplers longer to decrease the distance to the best likelihood of the best sampler, whereas in Neal's funnel, the initial sampling phase has the largest decrease in distance.
Additionally, in figure~\ref{fig:CH4_evo_Rosenbrock}, all samplers from the top six and upward have a log-likelihood ratio value above 1 compared to the best sampler. 
We do not see this for Neal's funnel; all samplers are within a log-likelihood ratio value of $10^{-2}$.

Notably, the snooker moves are ranked in both test cases near the lower end of the performance chart. The stretch moves cluster near each other in the ranking; however, the stretch move from \texttt{EMCEE} does have some distance to our implementation of the stretch move and the PCA stretch move. It is unclear where the difference directly originates from; we suspect that it arises from subtle differences in our implementation of the random number generation, meaning that the samplers traverse the landscape differently, but converge to the same target distribution. 

Between the blend moves, we see that the performance in the Rosenbrock case is much more spread than in Neal's funnel. The blend move with more differential evolution performs better in the Rosenbrock case, whereas in Neal's funnel, the performance of the blend moves is comparable.
From the differential evolution moves, we see that they perform well in the Rosenbrock toy model, but not so well in Neal's funnel.

\subsubsection{Application to multimodal landscapes}\label{sec:log_evo_multimodal}

\begin{figure*}
    \centering
    \includegraphics[width=0.8\linewidth]{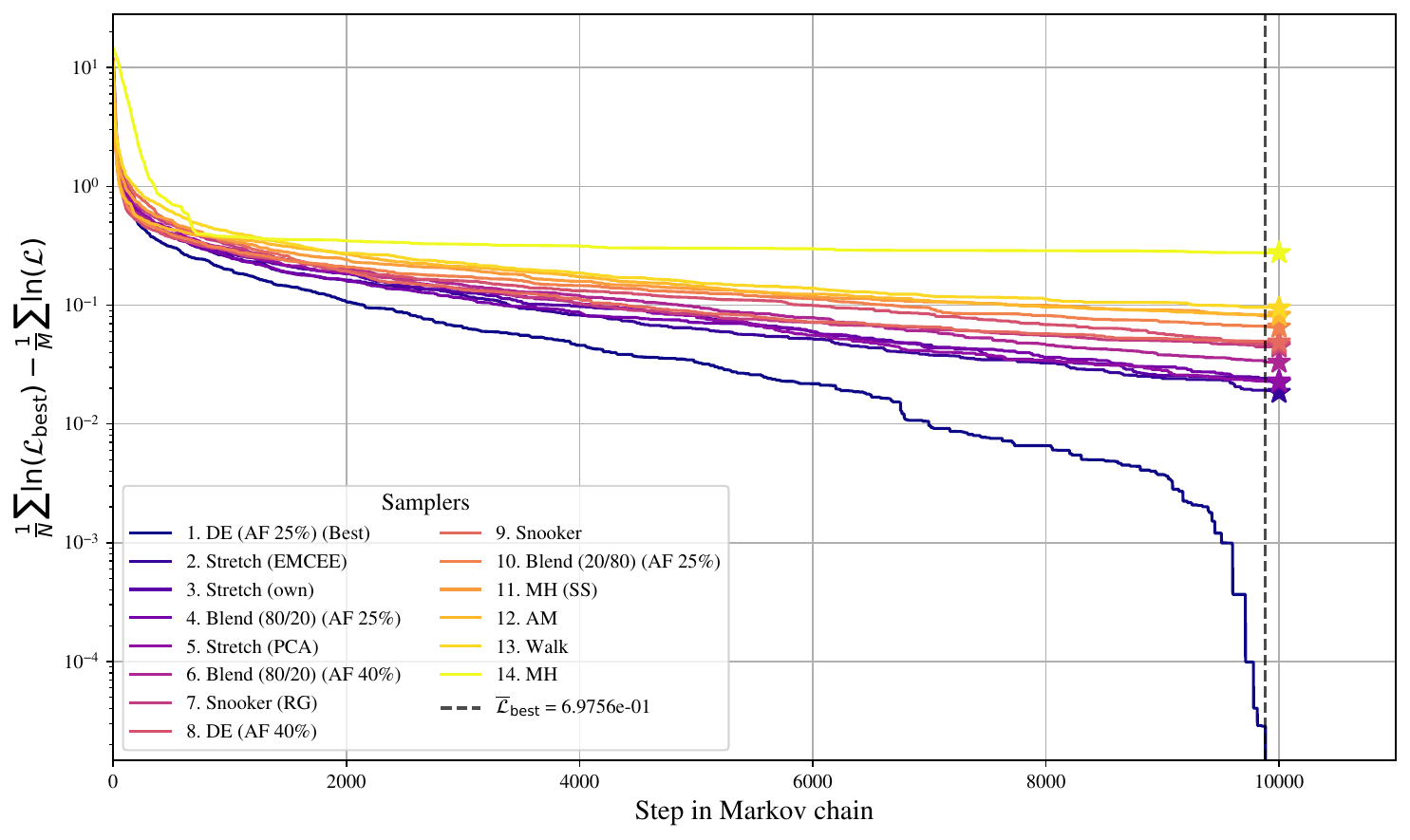}
    \caption{Sampler log-likelihood evolution compared to the best sampler of the 3D Gaussian toy model. The curves are generated by the difference between the average median log-likelihoods of samplers compared to the best sampler. The best sampler's best likelihood is indicated with a dashed line to highlight the position in the chain where this was reached on average.}
    \label{fig:CH4_evo_3D}
\end{figure*}

\begin{figure*}
    \centering
    \includegraphics[width=0.8\linewidth]{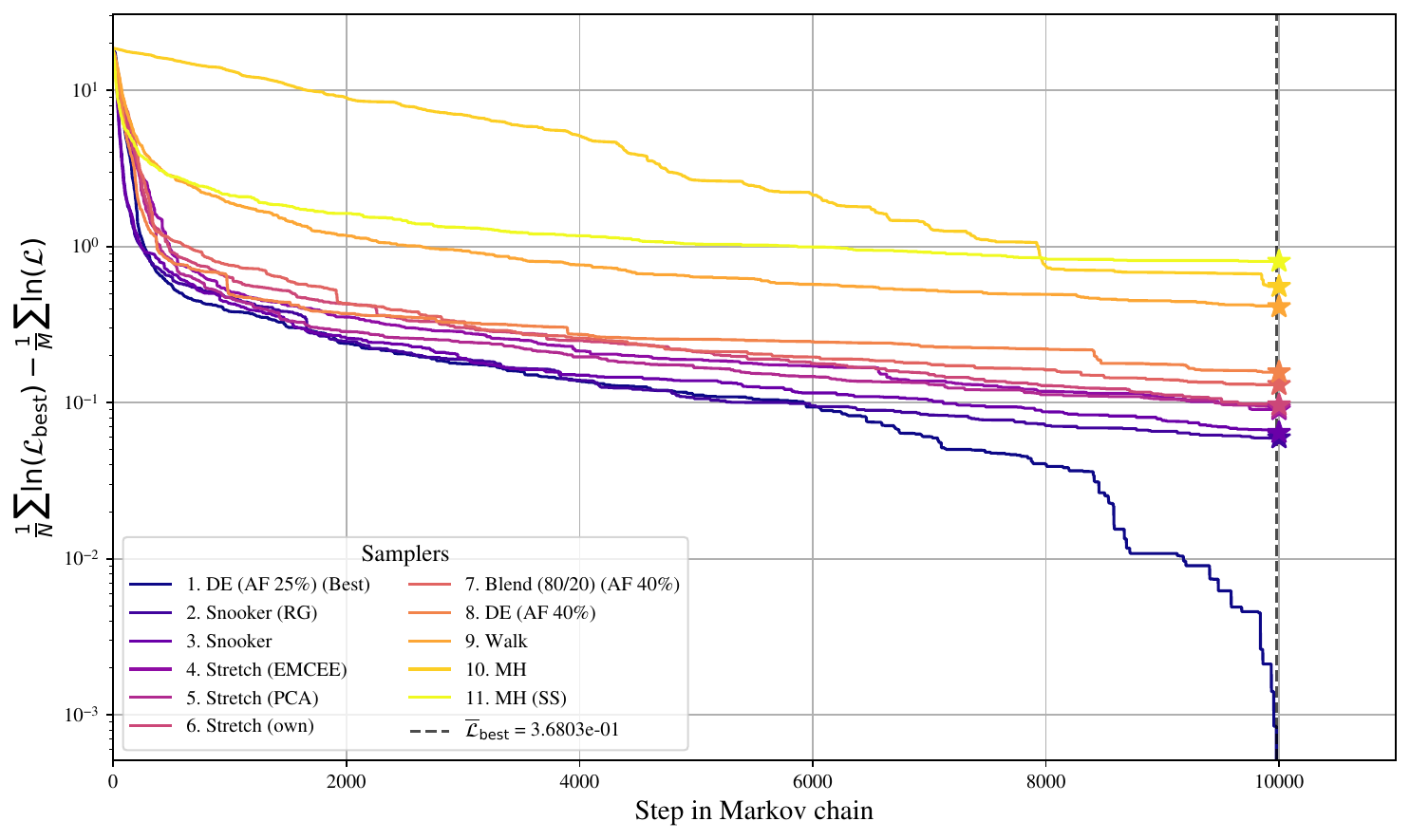}
    \caption{As figure~\ref{fig:CH4_evo_3D}, but for the 5D Gaussian toy model.}
    \label{fig:CH4_evo_5D}
\end{figure*}

\begin{figure*}
    \centering
    \includegraphics[width=0.8\linewidth]{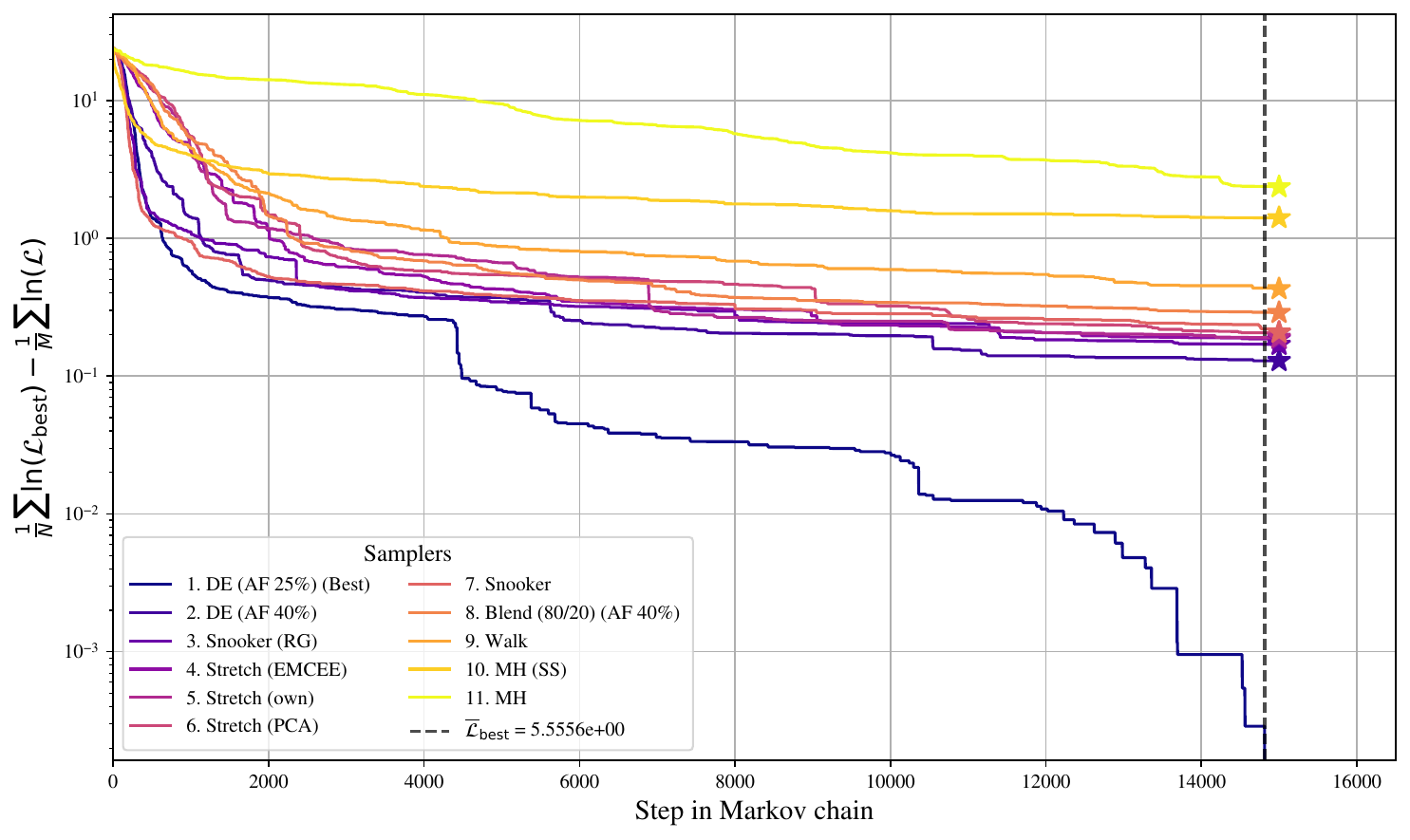}
    \caption{As figure~\ref{fig:CH4_evo_3D}, but for the 8D Gaussian toy model.}
    \label{fig:CH4_evo_8D}
\end{figure*}

We present the results of the evolution of log-likelihoods for the multimodal toy models in figures~\ref{fig:CH4_evo_3D}, \ref{fig:CH4_evo_5D} and \ref{fig:CH4_evo_8D}. From these, we see that the differential evolution move with a target acceptance fraction of 25\% outperforms all samplers in each test case. Notably, the magnitude by which it outperforms the other moves increases with dimensionality. For 3D, the closest sampler is within a log-likelihood difference of $\sim 2\cdot10^{-2}$, whereas for 5D this is $\sim 6\cdot10^{-2}$ and $\sim 0.5\cdot10^{-1}$ for 8D. 

Interestingly, for increasing dimensionality, we see that the snooker moves gain ground compared to the other moves. For the 5D test case, it ranks second and third; in the 8D model also ranks third. The snooker move obtains curves below the stretch-based moves in the 5D and 8D test cases. 
Given the stretch moves, we see, in addition to the unimodal results, that the stretch move from \texttt{EMCEE} is the best-performing. It has the lowest curves among the stretch moves we tested, although the results are close.

Additionally, considering the Metropolis-Hastings-based algorithms, we see that standardised scaling has a head start compared to the regular MH algorithm. In the 3D and 8D cases, it maintains this advantage, resulting in a curve below the MH curve. In the 5D case, this does not occur.
Notably, the MH-based variants are isolated and at the top of the results for the 5D and 8D cases, followed by the walk move, until a jump is made to the ensemble methods that utilise two or three walkers in their proposals. The figures indicate that ensemble-based sampling outperforms classic MCMC in multimodal settings.

\subsection{The robustness of samplers}\label{sec:robustness_metric}

In addition to evaluating the average convergence behaviour of samplers, we also assess their \textit{robustness}. This metric captures the consistency and reliability of a sampler across multiple independent runs, providing insight into its stability. Contrary to the previous metric, we store the chains with the highest and lowest achieved likelihoods of a sampler run. These are then used to construct populations of best- and worst-performing chains through the number of runs. We then compute the average of the best and worst chain populations, and enforce likelihood monotonicity again for comparability of the profiles. The difference between these two averaged population chains reflects the extent to which a sampler's performance can vary. A small difference indicates that the worst chains behave similarly to the best ones, signifying high robustness. Conversely, a large gap between the two suggests that the sampler's performance is sensitive to initialisation or stochastic effects and thus leads to low stability. This measure is particularly relevant when computational resources are limited and only a small number of chains can be run. In such settings, a robust sampler increases the possibility of achieving satisfactory convergence in a single run. When interpreted in conjunction with the log-likelihood ratio evolution curves, this robustness metric helps to identify samplers that are not only performant but also reliably so. The ideal sampler excels in both metrics, having quality convergence across runs, while also converging quickly and efficiently.

\subsubsection{Application to unimodal landscapes}

\begin{figure*}
    \centering
    \includegraphics[width=0.8\linewidth]{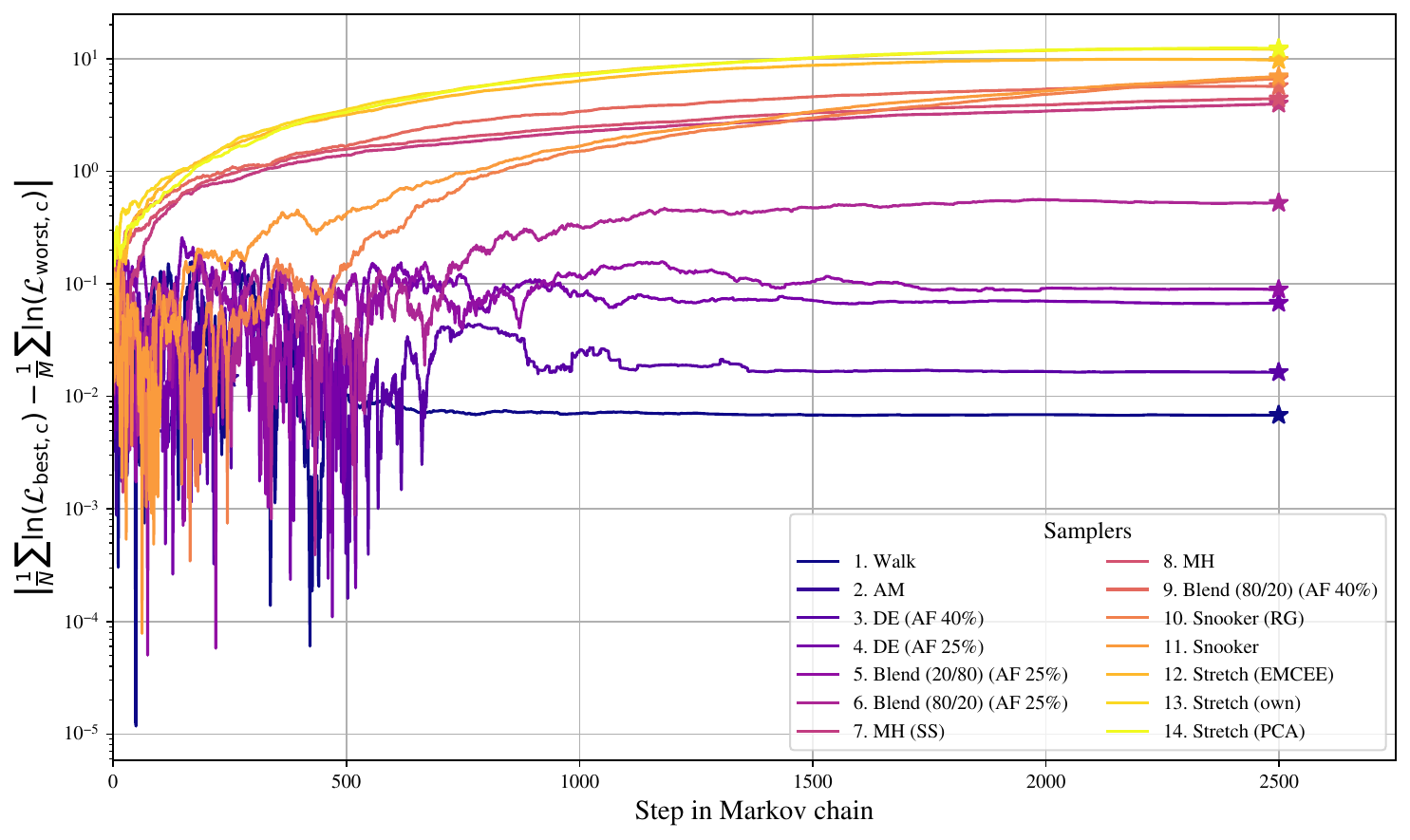}
    \caption{Robustness curves for the samplers applied to the Rosenbrock test case. Each curve is constructed by using the worst and best populations of chains across all runs. These are based on the overall best log-likelihood found in each run. The samplers are ranked according to their final robustness score and are presented in the legend. }
    \label{fig:CH4_robustness_Rosenbrock}
\end{figure*}

\begin{figure*}
    \centering
    \includegraphics[width=0.8\linewidth]{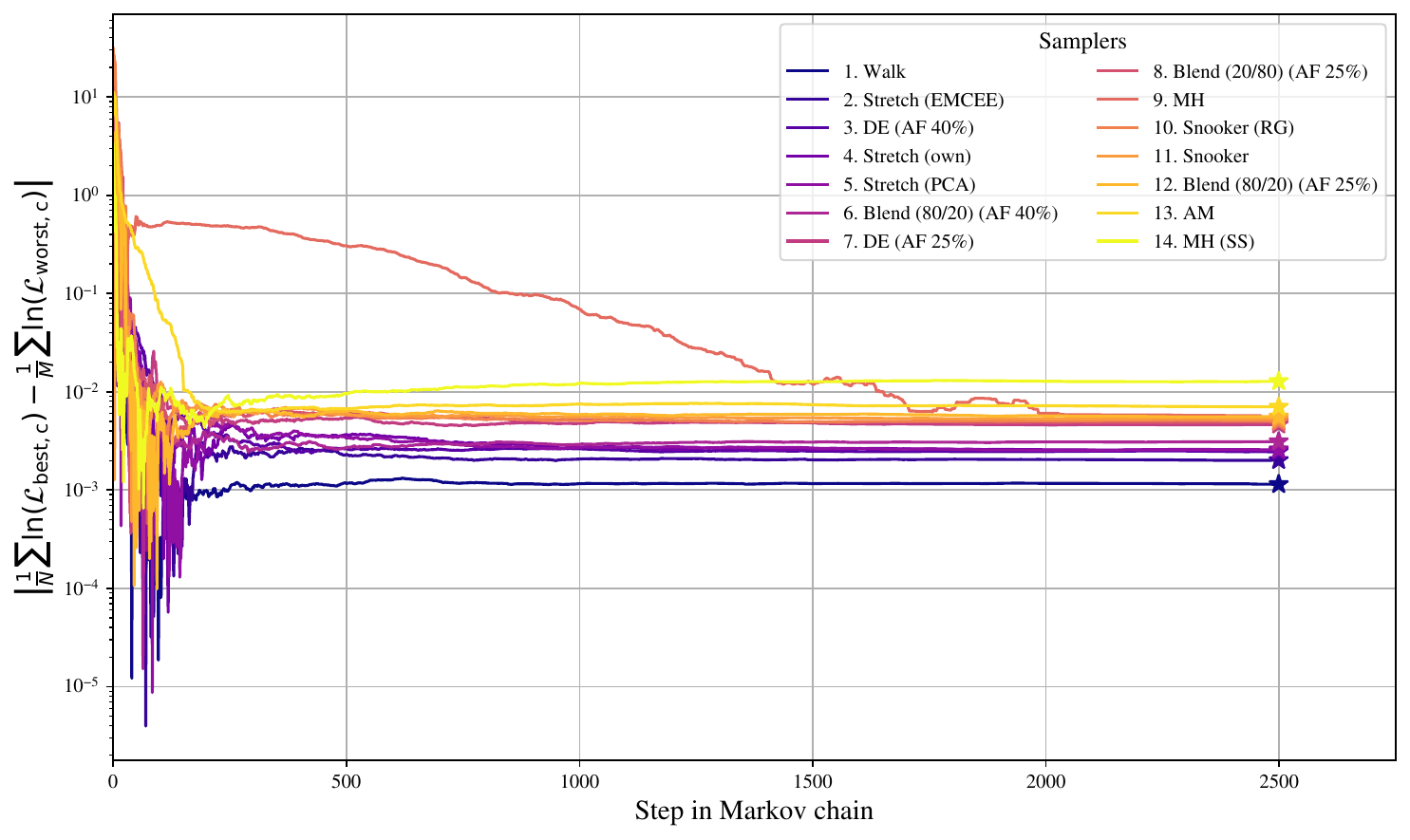}
    \caption{As figure~\ref{fig:CH4_robustness_Rosenbrock}, but for Neal's funnel test case.}
    \label{fig:CH4_robustness_Neal}
\end{figure*}

We present the results of the robustness metric of the Rosenbrock and Neal's funnel toy models in figures~\ref{fig:CH4_robustness_Rosenbrock} and \ref{fig:CH4_robustness_Neal}, respectively. From these, we see that the walk move has the largest robustness. It achieves a deviation of $\sim 3\cdot 10^{-2}$ in log-likelihood for the Rosenbrock test case and $\sim 10^{-3}$ in Neal's funnel. In the Rosenbrock case, the stretch-based moves take the lowest rankings. Notably, their curves increase over time, as opposed to the flattened levels achieved by the walk, adaptive Metropolis, and DE moves. Importantly, having a level curve is only a measure of good robustness whenever it is located at a low value. The worst-performing moves have a discrepancy above a log-likelihood of 1 in the Rosenbrock test case. In Neal's funnel, this is at most $10^{-2}$.

\noindent 
Given both figures, we see that there exists an initial phase in which the differences fluctuate frequently, after which the curves stabilise. The position of this varies per sampler, as seen in both figures. In the Rosenbrock figure, after roughly 700 steps, the samplers have relatively converged. In Neal's funnel, this occurs for around 250 steps in the Markov chain.

\newpage
\subsubsection{Application to multimodal landscapes}
\begin{figure*}
    \centering
    \includegraphics[width=0.8\linewidth]{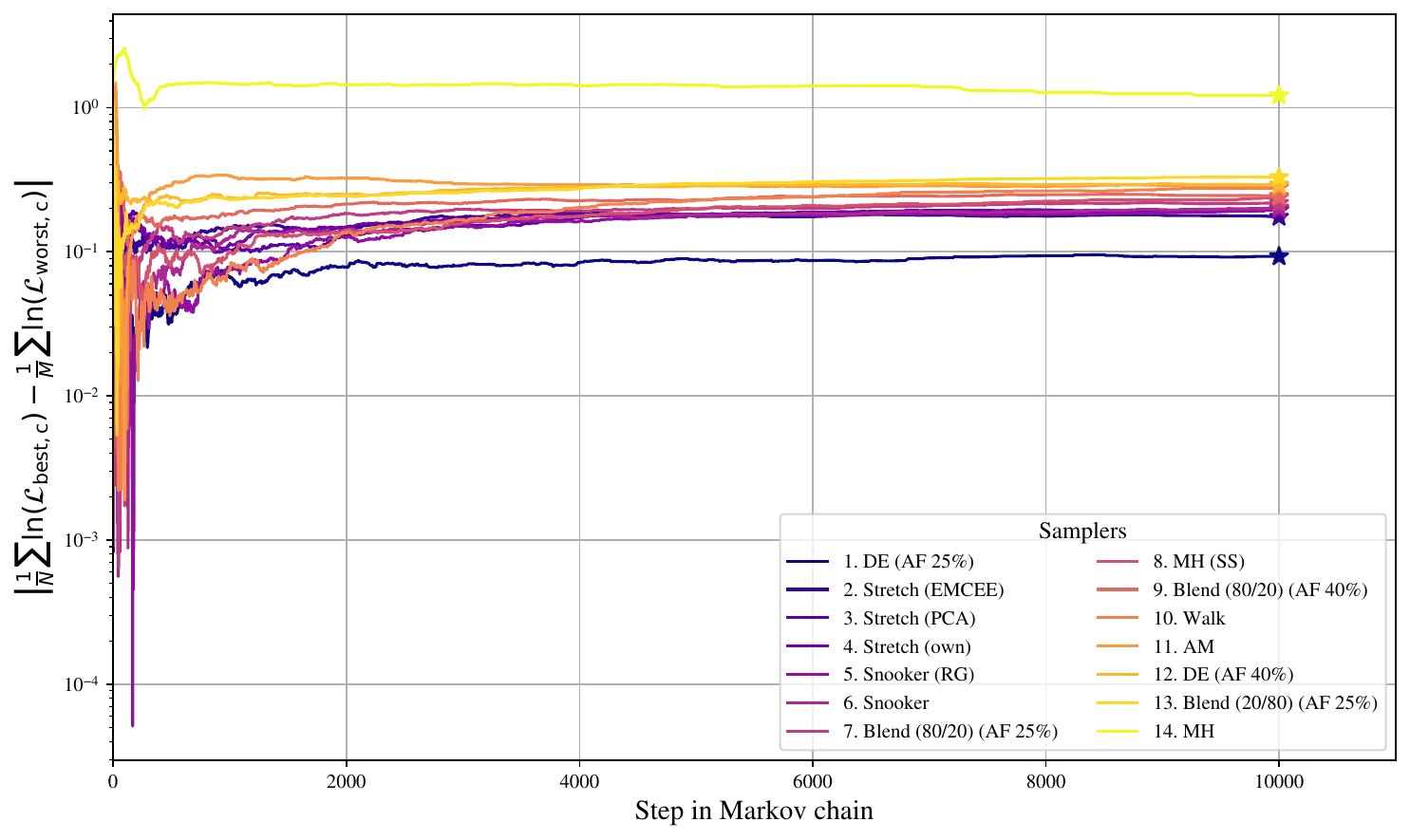}
    \caption{Robustness curves for the samplers applied to the 3D Gaussian toy model. The curves are constructed using the worst and best populations of chains across all runs. These are based on the overall best log-likelihood found in each run. The samplers are ranked according to their final robustness score and are presented in the legend.}
    \label{fig:CH4_robustness_3D}
\end{figure*}

\begin{figure*}
    \centering
    \includegraphics[width=0.8\linewidth]{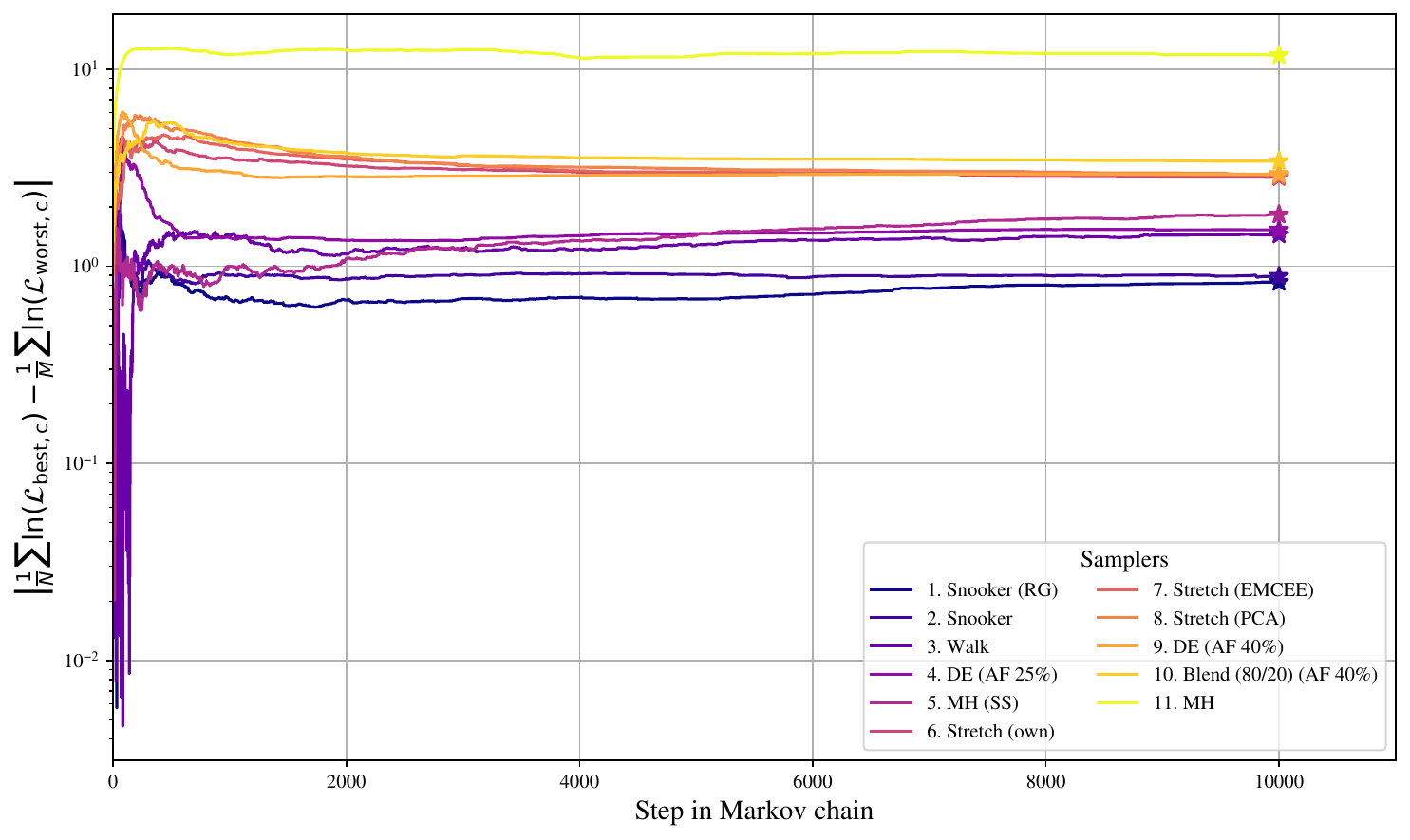}
    \caption{As figure~\ref{fig:CH4_robustness_3D}, but for the 5D toy model.}
    \label{fig:CH4_robustness_5D}
\end{figure*}

\begin{figure*}
    \centering
    \includegraphics[width=0.8\linewidth]{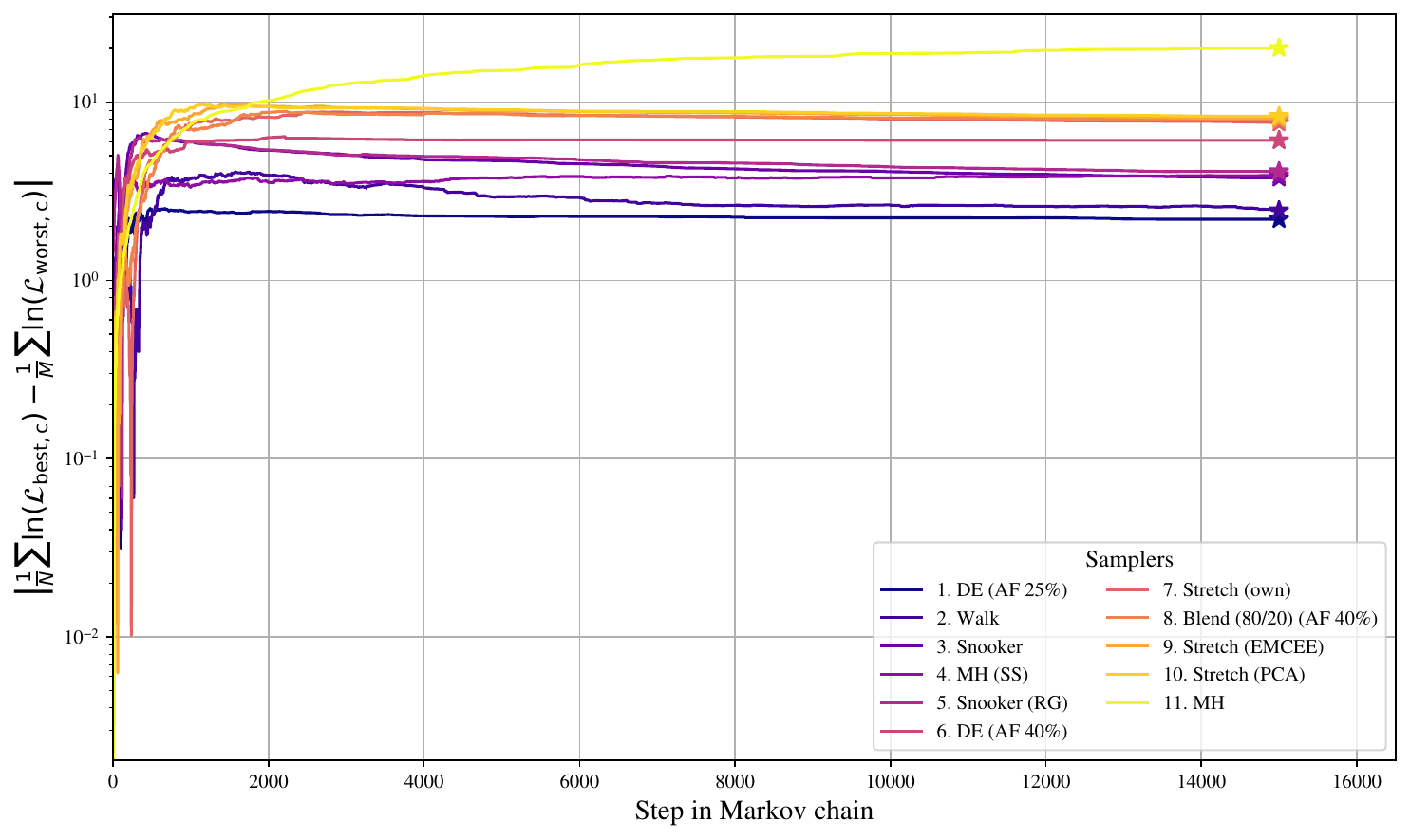}
    \caption{As figure~\ref{fig:CH4_robustness_3D}, but for the 8D toy model.}
    \label{fig:CH4_robustness_8D}
\end{figure*}

The robustness results of our three-, five-, and eight-dimensional test cases are presented in figures~\ref{fig:CH4_robustness_3D}, \ref{fig:CH4_robustness_5D}, and \ref{fig:CH4_robustness_8D}, respectively. From these, we see similar behaviour to the unimodal robustness results. Most samplers achieve a flattened regime after an initial phase of fluctuations.  The worst-performing sampler in all our multimodal test cases, in terms of robustness, is the Metropolis-Hastings sampler. Notably, for increasing dimensionality, the discrepancy between the worst and best chains grows for all samplers. In the 3D test case, the highest discrepancy is found for the Metropolis-Hastings algorithm at a log-likelihood difference value of $\sim 1$, whereas for the 5D and 8D cases, this is for $\sim 10$ and $\sim20$. The differential evolution move with a target acceptance fraction of 25\% performs well across all test cases; it never drops out of the top 5. For the 3D and 8D cases, it is the best-performing sampler. However, this does not apply to the 40\% version. This version is located near the lower-performing end of the robustness.

Interestingly, the snooker moves perform well in the 5D and 8D test cases. In the 5D case, it comes on top of the robustness ranking. In the 8D case, it performs as the third and fifth best. For the 3D model, it is outperformed by all stretch moves and the DE 25\% algorithm; nevertheless, it is frequently located in or around the top 5.
Moreover, from the figures, we see that standardisation results in an improvement of robustness compared to the default MH algorithm. In each multimodal test case, the standardised scaling approach has a curve located lower than the regular MH sampler. The robustness of the walk move in the unimodal test cases does not transfer directly to the multimodal cases, as it is no longer outperforming the other algorithms with some margin. Nevertheless, it still appears to be more stable than the Adaptive Metropolis and Metropolis-Hastings algorithms.

\newpage
\subsection{Improving on the best set of parameters}\label{sec:improving_best_estimates}
Once MCMC sampling has been performed for a given likelihood landscape, a natural follow-up question arises: \say{Can we further improve upon the best parameter set identified during sampling?}. This question is particularly relevant for complex models, where even marginal improvements in the parameter set can yield noticeably better model fits. While MCMC methods are primarily designed to approximate the posterior distribution rather than to optimise the likelihood explicitly, they often produce samples that lie close to the global optimum. However, due to their stochastic nature and finite sampling resolution, these methods may not land exactly on the optimal likelihood point. This observation motivates the use of optimisation techniques as a post-processing step to refine the best sample obtained. To pursue this refinement, we apply a deterministic optimisation algorithm starting from the best-likelihood sample identified across all samplers. Specifically, we use the Downhill Simplex (DHS) method, also known as the Nelder–Mead algorithm \citep{neldermead1965}. This is a derivative-free optimiser that is well-suited for noisy, non-smooth, or non-differentiable objective functions. It aims to climb the final stretch of the likelihood surface from a promising initial point, potentially uncovering a better-fitting parameter configuration. In our experiments, we apply the DHS algorithm to our multimodal toy landscapes. We exclude the unimodal cases from this step, assuming that their relatively simple structure allows MCMC to sample already close enough to the optimum.

\begin{table}
\centering
\caption{Configuration parameters for the Downhill Simplex algorithm.}
\label{tab:dhs_settings}
\begin{adjustbox}{width=0.32\textwidth}
\begin{tabular}{@{}p{3.5cm} p{1.4cm}@{}}
\toprule
\textbf{Parameter} & \textbf{Value} \\
\midrule
Initial step size & 0.005 \\
Convergence tolerance & $10^{-10}$ \\
Maximum number of iterations & 10\,000 \\
Objective function & $\mathcal{L}$ as equation~\eqref{eq:NdimGaussianLhood} \\
\bottomrule
\end{tabular}
\end{adjustbox}
\end{table}

\subsubsection{Application to toy models}\label{sec:DHS_application}
For our Downhill Simplex algorithm, we use the settings as described in table~\ref{tab:dhs_settings}. It is applied to the best set of parameters from the samples of the differential evolution algorithm with a target acceptance fraction of $25\%$. This sampler was chosen based on the results from sections~\ref{sec:log_likelihood_performance} and \ref{sec:robustness_metric}. The results are presented in table~\ref{tab:dhs_results_all}. From these, we see that the best-found likelihood is always an improvement of the optimum found from MCMC. The largest log-likelihood improvement is found for the 8D toy model, improving the log-likelihood from $\sim5.82$ to $\sim 6.19$. The smallest improvement is found for the 5D test case, improving from a log-likelihood of $\sim0.666$ to $\sim0.677$. Additionally, we can see that for increasing dimensionality, the number of required iterations for convergence increases. This is expected as the complexity increases for higher dimensionality. Importantly, this can halt the value of applying an optimisation algorithm since higher-dimensional likelihood functions are more costly to evaluate. Nevertheless, these results indicate that there is a potential room for improvement on the best found set of parameters.

\begin{table}
\centering
\caption{Results of DHS refinement across Gaussian landscapes of increasing dimensionality.}
\label{tab:dhs_results_all}
\begin{adjustbox}{width=0.45\textwidth}
\begin{tabular}{@{}lcccc@{}}
\toprule
\textbf{Dimensionality} & \textbf{Initial log$(\mathcal{L})$} & \textbf{Refined log$(\mathcal{L})$} & \textbf{Improvement} & \textbf{Iterations} \\
\midrule
3D & 0.748156 & 0.773032 & 0.0248766 & 75 \\
5D & 0.666361 & 0.677412 & 0.0110508 & 175 \\
8D & 5.82135  & 6.19364  & 0.372287  & 1\,270 \\
\bottomrule
\end{tabular}
\end{adjustbox}
\end{table}

\section{Discussion}\label{sec:CH5}
\subsection{Toy models}
For the construction of our multimodal landscapes, we use the Gaussian function as described in equation~\eqref{eq:NdimGaussian}. It is worth noting that alternative functions to the Gaussian exist and could potentially improve computational efficiency in higher dimensions. An interesting alternative could involve constructing landscapes based on functions that avoid the exponential calculation inherent to Gaussians. However, the bottleneck of our likelihood function is in the logarithm of the sum of individual Gaussians; therefore, improvements should primarily focus on this operation.

Additionally, it would be worthwhile to extend the dimensionality of our test cases to higher cases since these would be particularly beneficial for understanding the behaviour of each sampler in more complex problems. This, in turn, could help users select the most suitable sampler based on the dimensionality of their application. Given our current highest dimensionality of eight, our results might be limited; a higher dimensionality test case would give more weight to our outcomes. However, given the exponential increase in computational cost with dimensionality, our test would need to be done with a significantly smaller number of chains and/or re-runs.

\subsection{Ergodicity metric}
The ergodicity metric provides a diagnostic for evaluating how well different MCMC samplers can recover the expected distribution of samples across a discretised likelihood landscape. When comparing the results across the three toy models of increasing dimensionality, several consistent trends and notable deviations emerge. In the 3D case, nearly all ensemble-based samplers demonstrate close agreement with the expected occupation values, indicating highly ergodic behaviour. Additionally, both the original and our implementation of the stretch move closely follow the theoretical expectations, validating the correctness of our implementation and suggesting that the stretch move is highly ergodic in low dimensions. Moreover, newly introduced moves, such as the PCA stretch and blend moves, display behaviour comparable to the more established ensemble moves, reinforcing the correctness of their implementation and potential usefulness. Among the blend moves, the (20/80) blend with a 25\% target acceptance fraction stands out as the most ergodic, effectively balancing the exploratory strength of the DE with the regularising influence of the stretch move. These observations highlight the value of combining move types to modulate the exploratory-exploitative trade-off. The Metropolis-Hastings (MH) sampler, however, shows clear deviations, with several cells being oversampled relative to their expectation values. This suggests that the MH sampler can become trapped in local optima more easily, leading to prolonged sampling in specific regions and an inflated local density.

Moving to five dimensions, the sampling challenge increases, and the differences between samplers become more pronounced. The MH algorithm now underestimates the occupation of nearly all cells, markedly worse than in 3D. This confirms that the MH struggles to maintain coverage as dimensionality increases, or at least that the length of its burn-in phase increases more strongly with dimensionality than for ensemble methods. Interestingly, most ensemble samplers begin to exhibit some degree of overestimation in high-expectation cells. For instance, the DE with a 25\% target acceptance fraction, as well as the walk move, significantly oversample certain high-likelihood regions, with counts that exceed the expectation values by a factor of two or more. These high-likelihood modes act as strong attractors, particularly in complex, multimodal landscapes, where exploration becomes more difficult, and the chain can remain stuck. The stretch and snooker moves retain reasonable ergodicity but also begin to show small deviations compared to their strong performance in 3D. 

In eight dimensions, ergodic behaviour is compromised across all samplers. The MH sampler, in particular, fails to capture the expected distribution almost entirely, with the sparsest histogram coverage and minimal overlap with the expectation values. This underlines its ineffectiveness for high-dimensional, multimodal landscapes. Most samplers exhibit widespread underestimation of cell occupation of regions with significant likelihood, visible through a predominance of blue bins. This is particularly evident in samplers like stretch and PCA stretch, which lack coverage of the ergodic domain. The DE move with a 25\% acceptance target remains the only sampler to still overestimate certain high-likelihood regions, reinforcing the earlier observation that exploration-heavy configurations can lead to local overconcentration. Nevertheless, the snooker and DE (25\%) moves achieve the highest coverage of the expectation values in this setting, indicating that, despite some local overestimation, they provide the most ergodic behaviour under fixed settings. The reduced performance of stretch-based methods at high dimensionality suggests that they are less suited for strongly multimodal or high-dimensional exploration, where broader proposals or differential information are more beneficial.

Across all test cases, the observed ergodicity trends affirm that no sampler is universally optimal, but some are consistently more robust, in terms of ergodicity, than others. Importantly, in our analysis, we did not remove any burn-in phase. This choice was made since the size of the burn-in depends per testcase and the optimal size is not known exactly. However, by removing a burn-in period from the chains, the retained samples are expected to better represent the target distribution and therefore obtain a better ergodicity measure, since these early samples are strongly influenced by the initial conditions of the chains.

\subsection{Performance metric}
\subsubsection{Unimodal landscapes}
We observe from figures~\ref{fig:CH4_evo_Rosenbrock} and \ref{fig:CH4_evo_Neal} that the walk move is the most effective across both unimodal landscapes. This result can be attributed to its ensemble-based proposal strategy, which leverages the full state of the population. This allows it not only to rapidly approach the global optimum but also to iteratively refine its estimate as all walkers concentrate around the high-likelihood region. The continued improvement in its curve reflects this adaptive refinement.

In contrast to this, the Metropolis-Hastings (MH) based samplers generally underperform. In the Rosenbrock case, they occupy the lowest rankings, likely due to the local, uninformed nature of their proposals, which struggle with the sharp curvature and narrow optimum region. Interestingly, in Neal’s funnel, the plain MH algorithm ranks among the top three samplers. This divergence in performance suggests that while MH may initially struggle to explore complex regions, once it locates the optimum, its consistent local exploration enables further refinement, resulting in a noticeable late-stage improvement in its log-likelihood curve.

The difference in convergence behaviour across the two test cases is interesting. In Neal’s funnel, most of the reduction in log-likelihood ratio occurs early in the sampling process. This suggests that the global optimum is easier to locate in this case, likely due to its broader and more accessible geometry. In contrast, the Rosenbrock function’s narrow valley and strong curvature delay convergence, making it more challenging for samplers to approach the optimum early on. This difficulty is also reflected in the absolute log-likelihood ratios. In figure~\ref{fig:CH4_evo_Rosenbrock}, all samplers outside the top five exhibit log-likelihood ratios greater than one, implying their maximum likelihoods are at least a factor of $e$ lower. Neal’s funnel shows no such discrepancy; all samplers converge within a range of $10^{-2}$.

Looking at the performance across move types, the snooker moves consistently rank near the bottom in both landscapes, suggesting limited performance across these relatively easy test cases. The stretch moves, while similarly clustered, reveal subtle implementation differences. The \texttt{EMCEE} version achieves slightly better performance, possibly due to small deviations in tuning or algorithmic detail compared to our versions. Blend moves demonstrate more varied behaviour. In the Rosenbrock landscape, the variant incorporating more differential evolution performs notably better. In Neal’s funnel, however, performance across blend variations is more uniform. We note that differential evolution-based samplers are particularly effective on the Rosenbrock landscape but less so on Neal’s funnel. This contrast highlights the dependency of sampler performance on landscape structure; strategies that excel at exploring narrow, curved optima may not offer the same advantages in broader, flatter regions, as in Neal's funnel.

\subsubsection{Multimodal landscapes}
The log-likelihood evolution results for the 3D, 5D, and 8D multimodal toy models, shown in figures~\ref{fig:CH4_evo_3D}, \ref{fig:CH4_evo_5D}, and \ref{fig:CH4_evo_8D}, indicate how sampler performance changes with dimensionality. Across all three test cases, the differential evolution (DE) sampler with a 25\% target acceptance fraction consistently achieves the highest likelihoods, and its relative advantage becomes increasingly pronounced with dimensionality. This scaling behaviour strongly suggests that DE-based samplers are particularly well suited to high-dimensional inference problems. Further investigation into their performance at even higher dimensions may yield valuable insight to understand the extent to which the DE sampler performs well.

Interestingly, snooker moves improve with dimensionality as well. While they lag in 3D, they consistently rank among the top performers in both the 5D and 8D cases. These results indicate that the snooker move may offer satisfactory exploration capabilities in higher dimensions. This and the results of the DE sampler are especially relevant given the continued popularity of the stretch move as implemented in \texttt{EMCEE}, which performs reasonably well but is outpaced by both these samplers in the higher-dimensional scenarios. Utilising these methods over the stretch move can result in better results. Adding onto the performance of \texttt{EMCEE}'s stretch move, their stretch move has a performance edge over our implementations; it is evident but minor. This suggests that while it remains a robust default, there remains room for improvement in our implementations of this move.

The Metropolis-Hastings variants generally perform poorly, particularly in higher dimensions, where their curves in figures~\ref{fig:CH4_evo_3D}, \ref{fig:CH4_evo_5D}, and \ref{fig:CH4_evo_8D} are isolated above those of the ensemble-based methods. The MH algorithm with standardised scaling fares especially poorly in the 5D and 8D test cases. In 3D, the MH SS variant maintains an advantage over the standard MH approach, but in 5D, this reverses. This is likely due to the increased multimodality and the oversized proposal width in the standardised version.

\subsection{Robustness metric}

\subsubsection{Unimodal landscapes}
The robustness metric offers a clear view into the consistency of sampler performance across runs. An ideal sampler would exhibit a small and stable difference between its best and worst-performing chains. Such a scenario would indicate that improvements are shared throughout the ensemble. In our unimodal results of figures~\ref{fig:CH4_robustness_Rosenbrock} and \ref{fig:CH4_robustness_Neal}, samplers such as the walk move and differential evolution show this desirable behaviour, as evidenced by their relatively flat robustness curves over chain length. Robustness curves that plateau over time indicate that both best and worst chains improve at a similar pace, maintaining a constant deviation. This balance is a hallmark of more reliable samplers and contrasts sharply with methods where the deviation continues to grow.

In contrast to this, the increasing curves seen for stretch-based moves, particularly in the Rosenbrock test case, suggest that while some chains continue to improve their log-likelihood, others stagnate. This widening gap between the best and worst chains points to walkers becoming locally trapped, failing to benefit from the ensemble’s collective progress. Such divergence undermines robustness, as it reflects uneven performance within the same sampler. Interestingly, differential evolution avoids this pattern. Its update mechanism, relying on three walkers, may reduce the risk of decoupling between ensemble members, effectively maintaining performance across chains. Similarly, the walk move leverages the full ensemble during proposals, which likely contributes to its strong robustness profile, especially in unimodal cases.

In practical terms, log-likelihood discrepancies exceeding one highlight significant instability across chains and should be treated with caution. Comparing the Rosenbrock and Neal's funnel test cases reveals a marked difference in difficulty. Rosenbrock poses a greater challenge, as indicated by its longer transient phase before robustness stabilises and the larger chain-to-chain discrepancies. Neal’s funnel, in contrast, reaches stability faster and exhibits minimal growth in robustness deviation, suggesting it is an easier landscape for methods to navigate.

\subsubsection{Multimodal landscapes}
The robustness results for the multimodal landscapes, in figures~\ref{fig:CH4_robustness_3D}, \ref{fig:CH4_robustness_5D} and ~\ref{fig:CH4_robustness_8D} reveal several differences compared to the unimodal settings. Notably, the robustness curves in higher-dimensional Gaussian models display less fluctuation in the early stages of the chain, suggesting that samplers exhibit more stable behaviour throughout their chain lengths. However, this apparent stability is offset by a pronounced increase in the discrepancy between the best and worst-performing chains as dimensionality increases. This widening gap indicates that individual chains are more prone to becoming trapped in local optima, with the overall sampling quality becoming increasingly uneven in higher dimensions. The Metropolis-Hastings algorithm is particularly sensitive to this effect; in each of the multimodal test cases, it shows the largest robustness deviation, far exceeding all other samplers.

Among the ensemble methods, the differential evolution move with a 25\% target acceptance fraction consistently performs well. It remains within the top five samplers in all test cases and reaches the top rank in both the 3D and 8D settings. This consistency highlights the importance of controlled exploration in achieving robustness: with a lower acceptance fraction, the 25\% variant maintains a broader proposal distribution, encouraging more global movement across the landscape. In contrast, the 40\% variant, favouring more localised, exploitative moves, suffers a degradation in robustness performance. This underlines the key role of target acceptance fraction tuning in balancing exploration and exploitation. Further investigation into the optimal choice of this parameter could enhance the performance of the algorithm further. 

The snooker update moves emerge as particularly effective in higher-dimensional settings. While their performance is less competitive in the 3D case, they achieve top-tier rankings in the 5D and 8D models, suggesting that their mechanics may be better suited to navigating high-dimensional multimodal spaces. This trend mirrors their improved behaviour in the log-likelihood evolution, supporting the idea that the snooker strategy benefits from the increased complexity and volume of higher-dimensional parameter spaces. 

Another interesting finding is the impact of standardised scaling in the Metropolis-Hastings sampler. While still outperformed by more modern methods, the standardised variant consistently shows improved robustness compared to the default MH version. This indicates that even relatively minor adjustments to scaling strategies can enhance traditional samplers. 

Finally, the walk move, which dominated the unimodal robustness results, does not carry the same advantage in multimodal landscapes. Although it still outperforms several methods, including Adaptive Metropolis and Metropolis-Hastings, its relative performance drops. This is likely due to the walk move’s global proposal mechanism, which becomes less effective in complex landscapes. When the target distribution has many modes or sharp local structures, proposals based on the full ensemble can lead to inefficient exploration, as individual walkers may be misled by globally averaged directions. The degradation in robustness is consistent with observations from the log-likelihood evolution, where the walk move also struggled to maintain its advantage in higher-dimensional settings.

\subsection{Improving on the best set of parameters}
The improvement in log-likelihood achieved by applying the Downhill Simplex algorithm illustrates that it is indeed possible to refine MCMC results through post hoc optimisation. This outcome aligns with expectations, as MCMC methods are primarily designed for sampling from the posterior rather than for locating its global optimum.

The magnitude of improvement varies notably across dimensions. The smallest gain occurs in the 5D case, likely due to its configuration: although it shares the same chain length as the 3D case, it employs twice the number of chains. This increased diversity in parameter space exploration may have already allowed the MCMC sampler to identify a solution close to the global optimum. In contrast, the 8D case shows a substantial improvement in log-likelihood, suggesting that the chain length may have been insufficient for adequately exploring the vastly larger parameter space of higher dimensions, leaving more room for post hoc optimisation to improve the result. Our results of the ergodicity metric for this test case also support this thought since all samplers were less ergodic than the lower-dimensional landscapes.

Additionally, the number of iterations required for convergence of the Downhill Simplex algorithm increases with dimensionality, and particularly so between 5D and 8D. It may indicate that the MCMC chain settings were insufficient to ensure convergence in high-dimensional spaces, making the optimiser's contribution more impactful in 8D than in 3D or 5D. Despite this, applying an optimiser to suboptimal MCMC settings can account for these and return us with better estimates for parameters through the likelihood function.

\section{Conclusions}
Our results show that, in terms of robustness, convergence speed, raw performance, and ergodicity, several modern samplers outperform the traditional MH algorithm. Among the tested methods, the differential evolution sampler with a 25\% target acceptance fraction emerged as the most effective, particularly on multimodal toy models. The walk move performed best in our unimodal cases, both in terms of performance and robustness. To compare samplers meaningfully, we introduced several metrics that quantify different aspects of sampler performance. These metrics, designed to be applicable across a broad range of likelihood landscapes, enable systematic benchmarking and could serve as a general-purpose framework for evaluating MCMC methods. We also explored the benefits of applying an optimisation algorithm to the final states of a Markov chain. This post-processing step proves fruitful for uncovering higher-likelihood parameter sets, thereby improving the predictive capabilities of models that rely on maximum-likelihood estimates. Beyond sampling, we demonstrated that collected MCMC samples can be repurposed to reconstruct the likelihood landscape using a quadtree-based approach. This reconstruction provides structural insight into the landscape and enables the identification of parameters within models.

While this work focused primarily on ensemble MCMC methods, future studies could explore non-ensemble alternatives within a gradient-free context. Comparing these approaches may reveal new advantages or highlight limitations not captured in this study. Further research should also investigate higher-dimensional test cases since our current toy models go up to 8 dimensions, while higher-dimensional cases could solidify our findings.


\section*{Data Availability}
All data used in this work was generated randomly. Instructions for repeating the generation can be found in section \ref{sec:multi_modal_landscapes}.



\bibliographystyle{mnras}
\bibliography{references} 




\appendix

\section{Stretch move}

\subsection{Affine invariance proof}\label{appendix:stretch_AI}
The proof of the affine invariance of the stretch move is as follows. Given the stretch move, equation~\eqref{eq:stretch_move}, a proposal is called affine-invariant if it follows the form of $Y=Ax+b$. If we  substitute the stretch move for $x$, we arrive at:
\begin{equation}
    Y_{i} = A(X_{j}+Z(X_{i}-X_{j}))+b.
\end{equation}
We can write out this expression, and add in factors of $b$:
\begin{equation}
    Y_{i} = AX_{j} + AZ(X_{i}-X_{j})+b =  (AX_{j}+b)+Z(AX_{i}-AX_{j}) 
\end{equation}
\begin{equation*}
    = (AX_{j}+b)+Z(AX_{i}-AX_{j} + b -b).
\end{equation*}
This can then be reshuffled to get:
\begin{equation}
    Y_{i} = (AX_{j}+b)+Z(AX_{i}+b-(AX_{j} + b)).
\end{equation}
From this point, we note that we can express all quantities in the form of $Y=Ax+b$, e.g. $Y_{j}=AX_{j}+b$; this gives us:
\begin{equation}
    Y_{i} = Y_{j}+Z(Y_{i}-Y_{j}),
\end{equation}
which is exactly the stretch move, but in the affine transformed space. Hence, the stretch move is affine-invariant.

\subsection{The stretch factor \textit{Z}}\label{appendix:sampling_z}
 In equation~\eqref{eq:stretch_move}, $Z$ is the stretch factor, which is sampled from a density $g$ of the form:
 \begin{equation}\label{eq:sampling_z_stretch_move_appendix}
 g(z) \propto
     \begin{cases}
         \dfrac{1}{\sqrt{z}}, & \text{for } z \in \left[\dfrac{1}{a},\, a\right] \\
        0, & \text{otherwise}.
     \end{cases}
 \end{equation}
\citet{goodman2010} discuss that $a$ should be larger than 1, but recommend to use $a=2$ for good MCMC performance. To sample $Z$ from the $g(z)$ distribution, we can use inverse transform sampling. We start by normalizing equation~\eqref{eq:sampling_z_stretch_move_appendix}, $g(z) = C/\sqrt{z}$ for a constant $C$. We solve $C$ by integrating $g(z)$. $g(z)$ is the probability density function (PDF) of $Z$ since the integral over the domain converges to 1:
\begin{equation}\label{eq:derivation_C_stretch}
    C\int_{1/a}^{a}\frac{1}{\sqrt{z}}dz = 1
\end{equation}
\begin{equation*}
    =C\int_{1/a}^{a}z^{-1/2}dz = C\left[2z^{1/2}\right]^{a}_{1/a}= 1
\end{equation*}
\begin{equation*}
    =C\left[2\sqrt{a}-\frac{2}{\sqrt{a}}\right]= 1
\end{equation*}
\begin{equation*}
    C = \frac{1}{2\left(\sqrt{a}- 1/\sqrt{a}\right)}.
\end{equation*}
To obtain the cumulative distribution function of $g(z)$, we integrate the PDF up to $z$, for $1/a\leq z\leq a$ instead of up to $a$ as in equation~\eqref{eq:derivation_C_stretch}. This results in: 
\begin{equation}
    G(z) = 2C\left(\sqrt{z}-1/\sqrt{a}\right),
\end{equation}
which, using the expression of $C$, turns into:
\begin{equation*}
    G(z) = \frac{\sqrt{z} - 1/\sqrt{a}}{\sqrt{a}-1/\sqrt{a}}.
\end{equation*}
To sample z, we use inverse transform sampling, based on the expression of $G(z)$. Since $0\leq G(z)\leq 1$, we define $u\sim \mathcal{U}(0,1)$, such that: 
\begin{equation}
    u = G(z) = \frac{\sqrt{z} - \frac{1}{\sqrt{a}}}{\sqrt{a}-\frac{1}{\sqrt{a}}}
\end{equation}
which we solve for $z$: 
\begin{equation}\label{eq:z_expression_stretch}
    z=u^{2} \left(a+\frac{1}{a}-2\right)+2u\left(1-\frac{1}{a}\right)+\frac{1}{a}
\end{equation}
We can validate the values of $z$ by filling in the upper and lower limits of $u$; these should give us the limits of the density $g$ as seen in equation~\eqref{eq:sampling_z_stretch_move}. We use a value of $a=2$ as recommended by \citet{goodman2010}. This results, for $u=0$, in $z=1/2$. For $u=1$ we obtain $z=2$, therefore, we see that $1/2\leq z\leq 2$, as expected.

\subsection{Detailed balance}\label{appendix:detailed_balance_stretch_move}
To satisfy detailed balance for the stretch move, we require a modification of the Metropolis-Hastings acceptance criterion. From \citet{goodman2010}, a $Z^{N-1}$ factor is introduced to ensure this condition is met. This results in equation~\eqref{eq:stretch_criteria}:
\begin{equation}\
\alpha = \min\left(1, Z^{N-1}\frac{\pi(X_{i}^{\text{new}})}{\pi(X_{i})}\right),
\end{equation}
where $N$ is the dimensionality of the parameter space, and $\pi(X)$ denotes the target density.

We can derive this criterion, starting from the Metropolis-Hastings acceptance criterion, equation~\eqref{eq:Metropolis_acceptance_criterion}. For the stretch move, the proposal distribution is asymmetric, which implies that:
\begin{equation}
\frac{q(X_{i}|X_{i}^{\text{new}})}{q(X_{i}^{\text{new}}|X_{i})} \neq 1.
\end{equation}
The ratio of proposal densities evaluates to:
\begin{equation}\label{eq:q_ratios_stretch}
\frac{q(X_{i}|X_{i}^{\text{new}})}{q(X_{i}^{\text{new}}|X_{i})} = \frac{|X_{i}^{\text{new}} - X_j|^{N-1}}{|X_{i} - X_j|^{N-1}} = Z^{N-1},
\end{equation}
where $X_j$ is the complementary walker and $X_{i}^{\text{new}}$ is defined via the stretch move, equation~\eqref{eq:stretch_move}. Substituting this into the proposal ratio yields the $Z^{N-1}$ factor, and thus we obtain equation~\eqref{eq:stretch_criteria}.

An intuitive geometric explanation supports this adjustment. The stretch move proposes a point along the line connecting the current walker $X_i$ and the complementary walker $X_j$. This can be parametrised as
\begin{equation}
X_{i}^{\text{new}}(\lambda) = X_j + \lambda (X_i - X_j).
\end{equation}
The conditional density of points along this line segment is proportional to

\begin{equation}
    ||X_{i}^{\text{new}}-X_{j}||^{N-1}\pi(X_{i}^{\text{new}}),
\end{equation}
which accounts for the volume element in $N$-dimensional space \citep{goodman2010}. For the reverse move from $X_{i}^{\text{new}}$ to $X_i$, the corresponding density is
\begin{equation}
    ||X_{i}-X_{j}||^{N-1}\pi(X_{i}),
\end{equation}
leading to the proposal ratio in equation~\eqref{eq:q_ratios_stretch} and justifying the $Z^{N-1}$ correction to maintain detailed balance.

\section{Differential evolution move}

\subsection{Affine invariance proof}\label{sec:DE_affine_proof}
Given the slightly different mathematical form of the DE move compared to the stretch move, we present here the proof of affine invariance for the DE move. We start from the DE move as described by equation~\eqref{eq:DE_move},
\begin{equation}
    X_{i}^{\text{new}} = X_{i} + \gamma (X_{k} - X_{j}).
\end{equation}
Which we will substitute into $Y=Ax+b$,
\begin{equation}
    Y_{i} = A(X_{i} + \gamma (X_{j} - X_{k})) + b = AX_{i}+b + A\gamma(X_{j}-X_{k}) 
\end{equation}
\begin{equation*}
    = Y_{i} + \gamma(AX_{j}-AX_{k}),
\end{equation*}
where we can use the trick of adding in $b$'s inside the brackets:
\begin{equation}
    Y_{i} = Y_{i} + \gamma(AX_{j}-AX_{k}) =  Y_{i} + \gamma(AX_{j}+b-AX_{k}-b) =  Y_{i} + \gamma(Y_{j}-Y_{k}).
\end{equation}
Hence, the differential evolution move is affine-invariant.

\section{Ergodicity metric}

\subsection{Integration of N-dimensional Gaussian landscapes}\label{appendix:NDIM_integration}

To evaluate the likelihood contribution to the total of individual cells within our Gaussian landscapes, we need an integration method capable of operating in \(N\)-dimensions. Fortunately, since our likelihood landscapes are constructed as sums of multivariate Gaussians, we can use an analytical form of integration based on the error function (\(\mathrm{erf}\)). This provides a highly efficient and accurate alternative. In one dimension, the integral of a Gaussian over a finite interval \([a, b]\) is given by:
\begin{equation}\label{eq:erf_integral_1d}
    \int_{a}^{b}{e^{-(x-\mu)^2/2\sigma^{2}}dx} = \frac{\sigma\sqrt{\pi}}{\sqrt{2}}\left[\mathrm{erf}\left(\frac{b-\mu}{\sigma\sqrt{2}}\right) - \mathrm{erf}\left(\frac{a-\mu}{\sigma\sqrt{2}}\right)\right].
\end{equation}
This result extends naturally to the multivariate case due to the separability of Gaussians. The likelihood value, within some hypercube with bounds $\vec{a}$ and $\vec{b}$, for our landscapes is defined as the sum over its \(M\) Gaussian components in \(N\)-dimensions:
\begin{multline}\label{eq:integration_gaussian_landscape}
\int_{\vec{a}}^{\vec{b}} \mathcal{L}\; dx_1 dx_2 \dots dx_N 
= \int_{\vec{a}}^{\vec{b}} \sum_{i=1}^{M} A_i 
  \Biggl( \prod_{d=1}^{N} e^{-(x_{i,d}-\mu_{i,d})^2 / 2\sigma_{i,d}^2} \Biggr) d\vec{x} \\ 
= \sum_{i=1}^{M} A_i 
  \Biggl( \prod_{d=1}^{N} \int_{a_d}^{b_d} e^{-(x_{i,d}-\mu_{i,d})^2 / 2\sigma_{i,d}^2} dx_d \Biggr) \\ 
= \sum_{i=1}^{M} A_i 
  \Biggl[ \prod_{d=1}^{N} \frac{\sigma_{i,d}\sqrt{\pi}}{\sqrt{2}} 
  \times \Bigl( \mathrm{erf}\Big(\frac{b_d-\mu_{i,d}}{\sigma_{i,d}\sqrt{2}}\Big) 
  - \mathrm{erf}\Big(\frac{a_d-\mu_{i,d}}{\sigma_{i,d}\sqrt{2}}\Big) \Bigr) \Biggr].
\end{multline}
To integrate using the error function, we use the \texttt{scipy.special.erf} function from the \texttt{SciPy} library\footnote{\href{https://docs.scipy.org/doc/scipy/reference/generated/scipy.special.erf.html}{https://docs.scipy.org/doc/scipy/reference/generated/scipy.special.erf.html}}.

\subsection{Hypercube division}\label{appendix:hypercube_subdivison}

 Each Gaussian landscape can be defined over a bounded domain represented by a hypercube, where the bounds on each axis correspond to the parameter ranges listed in tables~\ref{tab:landscape_3D}, \ref{tab:landscape_5D}, and \ref{tab:landscape_8D} for the 3D, 5D, and 8D models, respectively. To obtain cells of a hypercube, we divide each axis into equally spaced segments, which define the boundaries of each cell. This ensures that the entire hypercube is divided into cells of equal volume. The number of divisions per axis is defined dynamically:
\begin{equation}
    \text{divisions per axis} = \min\left(5 + 8 \cdot \exp(-0.4(N - 3)), \; 5\right),
\end{equation}
where \(N\) is the dimensionality of the landscape. This choice was made to ensure a smooth transition across dimensionalities while capping the maximum number of divisions per axis at 5 for higher-dimensional cases (\(N > 13\)) to preserve numerical performance. For reference, the resulting numbers of divisions per axis are 13, 8, and 6 for the 3D, 5D, and 8D landscapes, respectively. We define the total number of cells that compose the hypercube by:
\begin{equation}
    \text{number of cells} = (\text{divisions per axis})^{N}.
\end{equation}
To then compute the likelihood integral for any cell, we must also determine its boundaries. For this, we use an \(N\)-dimensional indexing scheme that maps a linear index \(i\), ranging from 0 to the total number of cells minus one, to its corresponding coordinates in the \(N\)-dimensional grid. To ensure uniform treatment of all dimensions, the axes are rescaled to the interval of \([0, 1]\), which effectively transforms each cell into a mini-hypercube in normalised space. The mapping from a flat index \(i\) to multidimensional indices \((x_1, x_2, \dots, x_N)\) follows a generalised form of low-dimensional modular arithmetic, generalised to any dimensionality \(N\) as:
\begin{equation}\label{eq:NdimIndexing}
    x_k = \left( i \;\text{mod} \; D^k \right) // D^{k-1}, \quad \text{for } k = 1, 2, \dots, N,
\end{equation}
where \(D\) denotes the number of divisions per axis. This indexing ensures that we can correctly identify the location and extent of each cell in parameter space based on its flat index \(i\). 


\bsp	
\label{lastpage}
\end{document}